\documentclass[traditabstract]{aa}  
\usepackage{natbib}

\usepackage{mathrsfs}

\usepackage{amsmath}

\usepackage{txfonts}

\usepackage{array}

\usepackage{color}

\usepackage{graphicx}

\usepackage{rotating}

\usepackage{lscape}

\usepackage{multirow}

\usepackage{blindtext}

\usepackage[table]{xcolor} 

\usepackage[switch]{lineno}

\def\lsimeq
{\hbox{\raise0.5ex\hbox{$<\lower1.06ex\hbox{$\kern-1.07em{\sim}$}$}}}

\begin{document}
\title{Young and middle age pulsar light-curve morphology: Comparison of \emph{Fermi} observations with $\gamma$-ray and radio emission geometries}
\author{M. Pierbattista\inst{1,2} 
\and A. K. Harding\inst{3} 
\and P. L. Gonthier\inst{4}
\and I. A. Grenier\inst{5,6} 
}

\institute{ 
	{Nicolaus Copernicus Astronomical Center, Rabia\'nska 8, PL-87-100 Toru\'n, Poland; \email{mpierba@gmail.com}}
\and{INAF-Istituto di Astrofisica Spaziale e Fisica Cosmica, 20133 Milano, Italy}
\and {Astrophysics Science Division, NASA Goddard Space Flight Center, Greenbelt, MD 20771, U.S.A.}
\and {Hope College, Department of Physics, Holland MI, U.S.A. }
\and {Laboratoire AIM, Universit\'e Paris Diderot/CEA-IRFU/CNRS, Service d'Astrophysique, CEA Saclay, 91191 Gif sur Yvette, France}
\and {Institut Universitaire de France}
}
\date{}

\abstract
{Thanks to the huge amount of $\gamma$-ray pulsar photons collected by the \emph{Fermi} 
Large Area Telescope since its launch in June 2008, it is now possible to constrain 
$\gamma$-ray geometrical models by comparing simulated and observed light-curve 
morphological characteristics. We assumed vacuum-retarded dipole (VRD) pulsar magnetic field and tested 
simulated and observed morphological light-curve characteristics in the framework 
of two pole emission geometries, Polar Cap (PC) and Slot Gap (SG), and one pole 
emission geometries, traditional Outer Gap (OG) and One Pole Caustic (OPC). Radio core plus 
cone emission was assumed for the pulsar of the simulated sample. We compared simulated 
and observed recurrence of class shapes and peak multiplicity, peak separation, 
radio-lag distributions, and trends of peak separation and radio lag as a function 
of observable and non-observable pulsar parameters. We studied how pulsar 
morphological characteristics change in multi-dimensional observable and 
non-observable pulsar parameter space.

The PC model gives the poorest description of the LAT pulsar light-curve morphology.
The OPC  best explains both the observed $\gamma$-ray peak multiplicity and shape
classes. The OPC and SG models describe 
the observed $\gamma$-ray peak-separation distribution for low- and high-peak 
separations, respectively. This suggests that the OPC geometry best explains the 
single-peak structure but does not manage to describe the widely separated peaks 
predicted in the framework of the SG model as the emission from the two magnetic 
hemispheres. The OPC radio-lag distribution shows higher agreement with 
observations suggesting that assuming polar radio emission, the $\gamma$-ray 
emission regions are likely to be located in the outer magnetosphere. 
Alternatively, the radio emission altitude could be higher that we assumed.
We compared simulated non-observable parameters with the same parameters 
estimated for LAT pulsars in the framework of the same models. The larger 
agreement between simulated and LAT estimations in the framework of the OPC
suggests that the OPC model best predicts the observed variety of profile shapes.
The larger agreement obtained between observations and the OPC model predictions 
jointly with the need to explain the abundant 0.5 separated peaks with two-pole 
emission geometries, calls for thin OPC gaps to explain the single-peak geometry 
but highlights the need of two-pole caustic emission geometry to explain widely separated
peaks.}

\authorrunning{Pierbattista et al. 2016}
\titlerunning{Young and middle age pulsar light-curve morphology}
\maketitle

\section{Introduction}
\begin{table*}
\centering
\begin{tabular}{| c || c | c | c | c | c | }
\hline
 		& PC & SG  & OG & OPC & Radio \\ 
\hline
\hline
$\sigma_{\nu}$ 		& 0.005 	& 0.9		& 0.2		& 0.2		& 0.2		\\
\hline
$\delta\sigma_{\nu}$ & 0.001	& 0.1		& 0.01	& 0.01	& 0.01	\\
\hline
$Th_0$ 			& 0.02	& 0.05	& 0.03	& 0.03	& 0.04	\\
\hline
$Th_{mx-mn}$ 		& 0.001	& 0.15	& 0.03	& 0.03	& 0.01	\\
 \hline
$W_H$ 			& $W(0.50(\max-\min))$ 	& $W(\min + 0.50(\max-\min))$ 	& $W(0.50(\max-\min))$ 	& $W(0.50(\max-\min))$ 	& $W(0.50(\max-\min))$ 	\\
\hline
$W_B$ 			& $W(0)$			 	& $W(\min + 0.15(\max-\min))$  &  $W(0)$			 	& $W(0)$			 	& $W(0)$			 	 \\
 \hline
\end{tabular}
\caption{Values adopted in each model for the light-curve smoothing $\sigma_{\nu}$ and $\delta\sigma_{\nu}$, the 
zero threshold $Th_0$ and maximum-minimum detection threshold $Th_{mx-mn}$, and the peak widths, 
$W_H$ and $W_B$ for modelled $\gamma$-ray and radio light curves. The parameters $W_H$ and $W_B$ are expressed 
as the width $W$ of the peak measured at a certain fraction $x$ of the peak maximum, $W(x)$. The terms ``$\min$'' 
and ``$\max$'' refer to light-curve minimum and maximum, respectively. All values are given in phase units.}
\label{TabThres}
\end{table*}

The successful launch of the \emph{Fermi Gamma-ray space telescope} satellite (hereafter \emph{Fermi}) in June 
2008 represents a milestone in pulsar astrophysics research. Owing to the observations performed with its main 
instrument, the \emph{Large Area Telescope} (LAT), \emph{Fermi} increased the number of known young or 
middle-aged $\gamma$-rays pulsars by a factor $\sim$30, discovered the new category of $\gamma$-ray millisecond 
pulsars, and obtained unprecedented high quality light curves for the major part of the observed objects.

The increasing number of $\gamma$-ray pulsars detected by \emph{Fermi}, which is now second just to the number 
of pulsars detected at radio wavelengths, offered the possibility, for the first time with high statistics, of studying the 
collective properties of the $\gamma$-ray pulsar population and to compare them with model predictions (e.g.
pulsar population syntheses; \cite{wr11,twc11}; light-curve fitting and morphology analysis; \cite{wrwj09,rw10}).
\cite{pie10} performed a comprehensive study of the \emph{Fermi} $\gamma$-ray pulsars detected by LAT during 
the first year of observations 
\citep[The first Fermi large-area telescope catalogue of $\gamma$-ray pulsars,][ hereafter PSRCAT1]{aaa+10}.
\cite{pie10} tested radiative and geometrical models against the observations according to three different approaches: 
a neutron star (NS) population synthesis to compare collective radiative properties of observed and simulated pulsar 
populations; a simulation of the observed pulsar emission patterns to estimate non-observable pulsar parameters and 
to study their relationship with observable pulsar characteristics in the framework of different emission geometries; 
and a comparison of simulated and observed  light-curve morphological characteristics in the framework of different 
emission geometries. 
The population synthesis and the estimate of non-observable LAT pulsar parameters have been presented in
Pierbattista et al. (2012; hereafter PIERBA12)\nocite{pghg12} and by Pierbattista et al. (2015; hereafter PIERBA15) \nocite{phg+15}, 
respectively.

PIERBA12\nocite{pghg12} synthesised a NS population, evolved this population in the galactic gravitational potential assuming a 
supernova kick velocity and birth space distribution, and computed, for each NS of the sample, 
$\gamma$-ray and radio radiation powers and light curves according to radiative $\gamma$-ray models and a radio 
model. The implemented radiative $\gamma$-ray models are: the
Polar Cap model \citep[PC;][]{mh03}, Slot Gap model \citep[SG;][]{mh04a}, Outer Gap model \citep[OG;][]{crz00}, 
and the One Pole Caustic model \citep[OPC;][]{rw10,wrwj09}, which is an alternative formulation of the OG that only differs 
in the computation of radiative power and width of the emission region. 
In the OPC model, both gap width and gap-width luminosity have been set to reproduce the relation 
$L_\gamma \propto \dot{E}^{0.5}$ observed in PSRCAT1\nocite{aaa+10}; since the gap width is assumed equal to the 
$\gamma$-ray efficiency, $w_\mathrm{OPC} \propto \dot{E}^{-0.5}$, the luminosity must follow $L_\gamma = \dot{E} w_\mathrm{OPC}$. 
Because of that, the OPC cannot be considered a full physical model like PC, SG, and OG, but a phenomenological 
formulation of the OG, built in to reproduce the observations. Each pulsar radio luminosity
has been computed according to a radio core plus cone model, following the prescriptions from 
\cite{gvh04}, \cite{sgh07}, and \cite{hgg07}. Simulation details can be found in PIERBA12\nocite{pghg12}. 
The $\gamma$-ray and radio light curves of the simulated pulsar have been obtained according to the 
geometrical model from 
\cite{dhr04}. Under the assumption that $\gamma$-ray and radio photons are emitted tangent to the 
poloidal magnetic field line in the reference frame instantaneously corotating with the pulsar, $\gamma$-ray and 
radio light curves have been computed by assuming location and size of the emission region in the framework of 
each $\gamma$-ray model and radio model. The directions of the photons generated at the emission point have
been computed as described in \cite{bs10}. The complete description of the method used to simulate $\gamma$-ray 
and radio light curves of the observed pulsars can be found in PIERBA15\nocite{phg+15}.

\cite{phg+15} simulated the emission pattern of the young or middle-aged LAT $\gamma$-ray pulsars 
detected after three years of observations and published in the second LAT $\gamma$-ray pulsars catalogue 
\citep[][ hereafter PSRCAT2]{aaa+13}, and 
used them to fit the observed profiles. The LAT pulsar emission patterns were computed in the framework
of the very same radiative models implemented by PIERBA12\nocite{pghg12}, namely PC, SG, OG, OPC, and radio 
core plus cone, and by assuming emission geometry according to \cite{dhr04}. For each LAT pulsar and each model, 
$\gamma$-ray and radio light curves were computed for a grid of values of magnetic obliquity $\alpha$ (angle between
the pulsar magnetic and rotational axes) and observer line-of-sight $\zeta$ (angle between the rotational axis and the 
direction of the observer), both stepped every 5$^\circ$ in the interval 1 to 90, for the actual pulsar spin period (P), and 
for the magnetic field (B) and width of the emission gap region (W) computed as described in PIERBA12\nocite{pghg12}. 
A best-fit light curve with relative best-fit parameters, $\alpha$, $\zeta$, and W, were found in the framework of each
model and for each pulsar by fitting the observed light curves with the emission patterns through a maximum likelihood 
criterion.

The aim of this paper is to further develop and improve the light-curve morphological analysis implemented by \cite{pie10} 
by testing the PC, SG, OG, OPC, and radio core plus cone emission geometries against the young and middle-aged pulsar 
sample published in PSRCAT2\nocite{aaa+13}. We computed 
a series of light-curve morphological characteristics, namely $\gamma$-ray and radio peak numbers, their phase separation, 
the lag between the occurrence of $\gamma$-ray and radio peaks in radio loud (RL) pulsars, among others, for 
both observed and simulated light curves and within each model. The simulated pulsar light curves are those synthesised by
PIERBA12\nocite{pghg12} in the framework of PC, SG, OG, OPC, and radio core plus cone models by assuming emission 
geometry from \cite{dhr04}. We build a light-curve shape classifications according to 
the recurrence of the morphological characteristics in observed and simulated profiles and compared simulated and observed 
shape classifications in the framework of each model. We compared observed and simulated distributions of morphological 
characteristics as a function of observable and non-observable pulsar parameters.
The non-observable pulsar parameters, namely $\alpha$, $\zeta$, and $\gamma$-ray beaming factor $f_{\Omega}$, are 
those estimated by PIERBA15\nocite{phg+15} in the framework of the implemented emission geometries. We obtained 
constraints on the emission geometry that best explains the observations.

The outline of this paper is as follows. In Section \ref{Data selection} we characterise both simulated and observed 
$\gamma$-ray pulsar samples. In Section \ref{Light curves shape classification} we describe the method used to classify
simulated and observed pulsars and compare simulated and observed $\gamma$-ray and the peak multiplicity of radio light
curves. In Section \ref{Determination peaks} we describe the criteria used to measure the light-curve morphological
characteristics for both observed and simulated objects and compare observed and simulated distributions and trends
of observable and non-observable/estimated pulsar parameters. 
Section \ref{AZpl} shows
the $\gamma$-ray peak multiplicity and peak separation in the $(\alpha,\zeta)$ plane for both simulated and observed
pulsars (estimated values). Section \ref{summary} summarises the results.

In Appendix \ref{ShapeCLS} templates of the  $\gamma$-ray and radio shape classes defined in Section 
\ref{Light curves shape classification} are shown in the framework of each model. In Appendix \ref{PSREmispat}
the pulsar $\gamma$-ray and radio emission patterns are shown to ease the interpretation of the results. In Appendix
\ref{AppC} we compare the recurrence of simulated and observed $\gamma$-ray light-curve multiplicity 
for radio quiet and radio loud pulsars and of simulated and observed $\gamma$-ray and radio shape classes.
In Appendix \ref{AppD} we give exhaustive maps showing the $\gamma$-ray peak multiplicity and $\gamma$-ray 
peak separation as a function of the spin period (PC) and width of the acceleration gap (SG, OG, and OPC) in Figure 
\ref{MultiAZ} and \ref{SepAZ}, respectively. Appendix \ref{ksapp} describe the one-and two-dimensional Kolmogorov-Smirnov
(KS) tests used to quantify the agreement between observed and simulated distributions.

\section{Data selection and simulated sample}
\label{Data selection}

\begin{figure}
\begin{center}
\includegraphics[width=0.263\textwidth]{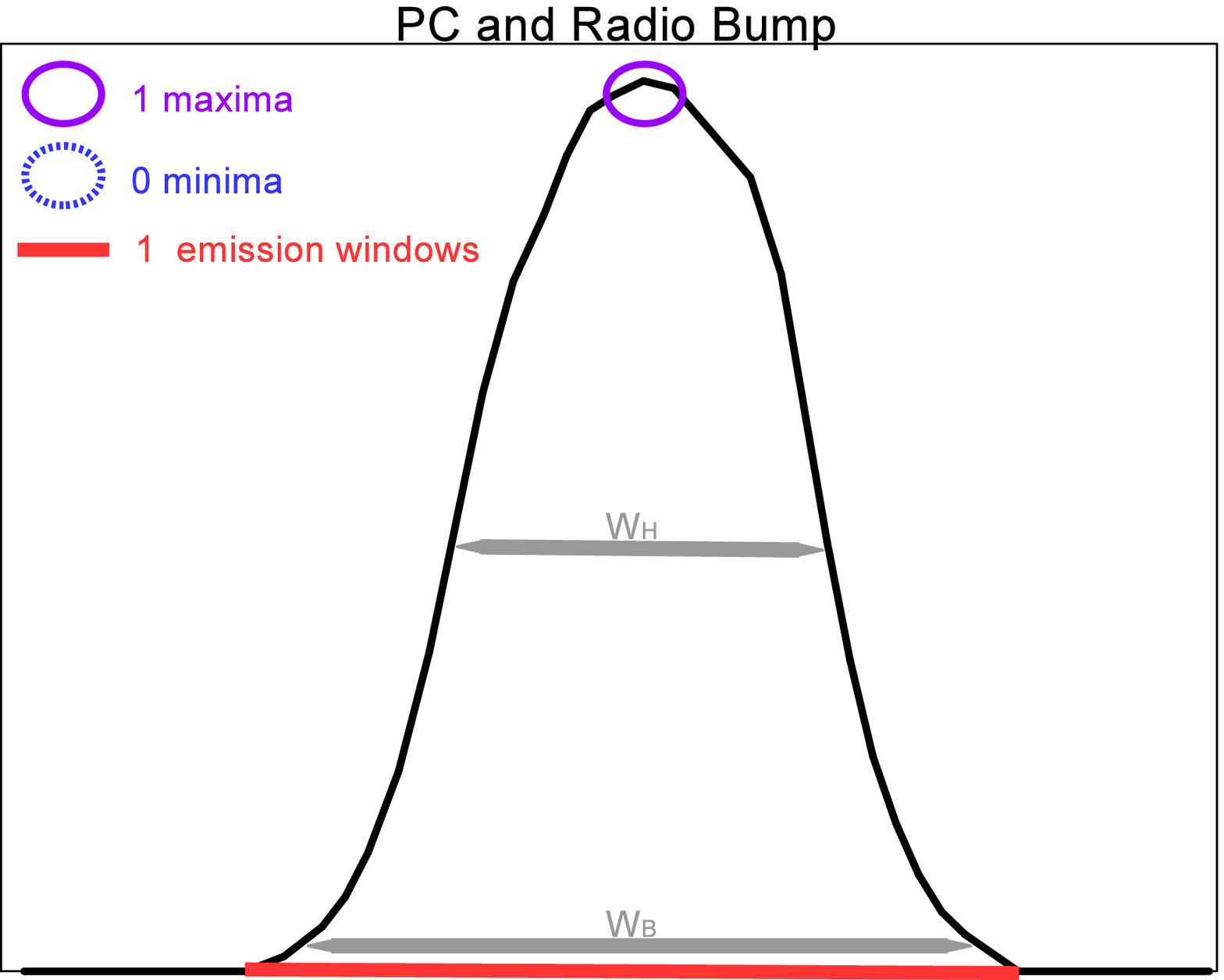} 
\includegraphics[width=0.263\textwidth]{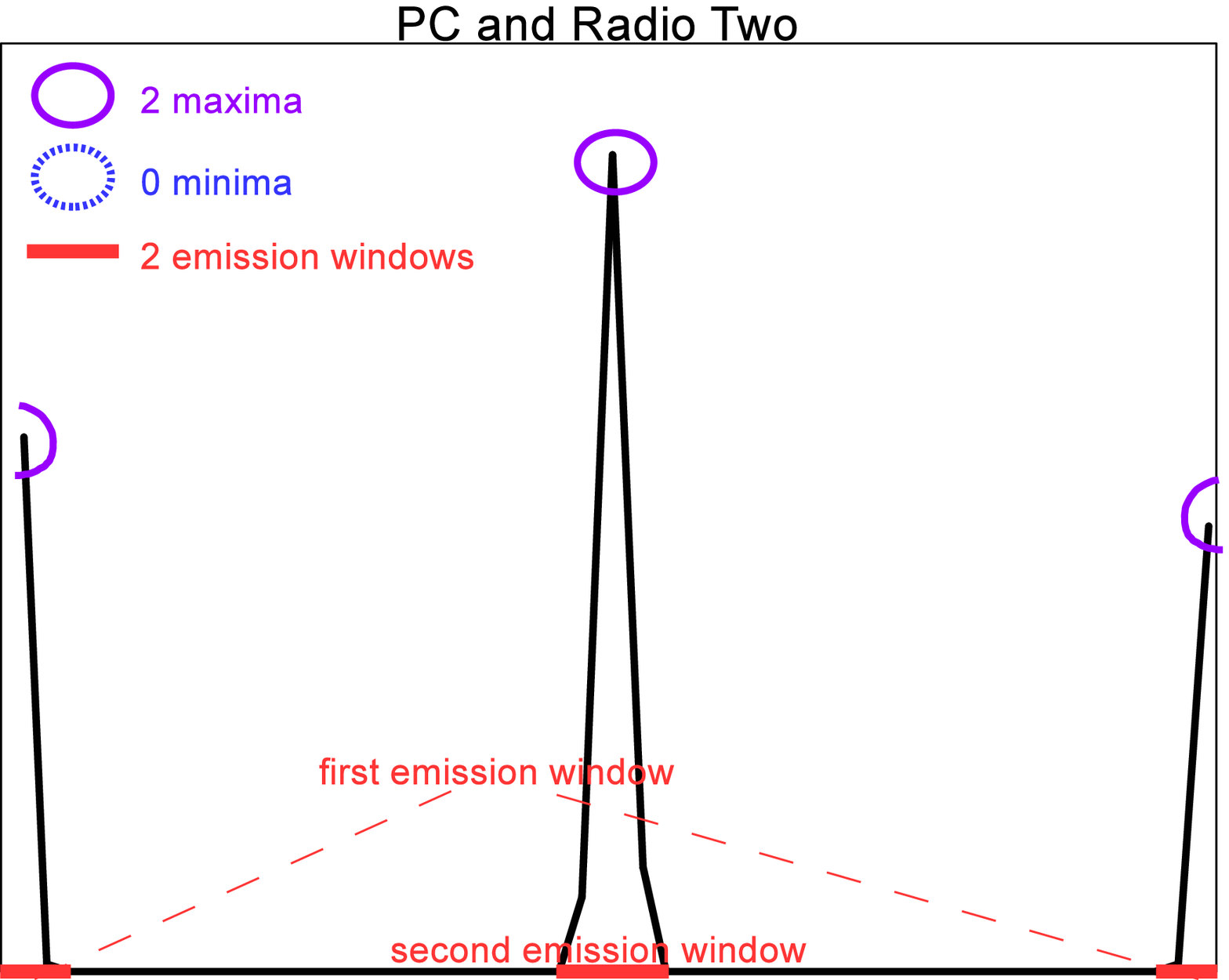} 
\includegraphics[width=0.263\textwidth]{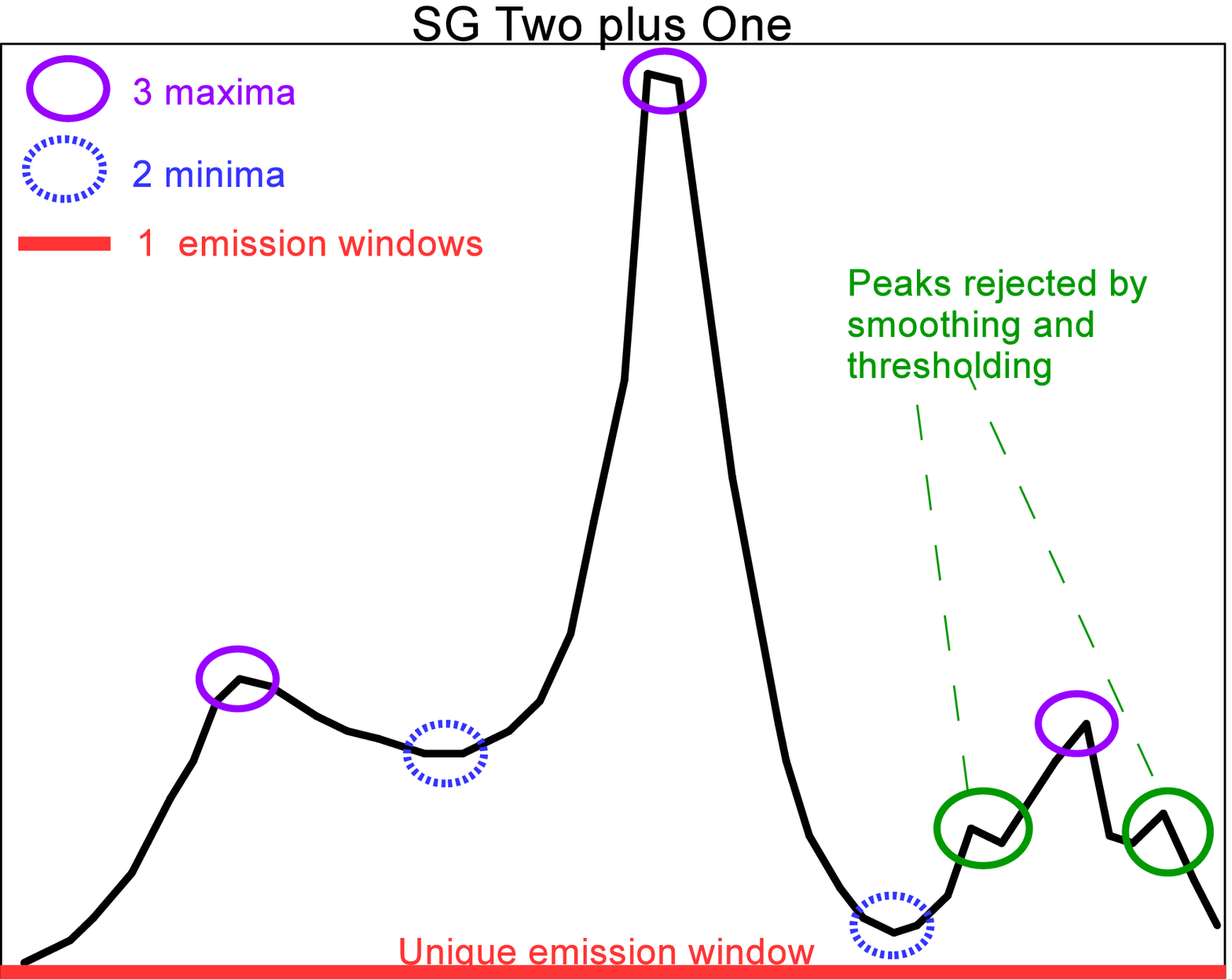} 
\includegraphics[width=0.263\textwidth]{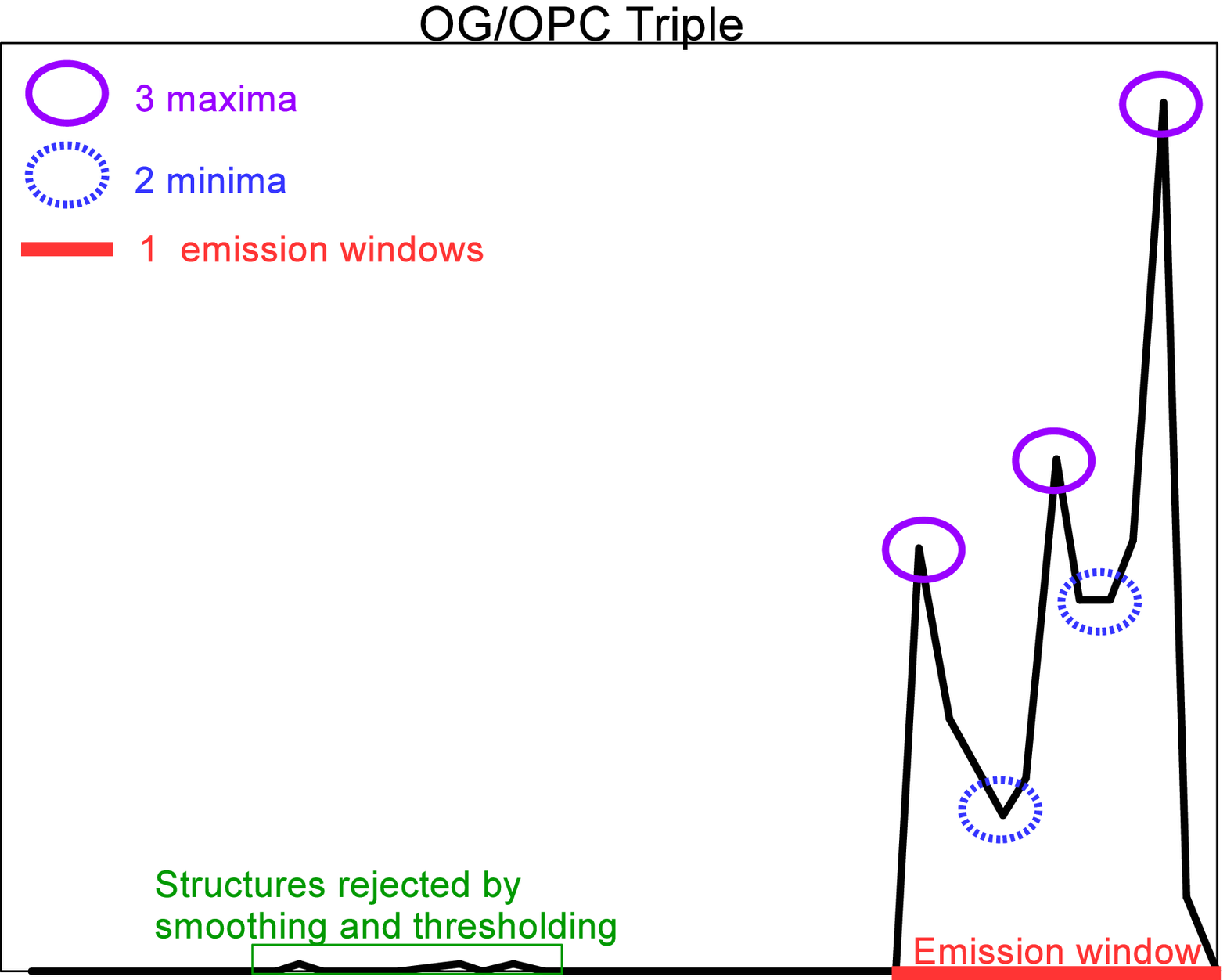} 
\caption{Examples of light-curve classification criteria used to classify the simulated  
	     light curves. From top to bottom are shown: PC or radio bump light curve; PC or radio
	     two peaks light curve; SG double plus single peak light curve; OG or OPC triple peak light
	     curve.}
\label{ClassEx1}
\end{center}
\end{figure}

We have analysed 77 young or middle-aged pulsars listed in PSRCAT2\nocite{aaa+13}. We obtained their $\gamma$-ray 
and radio light curves from the LAT catalogue data products public 
link\footnote{http://fermi.gsfc.nasa.gov/ssc/data/access/lat/2nd\_PSR\_catalog/}. 
We built each \emph{Fermi} $\gamma$-ray pulsar light curve with 3 years of LAT observations   
from 2008 August 4 to 2011 August 4 selecting only photons with energy larger than 100 MeV and belonging 
to the source class, as defined in the P7\_V6 instrument response function. We obtained the \emph{Fermi} 
$\gamma$-ray light curves as weighted light curves where each photon is characterised 
by its probability to belong to the pulsar with the photon weights following the Gaussian probability distribution 
(PSRCAT2\nocite{aaa+13}). We built the radio light curves of the analysed RL pulsars from observations 
mainly performed at 1400 MHz by the radio telescopes operating within the \emph{Fermi} Pulsar Search 
Consortium  \cite[PSC; ][]{rap+12} and the \emph{Fermi} Pulsar Timing Consortium \cite[PTC; ][]{sgc+08}, namely 
the Green Bank Telescope (GBT), Parkes Telescope, Nan\c cay Radio Telescope  (NRT), Arecibo Telescope, 
the Lovell Telescope at Jodrell Bank, and the Westerbork Synthesis Radio Telescope  \citep{sgc+08}. 
See PSRCAT2\nocite{aaa+13} for details about how the $\gamma$-ray and radio light curves were obtained.

The simulated radio and $\gamma$-ray light curves used in this paper are those obtained within the population 
synthesis  from PIERBA12\nocite{pghg12}. Those authors classified the simulated NSs as radio 
quiet  (RQ) or radio loud (RL) $\gamma$-ray pulsars according to emission-geometry visibility criteria ($\gamma$-ray and radio 
beam orientations) and according to the $\gamma$-ray flux and radio flux sensitivity. The emission-geometry 
visibility criteria classify a NS as a RL or RQ pulsar when the observer line of sight simultaneously intersects
$\gamma$-ray and radio emission beams or just the $\gamma$-ray emission beam, respectively. The $\gamma$-ray and 
radio flux sensitivity criteria adopted in PIERBA12\nocite{pghg12}  are scaled up to three years from one year of $\gamma$-ray LAT 
sensitivity and on the sensitivity of ten radio surveys. Different $\gamma$-ray sensitivities were used for pulsars discovered 
through blind search and for pulsar detected using a timing solution known from radio observations. 
The radio sensitivity for RL LAT pulsars has been obtained by recomputing the survey parameters of the ten radio surveys
covering the largest possible sky surface and for which the survey parameters were known with high accuracy. Details 
about $\gamma$-ray and radio visibility computation can be found in PIERBA12\nocite{pghg12}. 

In this paper we make use of the three-year LAT $\gamma$-ray pulsar sensitivity and on 
updated radio observations of the radio loud LAT pulsars. Since the morphological classification of the simulated light 
curves does not show significant variations between the whole simulated population and its visible subsample (as we 
show in section \ref{Shape classification and multiplicity}), we compare the LAT morphological characteristics
with the morphological characteristics of all simulated pulsars classified as visible according to emission geometry 
visibility criteria only.

\section{Shape classification of observed and simulated light curves}
\label{Light curves shape classification}

\subsection{Method}
\begin{table*}
\centering
\begin{tabular}{| l || c | c | c | c | }
\hline
 							& \multicolumn{2}{c|}{PC}									& \multicolumn{2}{c|}{SG} \\
\hline
\hline
\multicolumn{1}{|c||}{Shape Class}	& $(N_{E-W},Pk,Mn)$	& Peak Width Condition 				& $(N_{E-W},Pk,Mn)$	& Peak Width Condition 				\\
\hline
\hline
1- Bump 						& (1,1,0) 				& $W_B\ge0.40$ \& $HL_{R}\ge0.50$	& (1,1,0/1/2) 			&  $W_B\ge0.35$ \& $HL_{R}\ge0.50$  	\\
\hline
2- Sharp 						& (1,1,0) 				& $W_B<0.40$ \& $HL_{R}<0.50$		& (1,1,0/1/2) 			&  $W_B<0.35$  					\\
\hline
3- Shoulder 					& none				& /								& (1,1,0/1/2) 			& $W_B\ge0.35$ \& $HL_{R}\ge0.42$  	\\
\hline
4- Two 						&(1,2,0) 				& / 								& none				& / 								\\
\hline
5- Double 						& (1,2,1) 				& / 								& (1,2,2/3) 			& / 								\\
\hline
6- Double+Single				& (2,3,1) 				& / 								& (1,2,3) 				& / 								\\
\hline
7- Triple/Three 					& none				& / 								& none				& / 								\\
\hline
8- Two double 					& (2,4,2) 				& / 								& (1,4,4) 				& / 								\\
\hline
\end{tabular}
\

\

\begin{tabular}{| l || c | c | c | c |}
\hline
 							& \multicolumn{2}{c|}{OG/OPC}								& \multicolumn{2}{c|}{$\gamma$-ray LAT} \\
\hline
\hline
\multicolumn{1}{|c||}{Shape Class}	& $(N_{E-W},Pk,Mn)$	& Peak Width Condition 				& $(N_{E-W},Pk,Mn)$	&Peak Width Condition 			 	\\
\hline
\hline
1- Bump 						& (1,1,0) 				&  $W_B\ge0.20$ \& $HL_{R}\ge0.45$  	& (1,1,x)				&  $W_B\ge0.25$ \& $HL_{R}\ge0.40$    	\\
\hline
2- Sharp 						& (1,1,0) 				&  $W_B<0.10$ \& $HL_{R}<0.45$  		& (1,1,x) 				&  $W_B\ge0.25$ \& $HL_{R}\ge0.40$	\\
\hline
3- Shoulder 					& (1,1,0)  				&  $W_B\ge0.10$ \& $HL_{R}<0.45$ 	& (1,1,x) 				&  $W_B\ge0.25$ \& $HL_{R}<0.40$		\\
\hline
4- Two 						& (2,2,0) 				& / 								& (2,2,x)				&  / 		 						\\
\hline
5- Double 						& (1,2,0/1/2) 			& / 								& (1,2,x)				&  / 								\\
\hline
6- Double+Single				& none				& / 								& (2,3,x)				&  / 		 						\\
\hline
7- Triple 						& (1,3,1/2) 			& / 								& (1/3,3,x) 			&  / 								\\
\hline
8- Two double 					& (1,4,3) 				& / 								& (x,4,x)				&  / 								\\
\hline
\end{tabular}
\caption{Assignation of the $\gamma$-ray light-curve shape classes according to a different set of light-curve 
morphological characteristics for PC and SG simulated profiles (top table) and OG/OPC simulated profiles and 
$\gamma$-ray LAT  profiles (bottom table). `None' in correspondence of a shape class for a particular model,
indicates that this light-curve shape is not observed in the framework of this model. The `Peak Width Condition'
was uniquely used to discriminate between `Bump' and `Sharp' single-peak light curves. All values are given 
in phase units. Light-curve shape templates for the listed shape classes are shown in Table \ref{TabShapeR}.}
\label{TabAssG}
\end{table*}

We defined a number of light-curve morphological characteristics and assigned shape classes to both simulated 
and observed light curves according to the recurrence of those characteristics in the analysed profiles. They are:
\begin{itemize}
\item[1-] The number of light-curve phase windows with non-zero emission, $N_{E-W}$,  defined as the number 
of contiguous phase intervals in the analysed light curves that show non-zero emission (see the second and fourth 
panel of Figure \ref{ClassEx1} for examples of two and one non-zero emission windows, respectively);
\item[2-] The number of light-curve peaks, $Pk$ (defined as described in Section \ref{latlc} and \ref{simlc});
\item[3-] The number light-curve minima, $Mn$ (defined as described in Section \ref{latlc} and \ref{simlc});
\item[4-] The peak full width half maximum, $W_{H}$, and the width at the base of the peak, $W_B$, for single 
peak light curves, both expressed in pulse-phase units. Because of the differences in the off-peak emission predicted 
by the different models and in the observed light curves, both  $W_{H}$ and $W_B$ values have been optimised 
in the framework of each model and for observed profiles. The adopted values for $\gamma$-ray 
and radio light curves, both  simulated and observed, are given in Table \ref{TabThres}. An example of $W_{H}$ 
and $W_B$ is shown in the top panel of Figure~\ref{ClassEx1}. 

We used $W_{H}$ and $W_B$ to define a high-to-low peak width ratio, $HL_R=W_{H}/W_B$. $HL_R$ close to 0 
indicates a peak that gets rapidly pointed while $HL_R$ close to 1
indicates a peak with a more constant width along its vertical extension. The parameter $HL_R$ is used for the classification 
of the different kinds of single-peak light curves.
\end{itemize}
Examples of the light-curve maxima, minima, single-peak widths, and non-zero emission windows are shown, for 
each model, in Figure \ref{ClassEx1} while the shape classes used to classify $\gamma$-ray and radio light curves
are shown in Tables \ref{TabShapeG} and \ref{TabShapeR}, respectively. The listed morphological characteristics 
were computed for observed and simulated light curves according to different criteria.

\subsubsection{LAT $\gamma$-ray and radio light curves} 
\label{latlc}
\begin{figure}
\begin{center}
\includegraphics[width=0.49\textwidth]{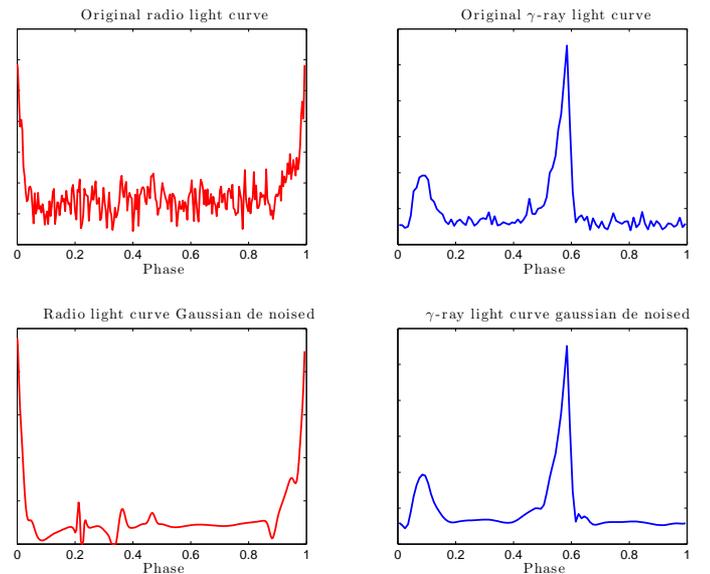} 
\caption{Examples of the wavelet Gaussian de-noising method used to classify the LAT pulsar light curves. The de-noising 
of radio and $\gamma$-ray light curves of pulsar J0205$-$6449 are shown on the left (red) and right (blue) 
panels, respectively.}
\label{deNoiseEx}
\end{center}
\end{figure}

The $\gamma$-ray and radio observed light curves are affected by observational noise that does not allow a clear identification
of the light-curve shape morphology and must be removed before the analysis. Both $\gamma$-ray and radio light curves of 
LAT pulsars have been de-noised using a wavelet transform with an iterative multi-scale thresholding algorithm and assuming 
Gaussian noise \citep{spr06}. We chose Gaussian noise because the statistics of the weighted $\gamma$-ray photons follows 
a Gaussian distribution (PSRCAT2\nocite{aaa+13}) and the radio light curves are well described by Gaussian statistics. 
Examples of the wavelet de-noising of $\gamma$-ray and radio LAT light curves for pulsar J0205+6449 are shown in Figure 
\ref{deNoiseEx} left (red) and right (blue) panels, respectively. 

The $\gamma$-ray light curve of LAT pulsars have been classified by adopting the very same peak multiplicity assigned by 
PSRCAT2\nocite{aaa+13} while the radio light curves of LAT pulsars have been classified on the basis of the analysis of the 
de-noised light curves. The de-noised $\gamma$-ray or radio light curves also allow us to highlight the presence of an eventual 
emission bridge between the peaks, which are fundamental for the peak-separation computation (see Section \ref{Determination peaks}).

\subsubsection{Simulated $\gamma$-ray and radio light curves}
\label{simlc}
The simulated light curves are characterised by computational fluctuations that affect the light-curve shape as 
real noise. The light curves of simulated pulsars were computed using the geometrical model from \cite{dhr04}. 
In the algorithm that implements the \cite{dhr04} model, the number of magnetic field lines where particles are 
accelerated and $\gamma$-rays are produced is an input parameter; a large magnetic field-line number yields  
smooth light curves with low computational fluctuations in a computational time that is
potentially prohibitive, while a low magnetic field-line density speeds up the computation but generates
light curves with a critical computational fluctuations. We chose a magnetic field line number optimised to have a 
reasonable computational time and moderated computational fluctuations and we reduced the computational 
fluctuations by smoothing all simulated light 
curves using a Gaussian filter. The smoothing consisted in computing  the fast Fourier transform (FFT) of a 
Gaussian function with standard deviation $\sigma_{\nu}$ and of the simulated light curve, convolving the two FFTs, 
and evaluating the inverse FFT of the convolved function. The result is a light curve with all fluctuations of frequency 
$\sigma_{\nu}$ smoothed. Since each model shows different computational noise, $\sigma_{\nu}$ was optimised 
in the framework of each model and for $\gamma$-ray and radio observed profiles. They are listed in Table \ref{TabThres}.
\begin{table}
\centering
\begin{tabular}{| l || c | c |}
\hline
 							& \multicolumn{2}{c|}{Radio Core plus cone / Radio LAT} 	\\
\hline
\hline
\multicolumn{1}{|c||}{Shape Class}	& $(N_{E-W},Pk,Mn)$	&Peak Width Condition 				\\
\hline
\hline
1- Bump 						& (1,1,0/1) 			&  $W_B\ge0.25$ \& $HL_{R}\ge0.50$  	\\
\hline
2- Sharp 						& (1,1,0/1) 			&  $W_B<0.25$ \& $HL_{R}<0.50$  		\\
\hline
3- Two 						& (2,2,0/1)  			& /								\\
\hline
4- Double 						& (1,2,1) 				& / 								\\
\hline
5- Double+Single 				& (2,3,1)	 			& / 								\\
\hline
6- Two double					& (2,4,2)/(4,4,0)			& / 								\\
\hline
7- Triple 						& (1,3,2)  	 			& / 								\\
\hline
8- Three 						& (3,3,0)				& / 								\\
\hline
9- Two triple					& (2,6,4) 				& / 								\\
\hline
\end{tabular}
\caption{Assignation of the radio light-curve shape classes according to a different set of light-curve morphological parameters 
for radio core plus cone simulated light curves and for observed radio light curves. The ``Peak Width Condition'' was uniquely 
used to discriminate between ``Bump'' and ``Sharp'' single-peak light curves. All width values are given in phase units. 
Light-curve shape templates for the listed shape classes are shown in Table \ref{TabShapeR}.}
\label{TabAssR}
\end{table}

The smoothed light curve is then analysed by the analysis script that detects and computes the morphological
light-curve characteristics $N_{E-W}$, $Pk$, $Mn$, $W_{H}$, $W_{B}$, and $HL_R$. In some cases the  
main light-curve peak is comparable to the computational fluctuations and the smoothing procedure could 
modify the peak shape. Moreover, some light curves are characterised by high off-pulse emission that does not allow a clear
identification of the light-curve emission windows $N_{E-W}$ (see Figure \ref{ClassEx1}).
In order to recognise computational noise maxima from light-curve peaks and not account for the off-peak emission 
in the number of emission-windows $N_{E-W}$, we defined two threshold criteria. 
The first criterion consists in placing at zero all light-curve intensities lower than a threshold value $Th_0$, usually few percent 
of the absolute light-curve maximum, which is different for each model. The second criterion consist in defining a maximum-to-minimum 
detection threshold $Th_{mx-mn}$ so that a light-curve maximum is classified as a peak only if the difference between its intensity 
and the intensity of the previous minimum is larger than the threshold value $Th_{mx-mn}$, which is different for each model (Figure \ref{ClassEx1} 
third panel). The definition of $Th_0$ and $Th_{mx-mn}$ for $\gamma$-ray and radio light curves is given in Table~\ref{TabThres}.

The association between the morphological parameters $N_{E-W}$, $Pk$, $Mn$, $W_{H}$, $W_{B}$, and $HL_R$ and the 
$\gamma$-ray and radio shape classes are shown in Tables \ref{TabAssG} and \ref{TabAssR}, respectively while templates 
of the correspondent light-curve shapes in the framework of each model are shown in Tables \ref{TabShapeG} and \ref{TabShapeR}. 
If the number of light-curve phase windows, the number of peaks, and the number of minima, ($N_{E-W}$, $Pk$, $Mn$), detected 
by the analysis script in a light curve do not correspond to any shape class listed in Tables \ref{TabAssG} and \ref{TabAssR}, the 
light curve is tagged as unclassified and the smoothing procedure is repeated by 
increasing $\sigma_{\nu}$ of a $\delta\sigma_{\nu}$. The re-smoothing of the analysed light curve is a procedure that is iterated 
by increasing $\sigma_{\nu}$ of a $\delta\sigma_{\nu}$ at each iteration up to a sure identification of the analysed profile. The 
increasing smoothing factors $\delta\sigma_{\nu}$ were optimised in the framework of each model and are listed in Table~\ref{TabThres}.

We built the $\gamma$-ray and radio light-curve shape classifications described in Tables \ref{TabAssG} and \ref{TabAssR}  
by merging all shape classes obtained in the framework of all observed and simulated $\gamma$-ray light curves and observed 
and simulated radio light curves, respectively. This explains why some shape classes are not observed in the framework of a particular 
model.
 \begin{figure}
\begin{center}
\includegraphics[width=0.49\textwidth]{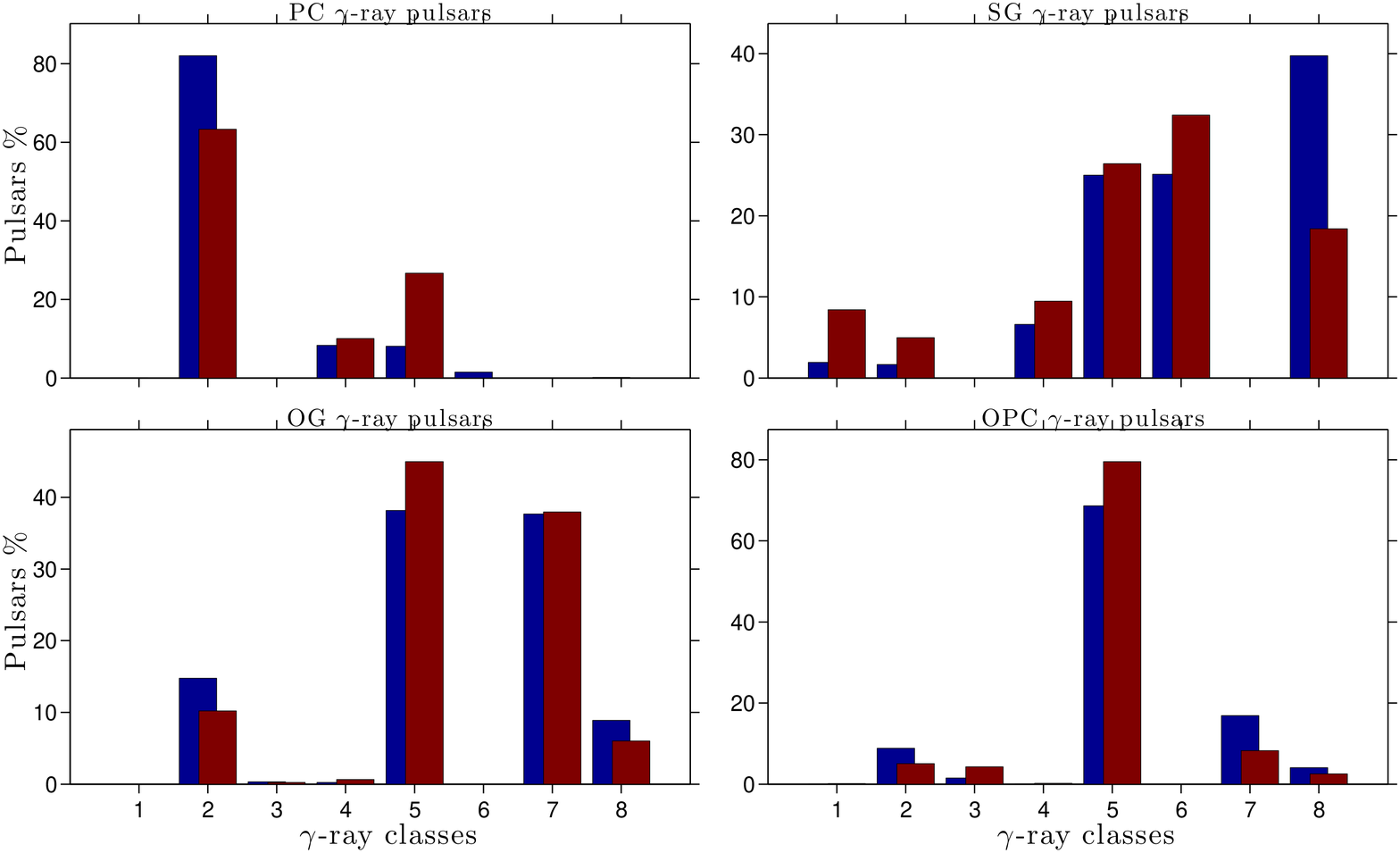} 
\includegraphics[width=0.45\textwidth]{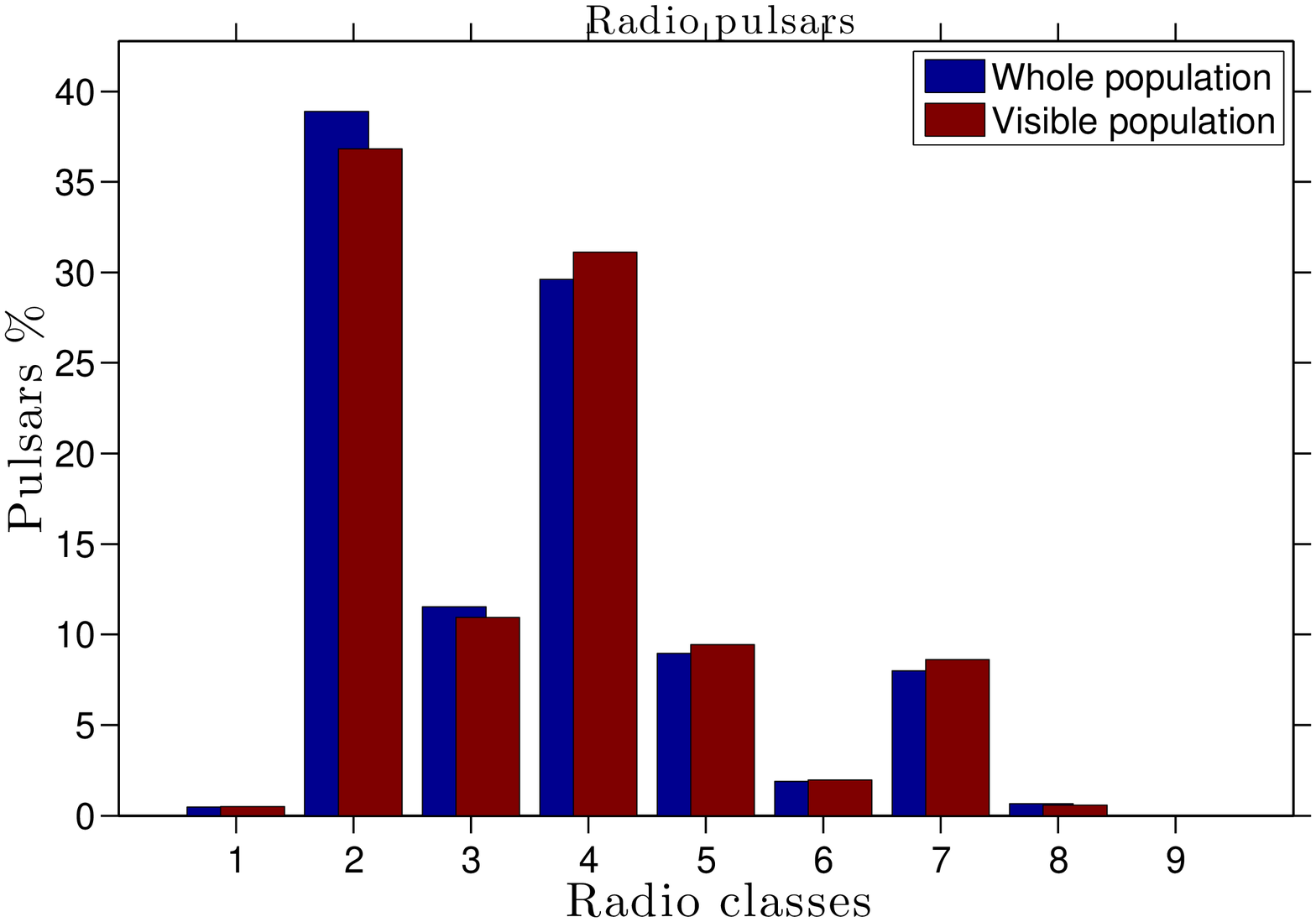}
\caption{Recurrence of the shape classes described in Tables \ref{TabAssG}  and \ref{TabAssR} and shown in Tables \ref{TabShapeG}
and \ref{TabShapeR} for the $\gamma$-ray pulsars (top panels), radio pulsars (bottom panel), and each model for the whole simulated 
pulsar population and its visible subsample, as synthesised by \cite{pghg12}.}
\label{G_AllVisHistoClas}
\end{center}
\end{figure}

\subsection{Shape classification and multiplicity: Comparison of model and observations}
\label{Shape classification and multiplicity}
\begin{table*}
\begin{center}
\begin{tabular}{| l | c  c | c  c | c  c | c  c |}
\hline
& \multicolumn{2}{c|}{PC} & \multicolumn{2}{c|}{SG}& \multicolumn{2}{c|}{OG}& \multicolumn{2}{|c|}{OPC}\\
\hline
& $D$ & $p_\mathrm{value}$ [\%] & $D$ & $p_\mathrm{value}$ [\%] & $D$ & $p_\mathrm{value}$ [\%] & $D$ & $p_\mathrm{value}$ [\%]  \\
\hline
\hline
Fig. \ref{G_VisHistoMulti}: $\gamma$-ray peak multiplicity  			& 0.66 	& 1e-28				& 0.51 	& 7e-16				&  \cellcolor{gray!30}0.32 	& \cellcolor{gray!30}2e-5	& \cellcolor{gray!60}0.12 	& \cellcolor{gray!60}23.3 \\
\hline
Fig. \ref{G_PeakSepHist}-left: $\Delta_\gamma$ distribution		 	& 0.42 	& 2e-8 				& \cellcolor{gray!60}0.18 	& \cellcolor{gray!60}2.5	& 0.54	& 5e-15				& \cellcolor{gray!30}0.29 	& \cellcolor{gray!30}3e-3\\
\hline
Fig. \ref{G_PeakSepHist}-right: $\Delta_\mathrm{Radio}$ distribution		& \cellcolor{gray!60}0.19    	& \cellcolor{gray!60}57 	& \cellcolor{gray!30}0.21 	& \cellcolor{gray!30}46	& 0.52 	& 3e-2				& 0.37 	& 2.6\\
\hline
Fig. \ref{RadLagHist}: $\delta$ distribution				 			&  0.81    	& 1e-5				& 0.50 	& 2e-7				& \cellcolor{gray!30}0.39 	& \cellcolor{gray!30}4e-4	&  \cellcolor{gray!60}0.22 	& \cellcolor{gray!60}2.7\\
\hline
\end{tabular}
\end{center}
\caption{Two-sample Kolmogorov-Smirnov statistics (2KS) and relative $p_\mathrm{value}$ between observed and simulated one-dimensional 
distributions shown in Figures \ref{G_VisHistoMulti} to \ref{RadLagHist} for each model. The 2KS statistics $D$ ranges between 0 and 
1 for distributions showing total agreement and total disagreement, respectively. The $p_\mathrm{value}$ is the probability to obtain 
the observed $D$ value under the assumption that the two distributions are obtained from the same distribution (null hypothesis). 
This is equivalent to rejecting the null hypothesis at a confidence level of 100-($p_\mathrm{value}$)\%. The 2KS test is described in 
Section \ref{Met2KS2}. The $D$ and $p_\mathrm{value}$ parameters relative to the first and second most consistent distributions are 
highlighted in dark grey and light grey cells, respectively.}
\label{Tab4}
\end{table*}

Figure \ref{G_AllVisHistoClas} shows the recurrence of $\gamma$-ray and radio shape classes described in Tables \ref{TabShapeG} 
and \ref{TabShapeR}  respectively, for the whole simulated population synthesised by PIERBA12\nocite{pghg12} and its visible subsample. 
The recurrence of each shape class does not change considerably from the whole population to its visible subsample showing that 
there are no important selection effects due to light-curve shapes. The most pronounced discrepancies are observed within the PC model
where the visible subsample shows a lack of sharp peaks (shape class 2) and an excess of double peaks (shape class 4) and in the
SG model where the double plus single peak (shape class 8) is less recurrent among visible objects while all other classes are more populated 
in the whole population. This is confirmed by KS statistics between total and visible subsample distributions of
0.19, 0.21, 0.05, 0.10, and 0.03 for PC, SG, OG, OPC, and radio models, respectively, with the KS statistics ranging from 0 for total agreement 
to 1 for total disagreement (see Appendix \ref{Met2KS2} for details). The consistency between the shape class recurrences in the whole simulated population and its visible subsample 
allows us to compare the collective properties of the whole simulated population by \cite{pghg12} with the same properties of the observed 
pulsar population.

\begin{figure}
\begin{center}
\includegraphics[width=0.49\textwidth]{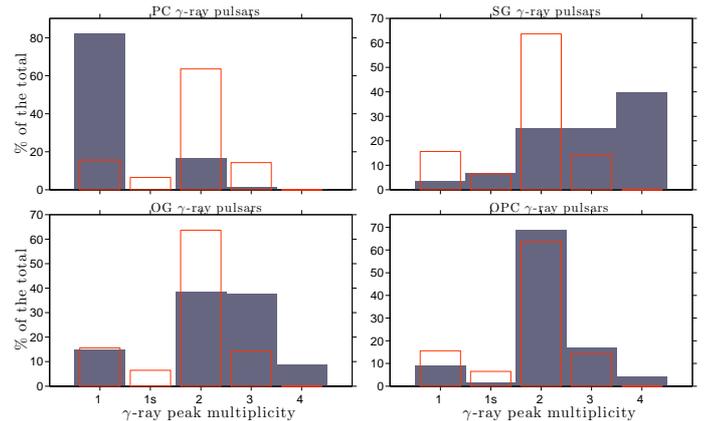} 
\caption{Recurrence of $\gamma$-ray peak multiplicity for the simulated (grey) and LAT (red)  $\gamma$-ray pulsar populations 
and each model. 1s refers to one peak with shoulder class 3 in Table \ref{TabShapeG}.}
\label{G_VisHistoMulti}
\end{center}
\end{figure}

We define the light-curve-peak multiplicity as the number of peaks detected in the radio or $\gamma$-ray light curve according to the method
described in Section \ref{Light curves shape classification}. The $\gamma$-ray and radio light curves multiplicities are associated with each shape
class as indicated in Tables  \ref{TabShapeG} and \ref{TabShapeR}, respectively.
Figure \ref{G_VisHistoMulti} compares the observed and simulated peak multiplicity distributions for the implemented $\gamma$-ray 
emission geometries.  
The statistical agreement between observed and simulated $\gamma$-ray peak multiplicity distributions was computed by performing
the two-sample Kolmogorov-Smirnov (2KS). Table \ref{Tab4} list the 2KS test results and suggests that  the outer magnetosphere models, OG 
and OPC, best explain the observations, with the OPC prediction showing the highest agreement with observations. The SG model 
predicts too many high-multiplicity profiles and too few double-peaked profiles. The PC model completely fails to explain the observed multiplicities.

The comparison of simulated and observed multiplicities of $\gamma$-ray-selected radio profiles is shown in Figure \ref{RMulti}. The different 
$\gamma$-ray model visibility criteria do not considerably affect the shape of the radio multiplicity distribution of simulated RL pulsar and do not 
allow us to discriminate the $\gamma$-ray model visibility that best explains the observations. The 2KS test results listed in Table \ref{C2ksTab}
show that all models poorly explain the observations with the SG model prediction rejected at a lower confidence level (CL).
One must note that 
even though the radio model is unique, the RL pulsars subsample changes within each $\gamma$-ray model as a function of the $\gamma$-ray 
visibility so the collective radio properties change within each $\gamma$-ray geometry. Hereafter we refer to each model RL objects as 
`$\gamma$-ray selected radio pulsars'. 

We studied how the $\gamma$-ray peak multiplicity changes for the RL and RQ subsamples of observed and simulated populations.
Figure \ref{G_HistoMultRLRQ} compares the recurrence of the $\gamma$-ray light-curve multiplicities in the framework of each model for RQ and RL
objects in the top and bottom panels, respectively. The $\gamma$-ray light-curve multiplicity of observed objects shows an increase of the single-peak 
light curves going from RQ to RL objects. None of the tested emission geometries manages to reproduce the observed behaviour of RQ and RL 
subsamples: From RQ to RL objects, the PC shows an increase of peak multiplicity two, while SG and OPC do not show significant changes in 
the predicted RQ and RL distributions. The OG model peak multiplicity changes with opposite trend with respect to observations: From RQ to RL 
objects, we observe an increase of the fraction of high-peak multiplicities 3 and 4 and a decrease of peak multiplicity 1.
The 2KS tests between simulated and observed distributions shown in Table \ref{C2ksTab} suggest that for both RQ and RL  $\gamma$-ray peak 
multiplicity distributions the OPC model shows the larger statistical agreement with observations, especially for RQ objects where the agreement 
reaches the 99.995\%.
\begin{figure*}
\begin{center}
\includegraphics[width=0.49\textwidth]{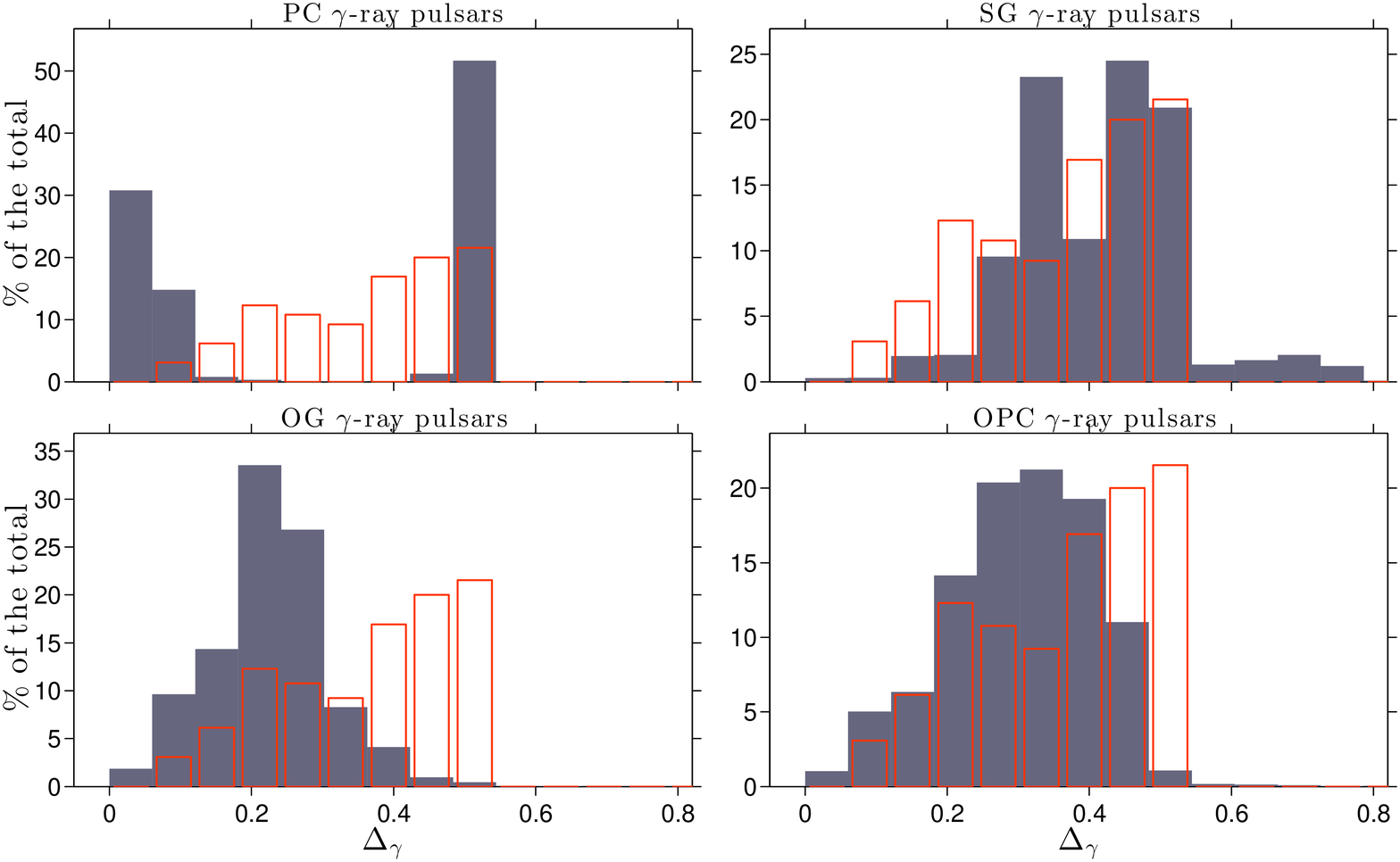} 
\includegraphics[width=0.49\textwidth]{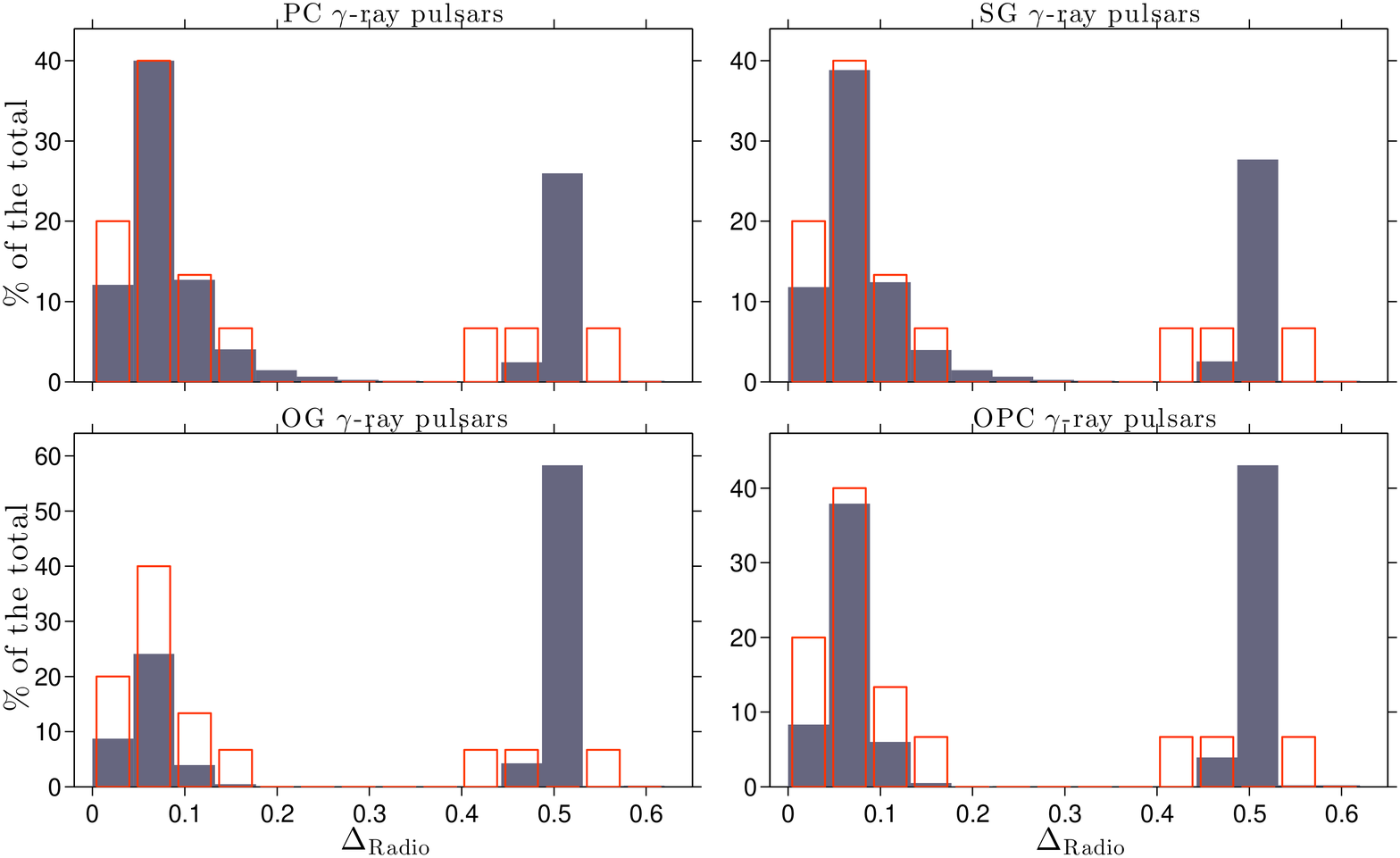} 
\caption{The $\gamma$-ray and radio peak separation for simulated samples (grey) and LAT sample (red) and each model in the left and right panels, respectively.}
\label{G_PeakSepHist}
\end{center}
\end{figure*}

In order to evaluate whether a model explains the observed shape recurrence, the peak multiplicity is a more objective indicator than
the shape classes. The number of peaks in the light curve is more easily recognisable and less biased by assumptions made to classify
observed and simulated profiles. Moreover the peak multiplicity was used instead of shape classes to compare model prediction and 
simulations by other studies, \citep[e.g.][]{wrwj09}. 
However, the shape classification defined in this paper enables a comparison of the observed and simulated light curves taking  
more light-curve details and degenerated classes into account, for example the multiplicity class number 2 is split in two shape 
classes, double peak and
two peaks, which recur differently within the simulated and observed samples. A comparisons of the recurrence of observed and simulated 
$\gamma$-ray and radio shape classes described in Tables \ref{TabShapeG} and \ref{TabShapeR}, respectively, are given in 
Figure \ref{G_HistoClas}, while the statical agreement  between simulated and observed distributions is given in Table \ref{C2ksTab}.
All models poorly describe the observation with SG and OPC showing larger statistical agreements with the data.

\section{Measure of the light-curve peak characteristics}
\label{Determination peaks}

We use the $\gamma$-ray and radio light curves classification described in Section \ref{Light curves shape classification} to fit observed and simulated 
light curves with a number of Gaussian and/or Lorentzian functions equal to the light-curve peak multiplicity. We tested Gaussian and Lorentzian 
distributions as possible shapes that best describe the simulated and observed light-curve peaks and we obtained that, in the majority of the cases, to fit 
the simulated peaks with Gaussian distributions gives a lower average $\chi^2$ value with respect to the Lorentzian fit. The Lorentzian distribution best 
explains some shape-class peaks in the framework of SG, OG, OPC and $\gamma$-ray observed light curves. The fit free parameters are: Gaussian 
standard deviation or Lorentzian $\gamma$ parameter (indicator of the peak width); the function amplitude (indicator of the peak intensity), the function mean 
value (indicator of the peak position), and an additive constant to account for the off-peak contribution. For each radio and $\gamma$-ray light curve we
obtained the peak phases, FWHMs, intensities, as well as the peak separation in light curves with peak multiplicity larger than one. To optimise 
the fitting procedure and speed up the computation, the fit-parameter intervals were optimised in the framework of each $\gamma$-ray and radio geometry.
For instance, the PC light curves show sharper peaks best fitted by low standard deviation Gaussians distributions while the SG peaks are wider and often 
best explained by broader Lorentzian distributions. Reference values of the Gaussian standard deviation or Lorentzian $\gamma$ parameters as indicators
of the pulse width are shown, in the framework of each model, in Table \ref{TabThres}.

Each shape-class light curve has been fit in the framework of each model with a number of Gaussian and/or Lorentzian functions as follows:
\begin{itemize}
\item {\bf PC, radio core plus cone, and LAT radio light curves}\\
All PC, radio core plus cone, LAT radio light curves are fitted with a combination of Gaussian functions equal to the light-curve peak multiplicity.

\item {\bf SG} \\
The majority of the SG light curves shown to be best described by a combination of Lorentzian functions with just the bump peak (shape class 1) 
and some double peaks (shapes class 2) best described by Gaussian functions. We noticed that the SG double-peak shapes class 5, enables two main
morphologies: the first one with smooth and bumpy peaks 
and the second one with more sharp and spiky peaks. Because of this shape dichotomy, the SG double peak (shapes class 5) is fit with both Lorentzian 
and Gaussian functions and the fit solution with the lower $\chi^2$ is considered.

\item {\bf OG and OPC} \\
OG and OPC light curves from classes 1 to 3 and 5 to 9 are best described by a single Gaussian or by a combination of Gaussian functions while the two peaks 
profiles (shape class 4) are fitted with Lorentzian functions. The different prescription for the width of the emission gap adopted by the 
OPC model \citep{rw10,wrwj09} allows for wider profiles for all shape classes: we observe OPC bump structures that are absent in the OG model and, for all 
high-peak multiplicity shape classes, the OPC light-curve peak appears wider than the OG peak.

\item {\bf LAT $\gamma$-ray light curves}  \\
The majority of the LAT $\gamma$-ray light curves is best described by a combination of Gaussian functions. The only case where a Lorentzian function 
best describes the observed peak shape is the shoulder (shape class 3). The shoulder structure has been fitted with two Lorentzian functions: one to account for the 
low plateau skewed on the left-hand side of the peak and the second to fit the main peak.
\end{itemize}

\subsection{$\gamma$-ray and radio peak separation}
\label{peak separation}

We estimate peak separation for all light curves with a peak multiplicity that is larger than one, both observed and simulated. When the light-curve 
peak multiplicity is equal to two, the peak separation is given by measuring the phase interval showing bridge emission.
In light curves with peak multiplicity larger than two, $\Delta_\gamma$ was computed according to the following criteria:
\begin{itemize}
\item[] {\centering \bf $\gamma$-ray light curves:}\\
\item {\bf Class 6: double plus single} \\
peak separation computed between the single peak and barycentre of the double peak
\item {\bf Class 7: triple}\\ 
peak separation computed between the two extreme peaks (only in OG and OPC cases)
\item {\bf Class 8: two double}\\ 
peak separation computed between the barycentres of the two double peaks. \\
\item[] {\centering \bf Radio light curves:}\\
\item {\bf Class 5: double plus single} \\
peak separation computed between the single peak and barycentre of the double peak
\item {\bf Class 6: two double}\\ 
peak separation computed between each double peak barycentre 
\item {\bf Class 7: triple}\\ 
peak separation computed between the two extreme peaks
\item {\bf Class 8 and 9: three peaks and more than four peaks}\\
peak separation computed between the two highest peaks.
\end{itemize}
Figure \ref{G_PeakSepHist} compares the $\gamma$-ray peak-separation distribution $\Delta_\gamma$ and the radio 
peak-separation distribution, $\Delta_\mathrm{Radio}$, for $\gamma$-ray-selected objects of all implemented models and 
LAT pulsars in the left and right panels, respectively. The observed $\Delta_\gamma$ distribution shown in figure 
\ref{G_PeakSepHist} left panel, ranges in the interval 0.1$\lsimeq\Delta_\gamma\lsimeq$0.55 and shows two peaks at 
$\Delta_\gamma\sim$0.2 and $\Delta_\gamma\sim$0.5. 
None of the proposed emission geometries manages to explain the observations, but the SG and OPC models predict the
observed distribution at high and low $\Delta_\gamma$ values, respectively; the 2KS test results listed in Table 
\ref{Tab4} shows that SG and OPC models explain the observations with the highest and second highest statistical 
significance, respectively. The PC model predicts $\Delta_\gamma$ 
just at low and high values: $\Delta_\gamma<$0.1 are generated by each magnetic pole hollow cone while 
$\Delta_\gamma\sim$0.5 are generated by the two emission cones from each magnetic pole separated by 0.5 in phases 
that start to be visible for high $\alpha$ and $\zeta$ (see Figure \ref{Em_Pat}). 
The OG distribution is antithetic to the observed distribution. This distribution shows a maximum at $\Delta_\gamma\sim$0.2 
which is consistent 
with the observations but goes to 0 at $\Delta_\gamma=$0.5 where the observed distribution shows its absolute maximum. 
The SG model explains the observations just for $\Delta_\gamma>$0.4, exhibits one peak at 0.35, which does not show 
up in the data, and clearly underestimates the observations for $\Delta_\gamma<$0.25. The OPC $\Delta_\gamma$ 
distribution shows a good consistency with observations for $\Delta_\gamma<$0.25  but completely fails to explain the 
observations at high $\Delta_\gamma$.
We note that SG and OPC are complementary in how they explain the observed $\Delta_\gamma$ distribution and 
how the different prescription for the gap width computation in the OPC model affects the peak-separation distribution. The 
OPC prescription for the gap width allows for wider light-curve double peaks since the gap widths are generally smaller than 
those of the OPC and this is evident by comparing the OG and OPC 
peak-separation distributions; the OG distribution peaks at 0.2 and quickly goes to zero at $\Delta_\gamma$=0.5, while the 
OPC distribution shows a broad peak in the interval 0.2 $<\Delta_\gamma<$ 0.4 and slowly decrease to a non-zero 
minimum at $\Delta_\gamma$=0.5. This partially solves the complete lack of $\Delta_\gamma$=0.5 in the OG model and 
allows the OPC to explain, even poorly, the observed trend. However the vast majority of observed $\Delta_\gamma$=0.5 
objects suggests that a SG-like two-pole caustic emission geometry is necessary to explain the wide separation of the
 observed $\gamma$-ray peaks, which, on the other hand, does not explain objects with $\Delta_\gamma<$ 0.25. 

The $\gamma$-ray peak-separation distribution was also studied in the framework of PSRCAT2\nocite{aaa+13}. Those authors 
performed a $\gamma$-ray light-curve analysis with the purpose to give the most precise possible measures of the peak positions 
and separations. Even though the purpose of our light-curve analysis is not to give precise measurements of peak positions and 
separations but to measure observed and simulated light curves under the same criteria to be compared, our $\Delta_\gamma$ 
distribution for LAT ordinary pulsars is consistent with the distribution obtained by PSRCAT2\nocite{aaa+13}.

The observed radio peak-separation distribution shown in the right panel of Figure \ref{G_PeakSepHist} ranges in two intervals, 
0$<\Delta_\mathrm{Radio}<$0.15, where it peaks at $\Delta_\mathrm{Radio}\sim$0.05, and 0.4$\lsimeq\Delta_\mathrm{Radio}\lsimeq$0.55. 
A comparison of the $\Delta_\mathrm{Radio}$ observed and simulated distribution in the framework of each model shows that 
all of the proposed emission geometries explain the observed trend for low peak separation $\Delta_\mathrm{Radio}<$0.15 while for
$\Delta_\mathrm{Radio}>$0.4 all models over predict the observations. This may be due to a radio cone used in our simulation
that is too large and that increases the probability that the line of sight intersects the emission from both poles.
The 2KS tests shown in Table \ref{Tab4} suggest
that all models explain well the observations with the two-pole emission geometries, PC and SG, best explaining the LAT distribution 
since they predict the observed proportion.
In analogy to the PC geometry, the $\Delta_\mathrm{Radio}\sim$0.1 are generated by double-peak structures typical of each radio 
emission beam, while $\Delta_\mathrm{Radio}\sim$0.5 are given by the distances of the two emission cones from each magnetic 
pole  for high $\alpha$ and $\zeta$ (see Figure \ref{Em_Pat}).

\subsection{Radio lag}
\label{Radio lag}

We consider the radio and $\gamma$-ray light curves of the same pulsar, both coherent in phase with the actual pulsar rotational phase.
The radio lag is then defined as the phase lag between a radio fiducial phase and the following $\gamma$-ray peak phase. The radio lag 
is considered as a tracker of the pulsar magnetosphere structure. The radio lag constrains the relative positions of $\gamma$-ray and 
radio emission regions
in the pulsar magnetosphere and can be used to discriminate the proposed $\gamma$-ray and radio emission geometries that best explain the 
observations. While it is relatively easy to produce radio and $\gamma$-ray light curves that are both coherent in phase  
with the pulsar rotation through timing techniques, the definition of the radio fiducial phase is more controversial. The commonly 
accepted definition for 
the radio fiducial phase is the phase of the radio peak following the pulsar magnetic pole that should be identified, case by case, 
by analysing the light-curve shape. 
However, this definition is strongly dependent on the light-curves quality, which might not be good enough to enable a robust identification of the 
magnetic pole, and might be biased by the radio model used to predict the magnetic pole phase in the observed light curve.

In PSRCAT2\nocite{aaa+13} the problem of finding solid criteria to assign a radio fiducial phase to observed light curves was solved by 
increasing the quality of the radio light curve with deeper radio observations of the analysed objects and by defining a series of morphological criteria.
When the light curve shows symmetric structures (double or higher multiplicity peaks) the pulsar fiducial phase is associated with the peak 
barycentre. When radio and $\gamma$-ray peaks are aligned (e.g. Crab pulsar), the radio emission cannot be explained by a conical beam 
generated above the magnetic poles but rather by caustic emission, as proposed for the Crab pulsar by \cite{vjh12}. In these cases the 
fiducial phase is associated with the phase of the radio precursor, which is a small light-curve feature 

The problem 
of finding a robust definition for the radio fiducial phase becomes critical when one compares simulations with observations. The magnetic-pole 
phase for simulated objects is known with high precision, it is independent of the assumed radio and $\gamma$-ray emission geometries, and 
is not affected by the light-curve quality. It is thus easy to identify the phase of the first peak after the magnetic pole. Moreover, the cone 
plus core radio emission geometry we adopt does not predict caustic radio
emission and no aligned radio and $\gamma$-ray peaks are possible. The obvious consequence is that observed and simulated radio-lag 
measurements are not completely consistent and the conclusions drawn from their comparison loses scientific reliability. 
The radio lag for simulated objects has been computed as the phase separation between the first radio peak detected by the analysis algorithm, the 
fiducial phase, and the phase of the following $\gamma$-ray peak. Both $\gamma$-ray and radio simulated light curves have been generated within the 
phase interval -0.5 to 0.5 with the pulsar fiducial plane (the plane containing magnetic and rotational axes) intersecting the line of sight at phases 0
and $\pm$0.5. Because of that, the fiducial phase was prevalently found at phase 0 and, for large $\alpha$ and $\zeta$ values, at phase -0.5 (see 
Figure \ref{Em_Pat}). In two-pole $\gamma$-ray emission geometries, PC and SG, each magnetic pole shines both in radio and in $\gamma$-rays,
and the radio lag is always measured between the fiducial phase and the $\gamma$-ray beam coming from the same magnetic hemisphere. 
In one-pole  $\gamma$-ray emission geometries there is just one magnetic hemisphere that shines in $\gamma$-rays and when, for large $\alpha$ 
and $\zeta$, radio emission from both magnetic poles starts to be visible in the light curve, cases of radio lag larger than 0.5 are possible; e.g. when 
radio emission beam starts shining at phase -0.5, the algorithm sets the fiducial phase equal to -0.5 and measures the radio lag as its distance from 
the $\gamma$-ray peak generated in the opposite magnetic hemisphere. All $\delta>$0.5 for OG and OPC models were subtracted of 0.5 to measure the radio 
lag as the phase lag between fiducial phase and  $\gamma$-ray peak in the framework of the same magnetic hemisphere.
\begin{figure}
\begin{center}
\includegraphics[width=0.49\textwidth]{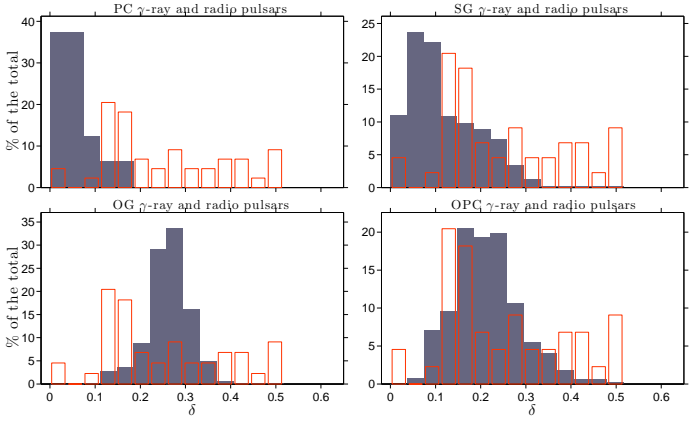}
\caption{Radio-lag distribution for simulated samples (grey) and LAT pulsars (red) and each model.}
\label{RadLagHist}
\end{center}
\end{figure}

We use the radio and $\gamma$-ray light curves published in PSRCAT2\nocite{aaa+13} to estimate the radio lags of RL LAT pulsars as the phase 
separation between the fiducial phase and the following $\gamma$-ray peak. 
We define the fiducial phase as the phase of the first peak, which appears in the radio light curve as detected from the algorithm described 
in Sections \ref{Light curves shape classification} and \ref{Determination peaks}, to measure the radio lag for RL LAT pulsars consistent 
with the simulated sample. When two peaks separated by 0.5 in phase are visible in the radio 
light curve, we assign the fiducial phase to the radio peak for which the radio lag is smaller than 0.5. In this way we avoid the assumption 
about the magnetic pole position and the radio precursor that is not modelled in the framework of the implemented radio-beam emission geometry.

Figure \ref{RadLagHist} compares the radio lag $\delta$ distribution of simulated and observed RL pulsars in the framework of each model. A 
comparison between our $\delta$ distribution and the distribution obtained by PSRCAT2\nocite{aaa+13} shows a total consistency both in the 
range of values and in the proportions. Our $\delta$ distribution ranges in the interval $0<\delta<0.5$, and raises steeply up to its maximum
at $\delta=0.15$ and shows a stable flat trend in the interval $0.2<\delta<0.5$. 
None of the tested models manages to explain the observed distributions over all the range of observed $\delta$ but the OPC shows higher 
agreement with the data. The 2KS test statistics given in Table \ref{Tab4} shows that the OPC model explains the observations with a CL at 
least four orders of magnitude larger than the other models and that the OG predictions explain the observations with the second highest CL.
Both PC and SG 
models predict emission beams most of the time overlapping the radio emission beam, with the PC emission beam tightly matching the radio 
beam both in size and pulse phase (see Figure \ref{Em_Pat}). This generates the excess of $\delta<$0.1 predicted by both PC and SG emission 
geometry with the SG geometry predicting larger $\delta$ values because of its wider emission beam. The OG model completely fails in 
predicting the observed distribution both in shape and proportions. The OPC is the model that best predicts the observed $\delta$ range and 
proportions. This model predicts a peak that is too broad in the range $0.15<\delta<0.25$ which partly overlap the observed peak but 
underestimates objects in 
the range $0.35<\delta<0.5$. Of particular interest is the difference between OG and OPC distributions. As also shown in Figure \ref{G_PeakSepHist}, 
the different prescription for the gap width adopted by the OPC enables broader light-curve peaks that occur closer, in phase, to the radio peak, 
thereby decreasing the radio lag. Under the assumption that the radio model used in this simulation is correct and that the adopted VRD magnetic 
field geometry is correct, this picture supports an outer magnetosphere location of the emission gap, points to the need to model 
broader light-curve $\gamma$-ray peaks, and highlights the importance of the gap width and B field structure choice in the light-curve modelling.
\begin{figure}
\begin{center}
\includegraphics[width=0.49\textwidth]{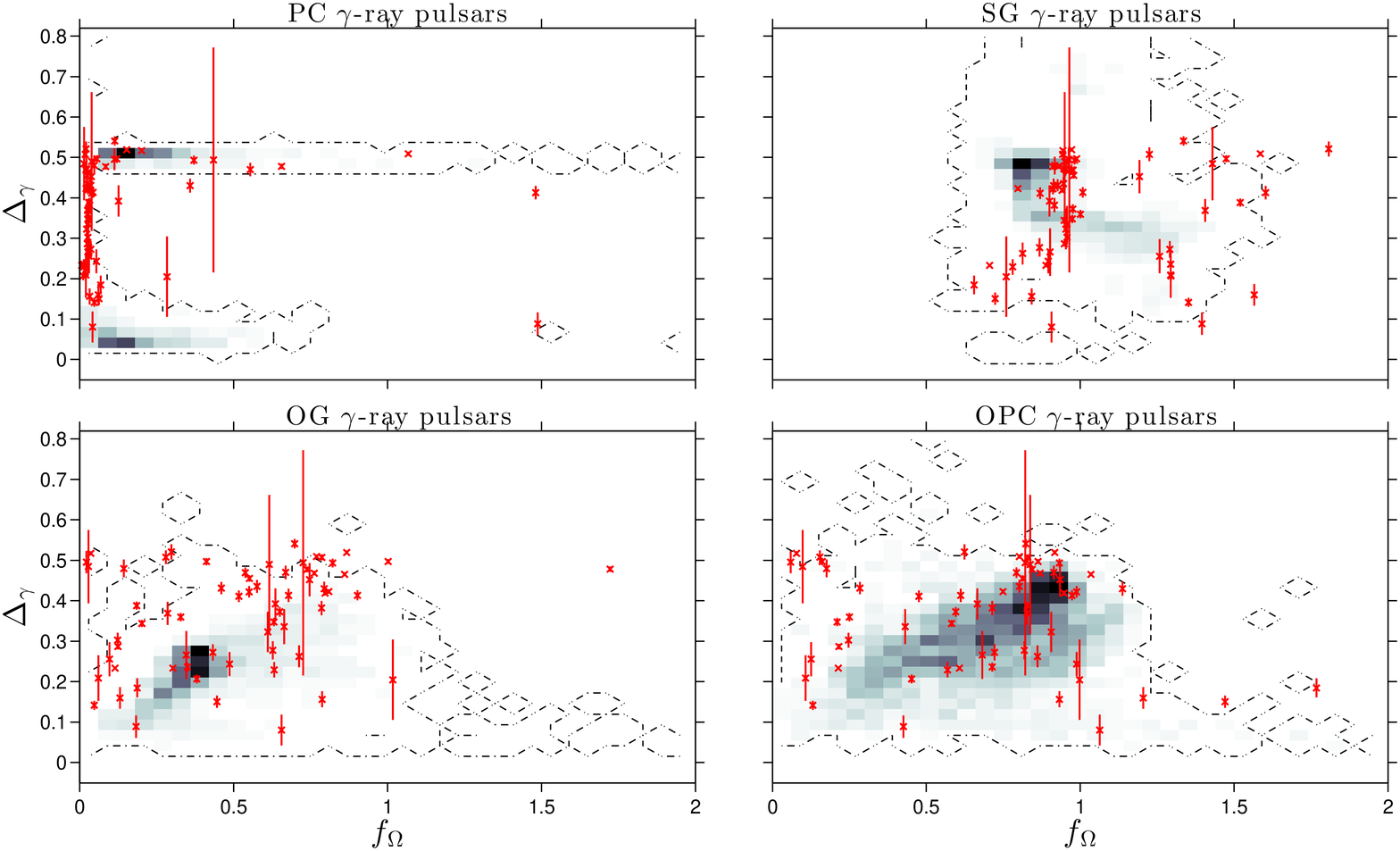} 
\includegraphics[width=0.49\textwidth]{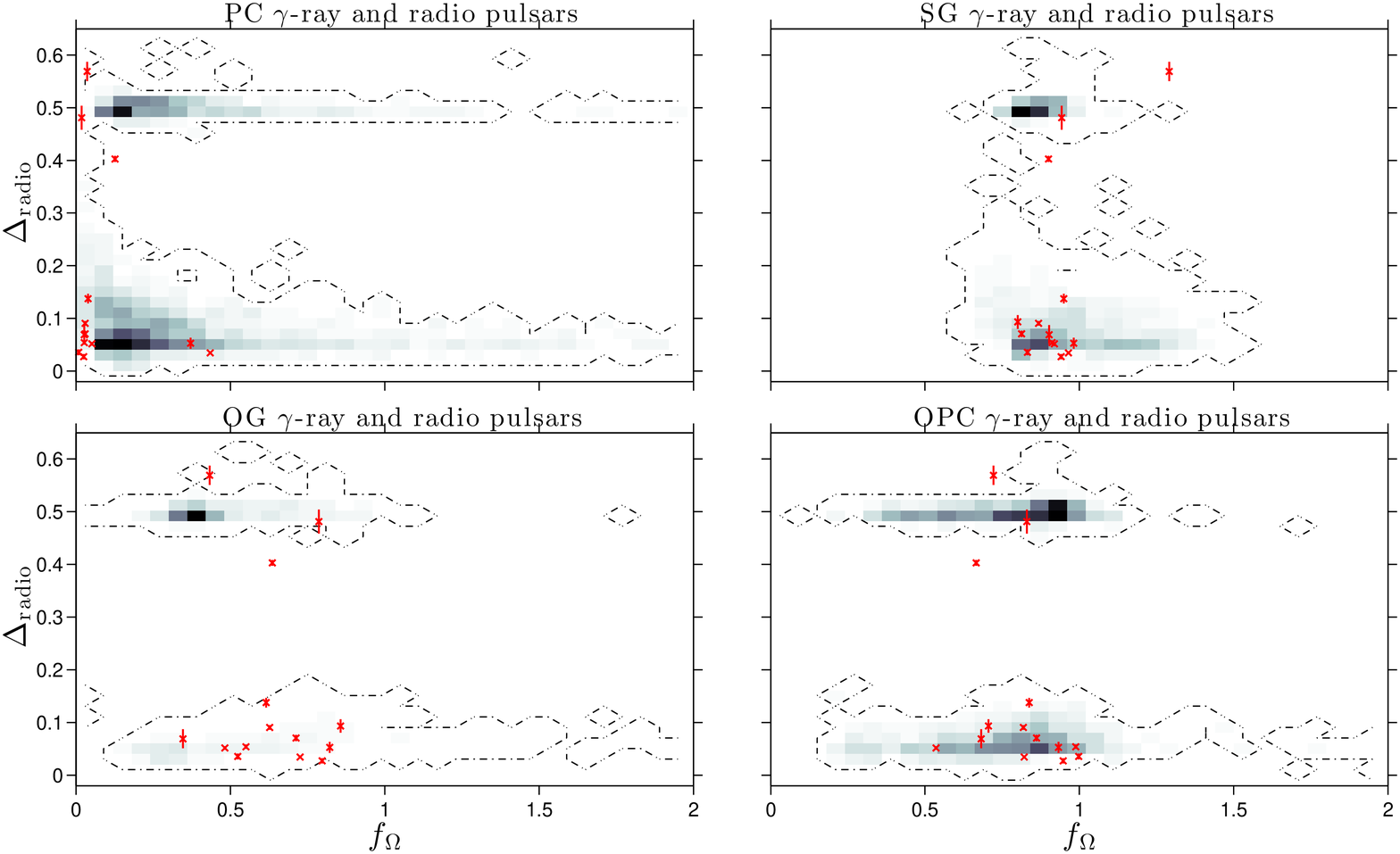} 
\caption{Distributions of the $\gamma$-ray and radio peak separation of simulated (grey) and observed (red) pulsars obtained, for each model, as 
a function of the $\gamma$-ray beaming factor are shown in the top and bottom panels, respectively.  The model distributions have been obtained as 
two-dimensional number-density histograms. The dash-dotted line highlights the minimum number density contour. The LAT measurements are given with 
1$\sigma$ error bar uncertainty.}
\label{G_PeakSepBeam}
\end{center}
\end{figure}

One has to note that the model radio lags obtained in this study are strongly dependent on the assumed  VRD magnetic field geometry. \cite{khkc12} 
compute the $\gamma$-ray light curves in a dissipative magnetosphere (DM) with conductivity ranging from 0 to $\infty$, corresponding to VRD magnetic 
field geometry and to a force-free electrodynamics (FFE) magnetic field geometry, respectively. Those authors conclude that VRD magnetospheres 
shows the smallest radio to $\gamma$-ray lags while more realistic DMs predict average larger lags \citep{khkc12}. All $\gamma$-ray model radio-lag distributions
computed in this study  would be shifted towards higher values if computed in the framework of a DM, and this would solve the excess of low radio-lag 
values predicted by the PC and SG models. Particularly interesting is the case of the SG model that shows the correct shape of the observed radio-lag 
distribution but shifted towards values that are too low; the implementation of the SG model within a DM may considerably improve the SG prediction.

\subsection{Non-observable pulsar parameters: $\Delta_\gamma$, $\Delta_\mathrm{Radio}$, and $\delta$ as a function of f$_\Omega$, 
$\alpha$, $\zeta$}
\label{NonObservable}

We compare simulated and estimated correlations between $\Delta_\gamma$, $\Delta_\mathrm{Radio}$, and $\delta$ and non-observable 
pulsar parameters, namely beaming factor $f_{\Omega}$, magnetic obliquity $\alpha$, and observer line-of-sight $\zeta$. The simulated 
distributions of $f_{\Omega}$, $\alpha$, and $\zeta$ are those synthesised in the framework of PC, SG, OG, and OPC emission models by PIERBA12 
\nocite{pghg12} via the emission geometry proposed by \cite{dhr04}. 
The same parameters for LAT pulsars have been estimated from fits to the observed pulsar light curves by 
PIERBA15 in the framework of the same emission models and using the same emission geometry by \cite{dhr04}. 

\begin{figure}
\begin{center}
\includegraphics[width=0.49\textwidth]{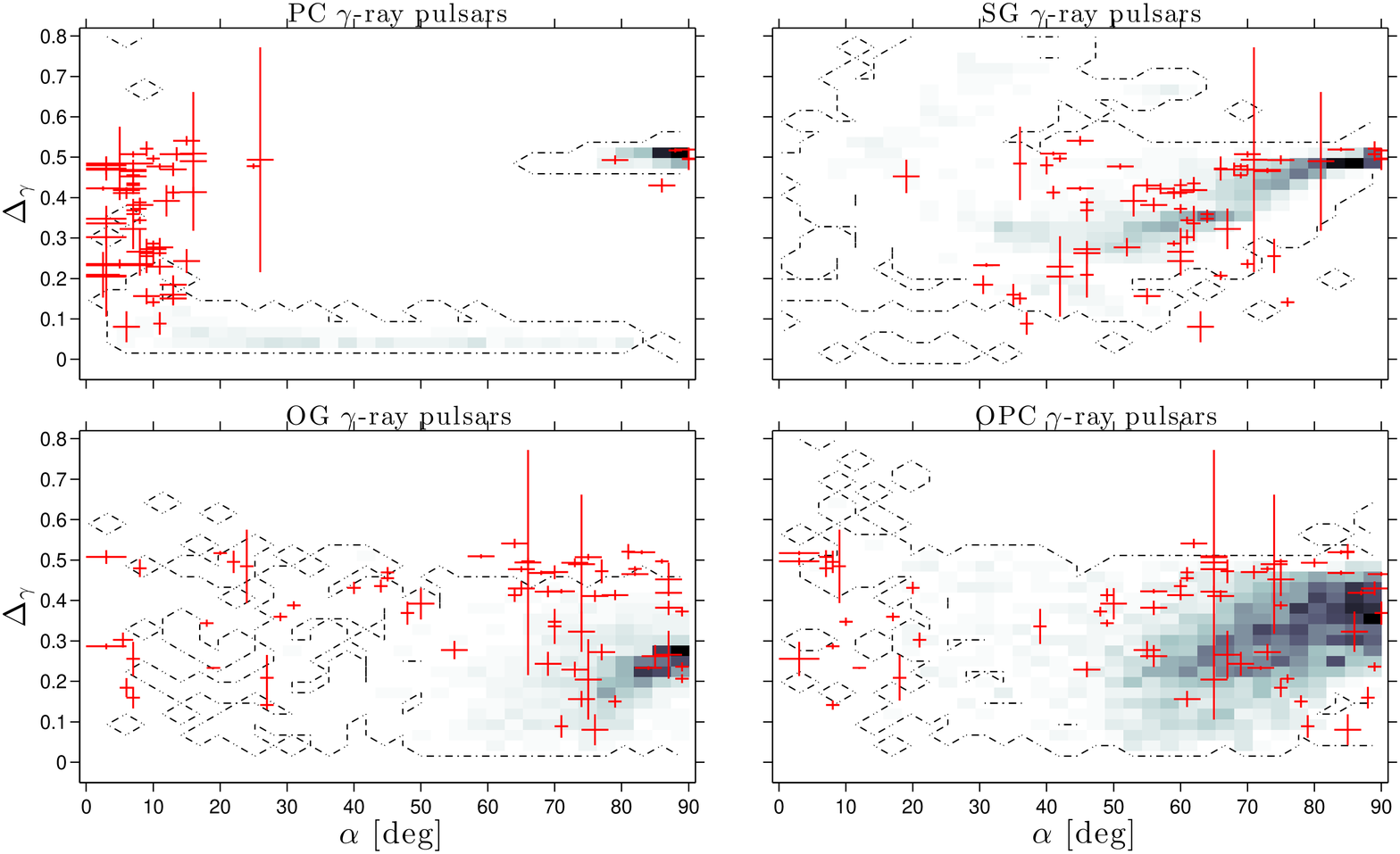} 
\includegraphics[width=0.49\textwidth]{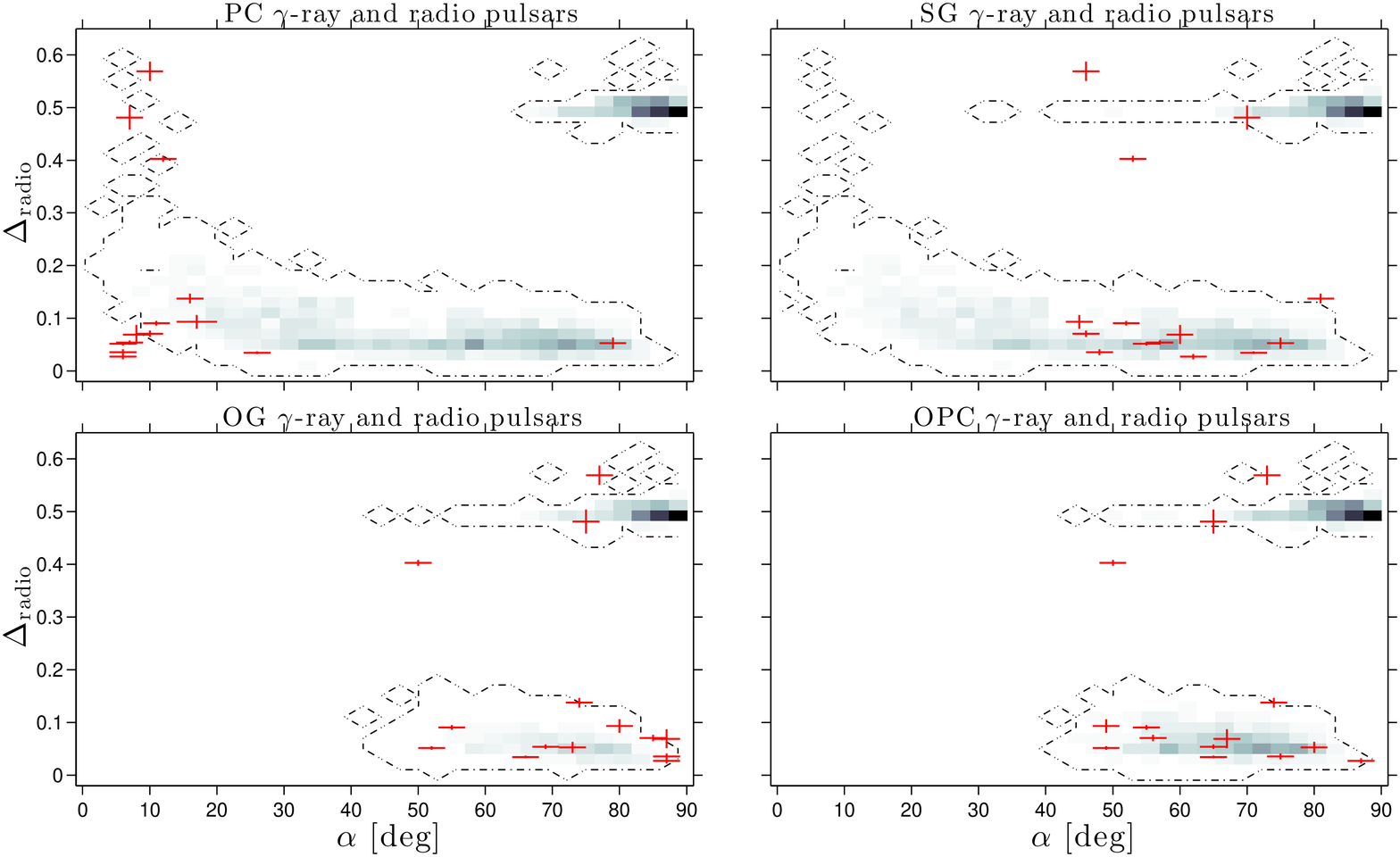} 
\caption{Distributions of the $\gamma$-ray and radio peak separation of simulated (grey) and observed (red) pulsars obtained, for each model, as 
a function of the magnetic obliquity are shown in the top and bottom panels, respectively.  The model distributions have been obtained as 
two-dimensional number-density histograms. The dash-dotted line highlights the minimum number density contour. The LAT measurements are given with 
1$\sigma$ error bar uncertainty.}
\label{G_PeakSepalpha}
\end{center}
\end{figure}

\begin{figure}
\begin{center}
\includegraphics[width=0.49\textwidth]{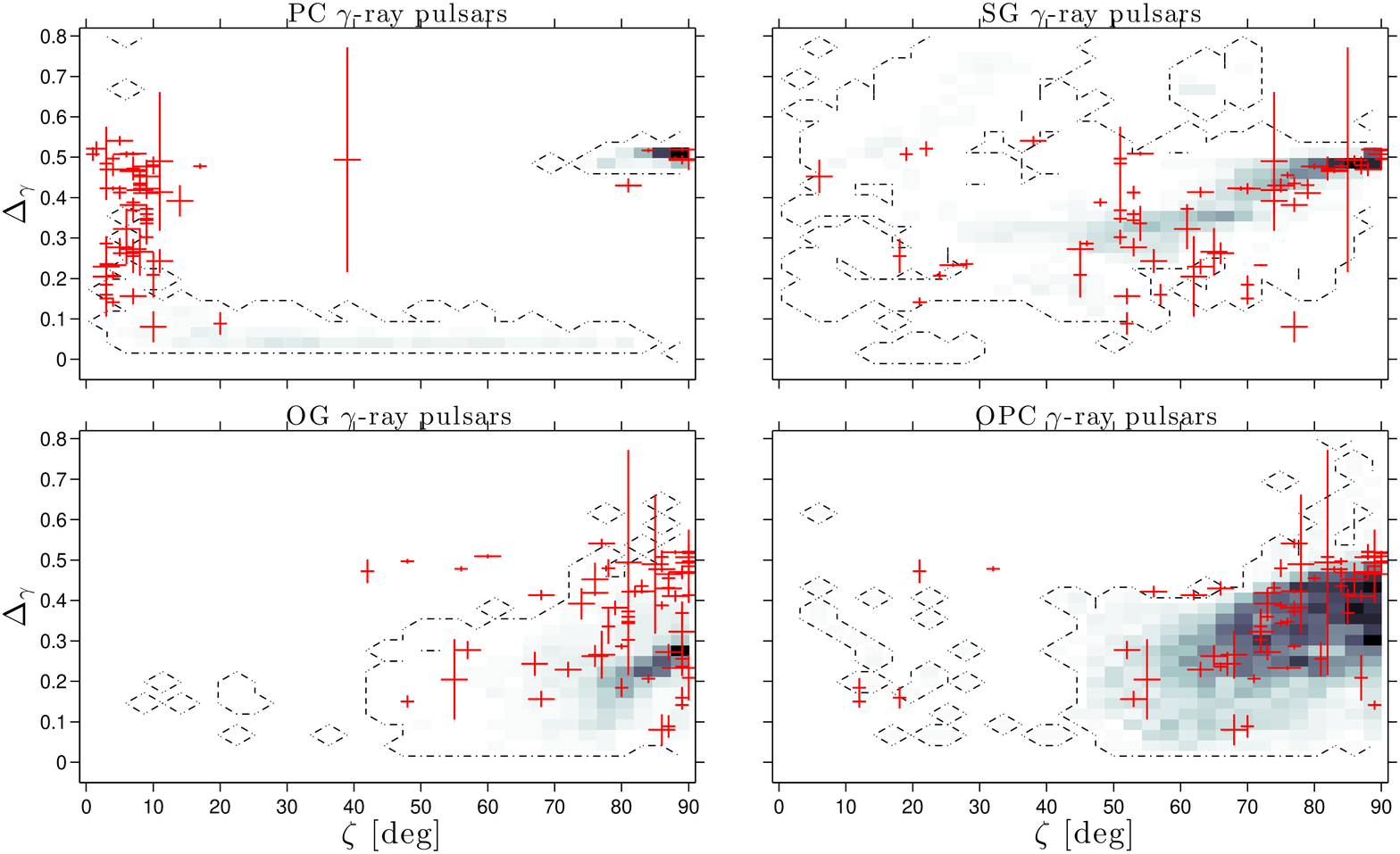} 
\includegraphics[width=0.49\textwidth]{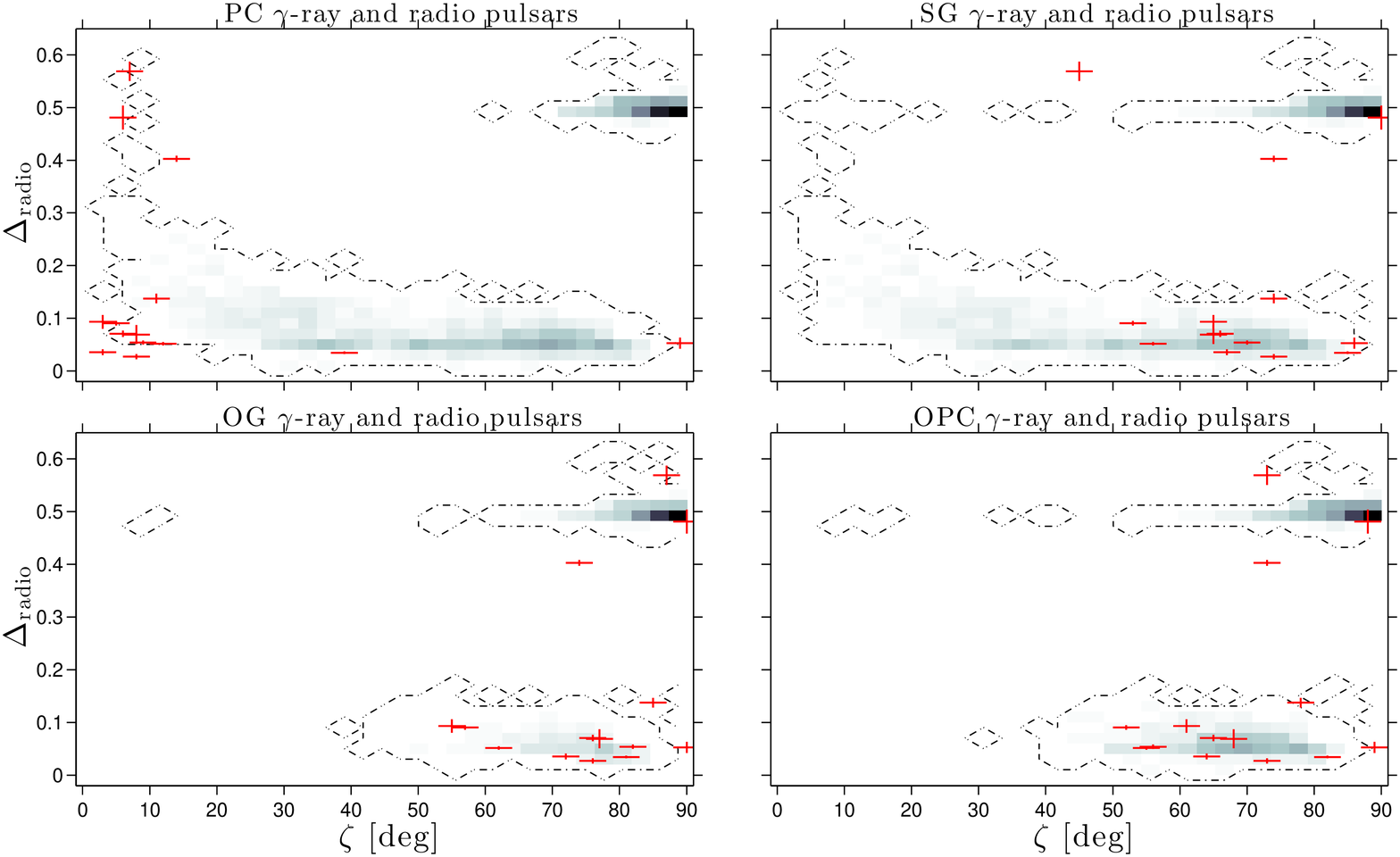} 
\caption{Distributions of the $\gamma$-ray and radio peak separation of simulated (grey) and observed (red) pulsars obtained, for each model, as 
a function of the observer line of sight are shown in the top and bottom panels, respectively.  The model distributions have been obtained as 
two-dimensional number-density histograms. The dash-dotted line highlights the minimum number density contour. The LAT measurements are given with 
1$\sigma$ error bar uncertainty.}
\label{G_PeakSepzeta}
\end{center}
\end{figure}

The simulation by PIERBA12\nocite{pghg12} gives the allowed range of values for the parameters $f_{\Omega}$, $\alpha$, and $\zeta$
within each model. Since $f_{\Omega}$, $\alpha$, and $\zeta$  for LAT pulsars were estimated by implementing the same geometrical model 
used by PIERBA12\nocite{pghg12} to synthesise the simulated-population light curves, we expect a match between estimations and simulations 
within the same model only if the model manages to reproduce the variety of the observed light-curve shapes.     
If the variety of observed light-curve shapes is not explained in the framework of a particular model, the best-fit parameters $f_{\Omega}$, $\alpha$, 
and $\zeta$ does not match the interval of the most likely values as simulated by PIERBA12\nocite{pghg12} in the framework of the same model. 
Because of that, a mismatch between estimated (PIERBA15\nocite{phg+15}) and simulated (PIERBA12\nocite{pghg12}) trends in the framework of 
a particular model suggests the inadequacy of that model in explaining the observations. 
An example of inadequacy of a model in explaining the observed light curves is given by the PC model and it is evident by looking at the PC 
panel at the top of Figure \ref{G_PeakSepalpha}. 
The vast majority of the PC light curves show sharp peaks and low off-peak emission that do not manage to predict the variety of the LAT light-curve 
shapes. This implies that the PIERBA15\nocite{phg+15} fitting algorithm selects as best-fit solutions only PC light curves with very low $\alpha$ and 
$\zeta$ angles, which are those characterised by broader peaks and more likely to explain the observed shapes (see Figure \ref{Em_Pat}). 
As a consequence, in the PC panel of Figures \ref{G_PeakSepalpha} and \ref{G_PeakSepzeta}, the estimated points are all grouped at low 
$\alpha$ and $\zeta$ values and do not match the $\alpha$ and $\zeta$ intervals predicted by the PC simulation. This suggests the inadequacy of the 
PC model in predicting the observations. 
The statistical agreement between observed/estimated and simulated distributions was quantified by computing the two-sample Kolmogorov-Smirnov (2KS) statistics for the two-dimensional distributions proposed by \cite{ptvf92} as described in Section \ref{Met2KS2}. The D and $p_\mathrm{value}$ values computed between each observed/estimated and simulated distribution are listed in Table \ref{Tab5}.

Figure \ref{G_PeakSepBeam} compares estimated and simulated distributions for $\Delta_{\gamma}$ and $\Delta_\mathrm{Radio}$ as a function of 
the beaming factor $f_{\Omega}$ and for each model. The $\Delta_{\gamma}$-$f_{\Omega}$ plane is a tracker of the magnetospheric 
region where the $\gamma$-ray emission is generated since both $\Delta_{\gamma}$ and $f_{\Omega}$ depends on the emission-region 
structure. Overall none of the distributions estimated in the framework of the proposed emission geometries matches the simulated ranges
but the SG and OG models show consistency with observations at the highest and second highest CL, respectively, as indicated in Table \ref{Tab5}.
Both OG and OPC models show highly dispersed distribution and suggest that, for the bulk of the simulated distributions, $\Delta_{\gamma}$ 
increases as $f_{\Omega}$ increases. The OPC estimates match the simulations only for 
$f_{\Omega}>$0.4 since they manage to reproduce the bulk of objects centred at $f_{\Omega}\sim$0.85 and  $\Delta_{\gamma}\sim$0.45 while 
the OG estimates do not match the bulk of the simulated objects. The PC and SG models do not predict any $\Delta_{\gamma}$ variation as 
$f_{\Omega}$ changes and, in both cases, the estimates do not match the bulk of the simulations.
In the $\Delta_\mathrm{Radio}$-$f_{\Omega}$ plane, all model estimates reasonably match the simulated trends and given the low number of observed 
objects, the relative $p_\mathrm{values}$ listed in Table \ref{Tab5} cannot be used to discriminate the model that best explains the data.

Figures \ref{G_PeakSepalpha} and \ref{G_PeakSepzeta} compare estimated and simulated trends of $\Delta_{\gamma}$ and $\Delta_\mathrm{Radio}$ 
as function of $\alpha$ and $\zeta$, respectively. The parameters $\Delta_{\gamma}$ and $\Delta_\mathrm{Radio}$ are expected to 
change with increasing $\alpha$ and $\zeta$ with different trends and in the framework of different emission geometries, as shown in 
Figure \ref{Em_Pat}. In the PC and radio emission geometries, the size of the conical emission beams decrease as $\alpha$ and $\zeta$ 
increase, therefore $\Delta_{\gamma}$ and $\Delta_\mathrm{Radio}$ are expected to decrease with increasing $\alpha$ and $\zeta$. This behaviour is barely 
visible in the PC $\gamma$-ray separation but it is evident in the radio cases for PC and SG-selected objects. The PC estimates populate the low 
$\alpha$ and $\zeta$ regions of the 
$\Delta_{\gamma/radio}$-$\alpha$ and $\Delta_{\gamma/radio}$-$\zeta$ planes, respectively, and do not match the major fraction of simulated 
points. The SG model predicts increasing $\Delta_{\gamma}$ when both $\alpha$ and $\zeta$ increase. As shown in figure \ref{Em_Pat}, 
the bright SG caustic widens when $\zeta$ increases for fixed $\alpha$ and when $\alpha$ increases, the caustic from the second magnetic 
pole starts to be visible in the light curve and $\Delta_{\gamma}$ tend to 0.5. In both $\Delta_{\gamma}$-$\alpha$ and $\Delta_{\gamma}$-$\zeta$ 
planes, the SG estimates match the simulated trends. The
OG and OPC models are explained in the framework of a one-pole caustic emission geometry; when $\alpha$ increases, no emission from the 
other pole shows up at high $\zeta$ and $\Delta_{\gamma}$ does not tend to 0.5 as $\alpha$ or $\zeta$ increase. This is less evident in 
the OPC model predictions where,
because of the different prescription for the gap width, wider peaks are possible and both $\Delta_{\gamma}$-$\alpha$ and $\Delta_{\gamma}$-$\zeta$
trends show broader distributions. Both figures \ref{G_PeakSepalpha} and \ref{G_PeakSepzeta} show
that $\Delta_{\gamma}$ increases for increasing $\alpha$ or $\zeta$ with the OPC trend showing higher dispersion with respect to the OG trend  
as a result of the different prescription used in the OPC to compute the width of the accelerator gap. In the plane 
$\Delta_{\gamma}$-$\alpha$, neither OG nor OPC estimates match the bulk of the simulated distribution; in both cases, the estimates show an excess 
at $\Delta_{\gamma}=0.5$, which is not predicted by the models. In the plane $\Delta_{\gamma}$-$\zeta$, the OG estimates over predict
 $\Delta_{\gamma}=0.5$ 
and fail again in matching the bulk of the simulated distribution, while the OPC estimates match the simulation explaining the simulated
increasing trend well. The results of the statistical test shown in Table \ref{Tab5} find that  in the plane $\Delta_{\gamma}$-$\alpha$, PC and 
SG models predict the observations with the highest and second highest CL while in the $\Delta_{\gamma}$-$\zeta$ plane, the OG and SG models 
give the highest and second highest CL predictions of the data. The best agreement obtained between observed/estimated and simulated 
distributions in the plane $\Delta_{\gamma}$-$\alpha$ is fictitious since the distributions are totally not overlapping and inconsistent.

In the $\Delta_\mathrm{Radio}$-$\alpha$ and $\Delta_\mathrm{Radio}$-$\zeta$ planes, as a consequence of the radio beam shrinking with increasing 
$\alpha$ or
$\zeta$, the simulated $\Delta_\mathrm{Radio}$ are expected to decrease with increasing $\alpha$ and $\zeta$, especially at low angles, as is evident
from Figure \ref{Em_Pat}. This trend is well visible in the PC- and SG-selected objects with $\alpha$ and $\zeta$ values ranging from 0 to 90 
degree while the trend is not appreciable for OG- and OPC-selected objects, which are mainly characterised by high $\alpha$ and $\zeta$. 
Radio objects 
selected in the framework of all $\gamma$-ray models show an excess for $\alpha$ and $\zeta$ $>70^\circ$ and $\Delta_{\gamma}$=0.5. As in 
the PC case, this excess is a consequence of the radio emission from the other pole that, for high $\zeta$ angles, starts to be visible for 
high $\alpha$ 
values and allows peak separations equal to the magnetic pole distance, 0.5. Overall, all model estimates match the simulated trends
with consistent CL with PC and OG showing the highest CL in the $\Delta_\mathrm{Radio}$-$\alpha$ and $\Delta_\mathrm{Radio}$-$\zeta$ 
planes, respectively (see Table \ref{Tab5}).

\begin{figure}
\begin{center}
\includegraphics[width=0.49\textwidth]{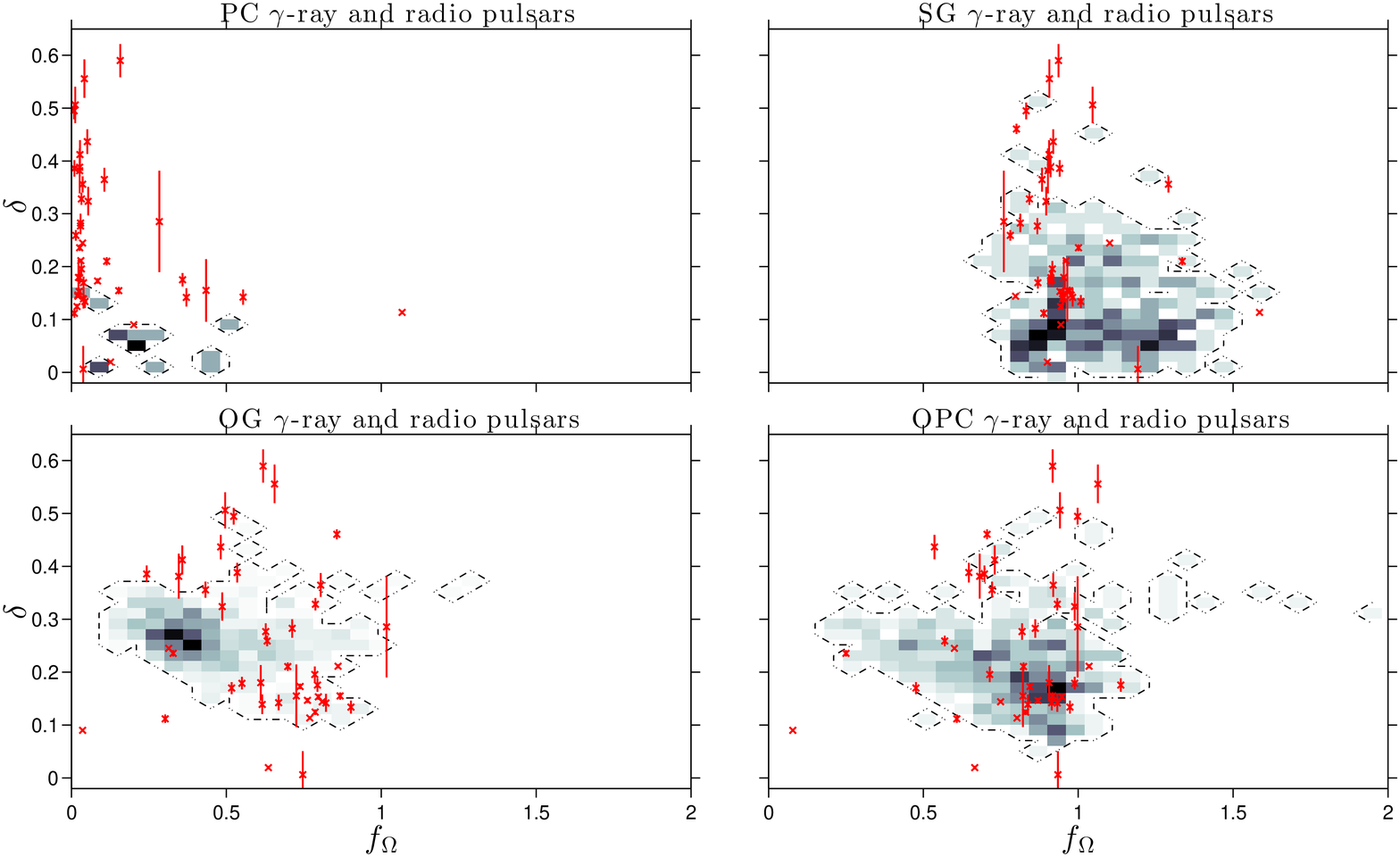}
\includegraphics[width=0.49\textwidth]{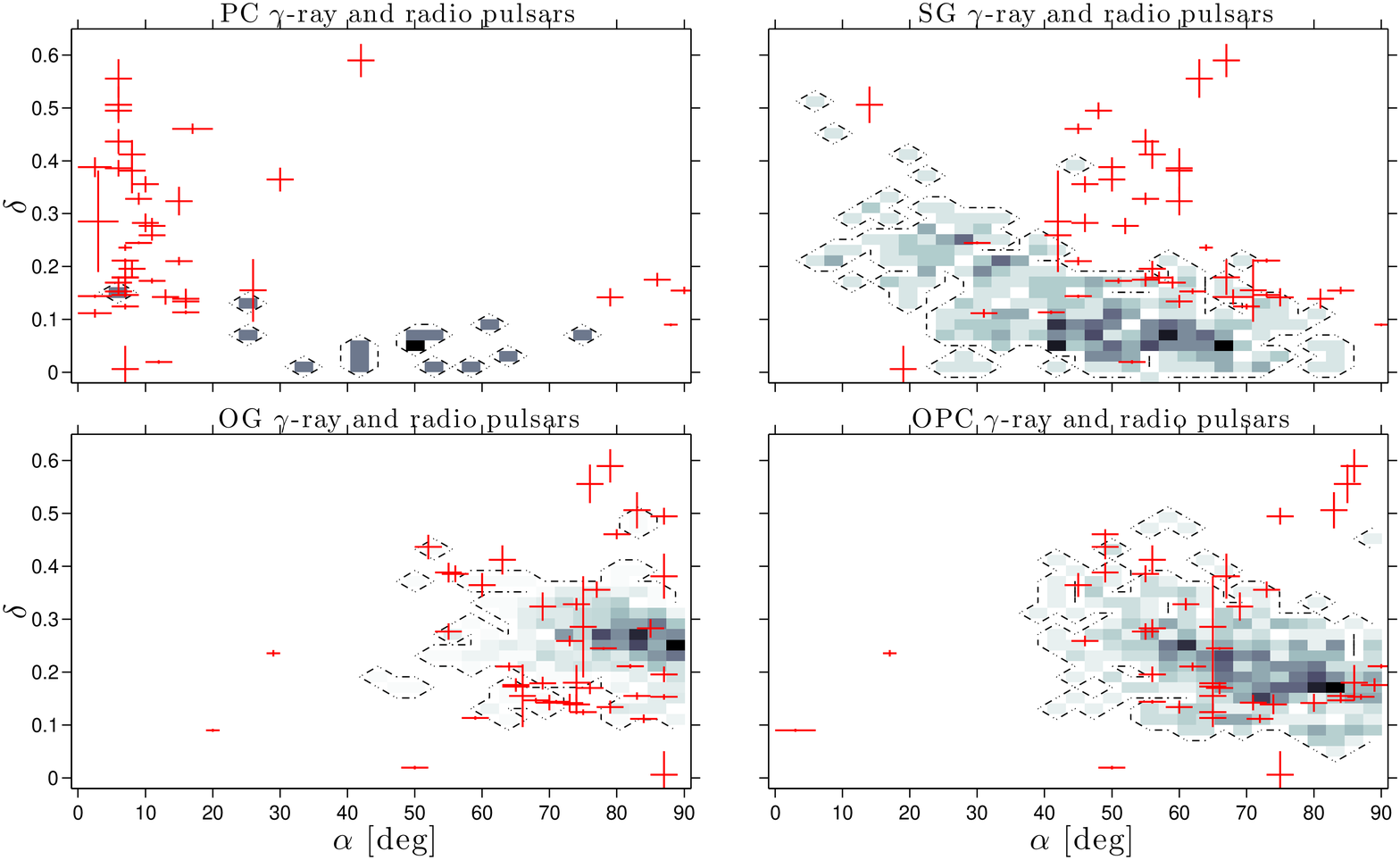} 
\includegraphics[width=0.49\textwidth]{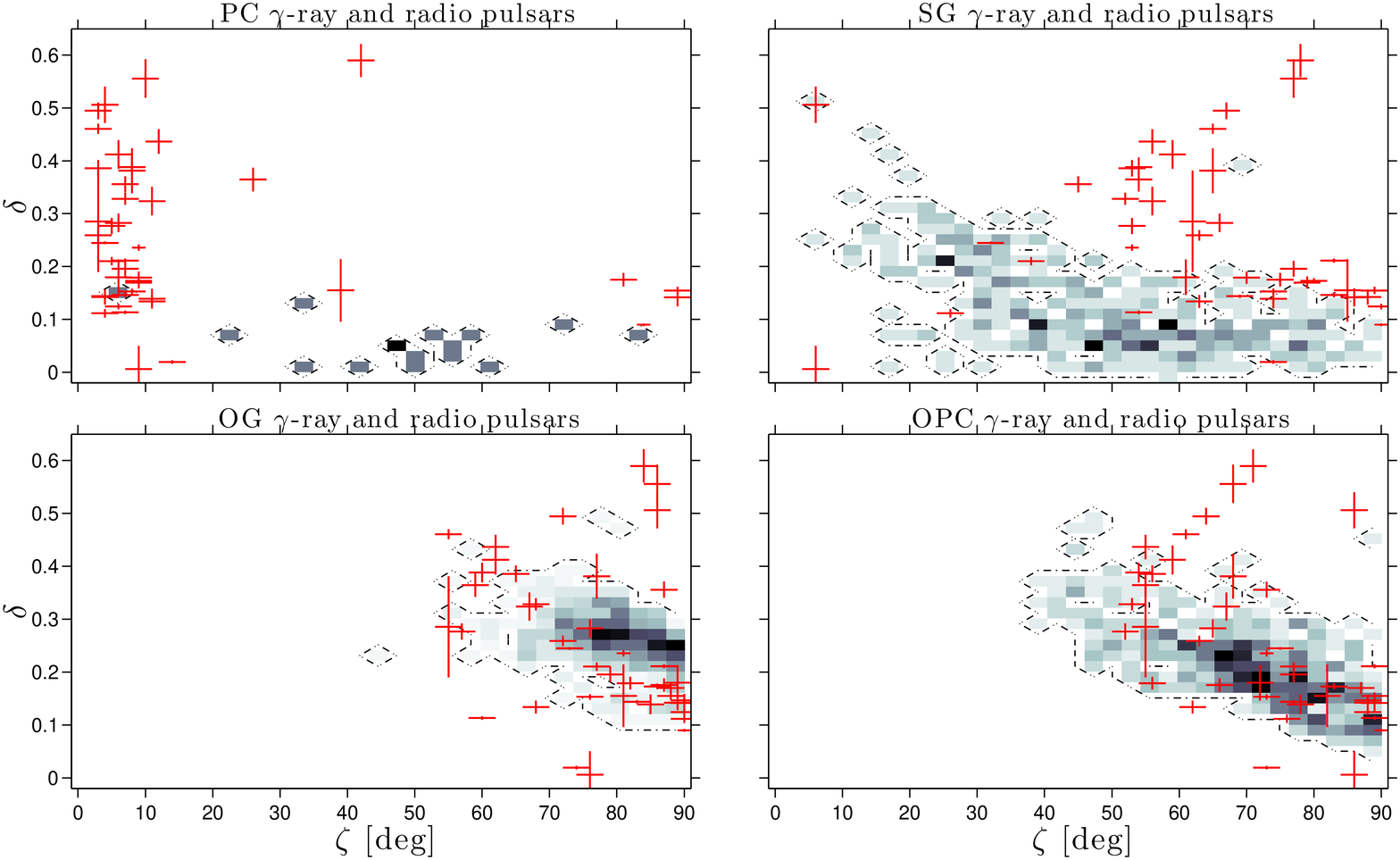} 
\caption{Distributions of the radio lag of simulated (grey) and observed (red) pulsars obtained, for each model, as a function of the $\gamma$-ray 
beaming factor, magnetic obliquity, and observer line of sight are shown in the top, central, and bottom panels, respectively. The model distributions 
have been obtained as two-dimensional number-density histograms. The dash-dotted line highlights the minimum number density contour. The LAT 
measurements are given with 1$\sigma$ error bar uncertainty.} 
\label{RadLagBAZ}
\end{center}
\end{figure}

Figure \ref{RadLagBAZ} compares estimated and simulated trends for $\delta$ as functions of $f_{\Omega}$,  $\alpha$, and $\zeta$, in the 
top, middle, and bottom panels, respectively. In the $f_{\Omega}$-$\delta$ plane, the two-pole emission geometries PC and SG, do not predict any trend 
for $\delta$ changing with increasing $f_{\Omega}$ while the one-pole caustic models, OG and OPC, predict mild decreasing $\delta$ as $f_{\Omega}$ 
increases. The estimates obtained in the framework of PC and SG emission geometries fail to match the model prediction and do not match the bulk 
of the objects modelled while OG and OPC estimates match the bulk of the model prediction and seem to best represent the modelled behaviour showing,
the highest and second highest $p_\mathrm{value}$ (see Table \ref{Tab5}).

In the $\alpha$-$\delta$ and $\zeta$-$\delta$ planes, a decreasing trend of $\delta$ as $\alpha$ and $\zeta$ increase clearly shows up in the 
simulations for all models except for the PC. This behaviour can be explained in the light of figure \ref{Em_Pat}; as $\alpha$ increases, 
the bright caustic approaches the 
radio emission beam, thereby decreasing $\delta$. The very same behaviour is observed in each phase-plot panel when $\zeta$ increases for fixed $\alpha$ 
values; in both SG and OG/OPC emission geometry the bright caustic gets closer to the radio emission beam when $\zeta$ increases. The same decreasing 
trend should also be observed in the PC model prediction since $\gamma$-ray and radio emission beams get closer when $\alpha$ increase but the trend 
does not show up because of the paucity of simulated PC objects. The outer and intermediate-high magnetosphere model estimates from OPC and SG best 
match the modelled trends showing the largest $p_\mathrm{value}$ values (see Table \ref{Tab5}.)
\begin{figure}
\begin{center}
\includegraphics[width=0.49\textwidth]{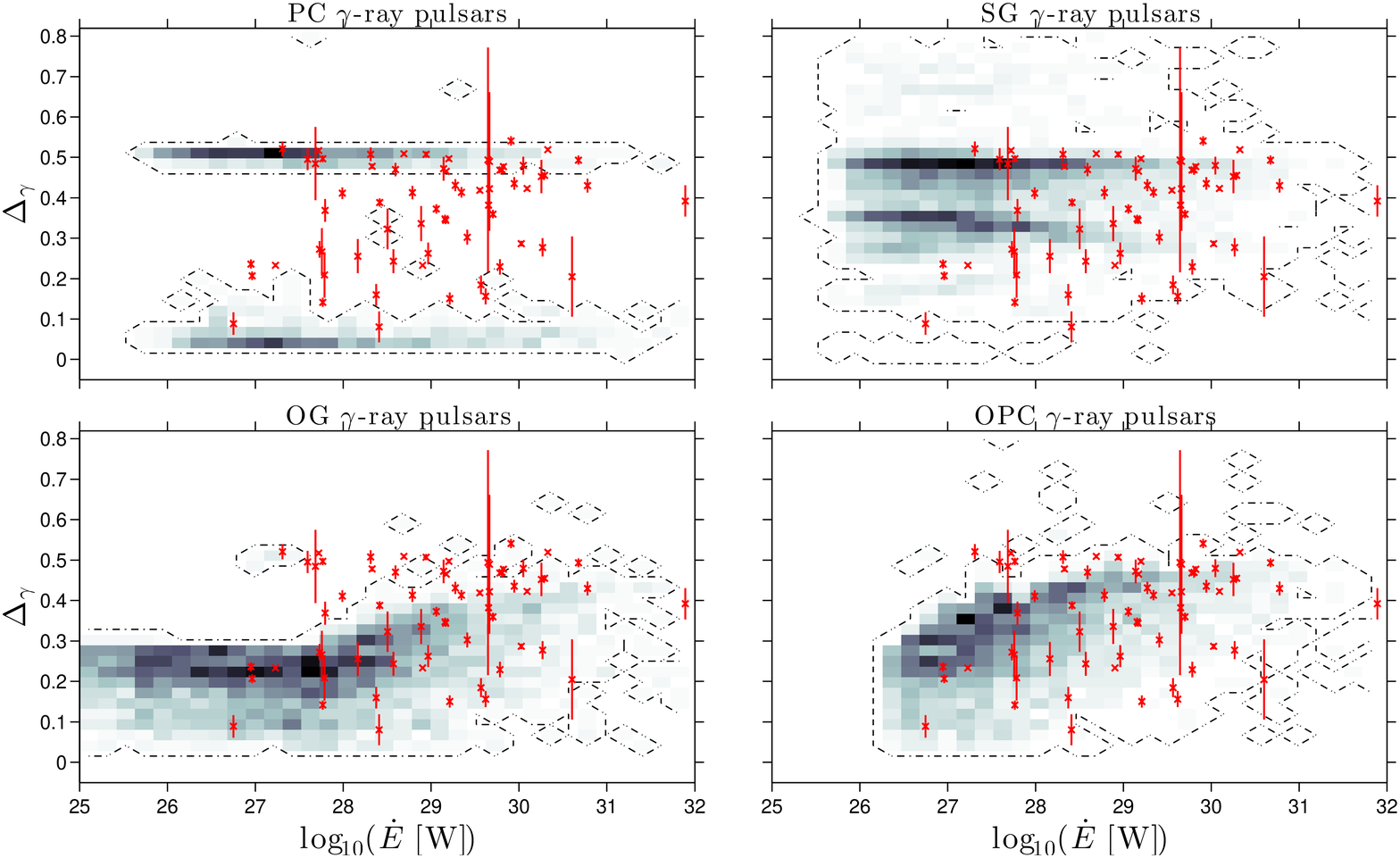} 
\includegraphics[width=0.49\textwidth]{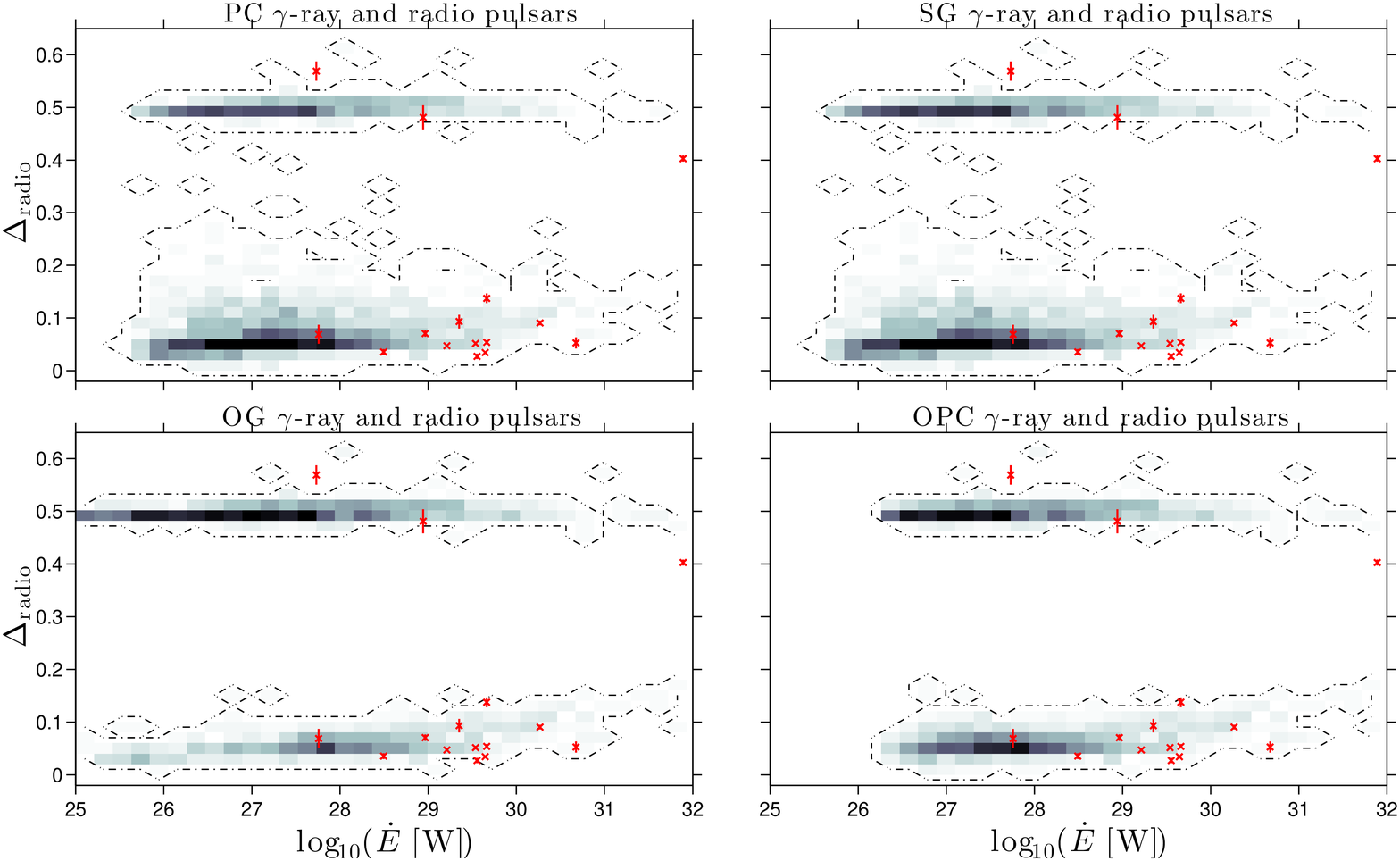} 
\includegraphics[width=0.49\textwidth]{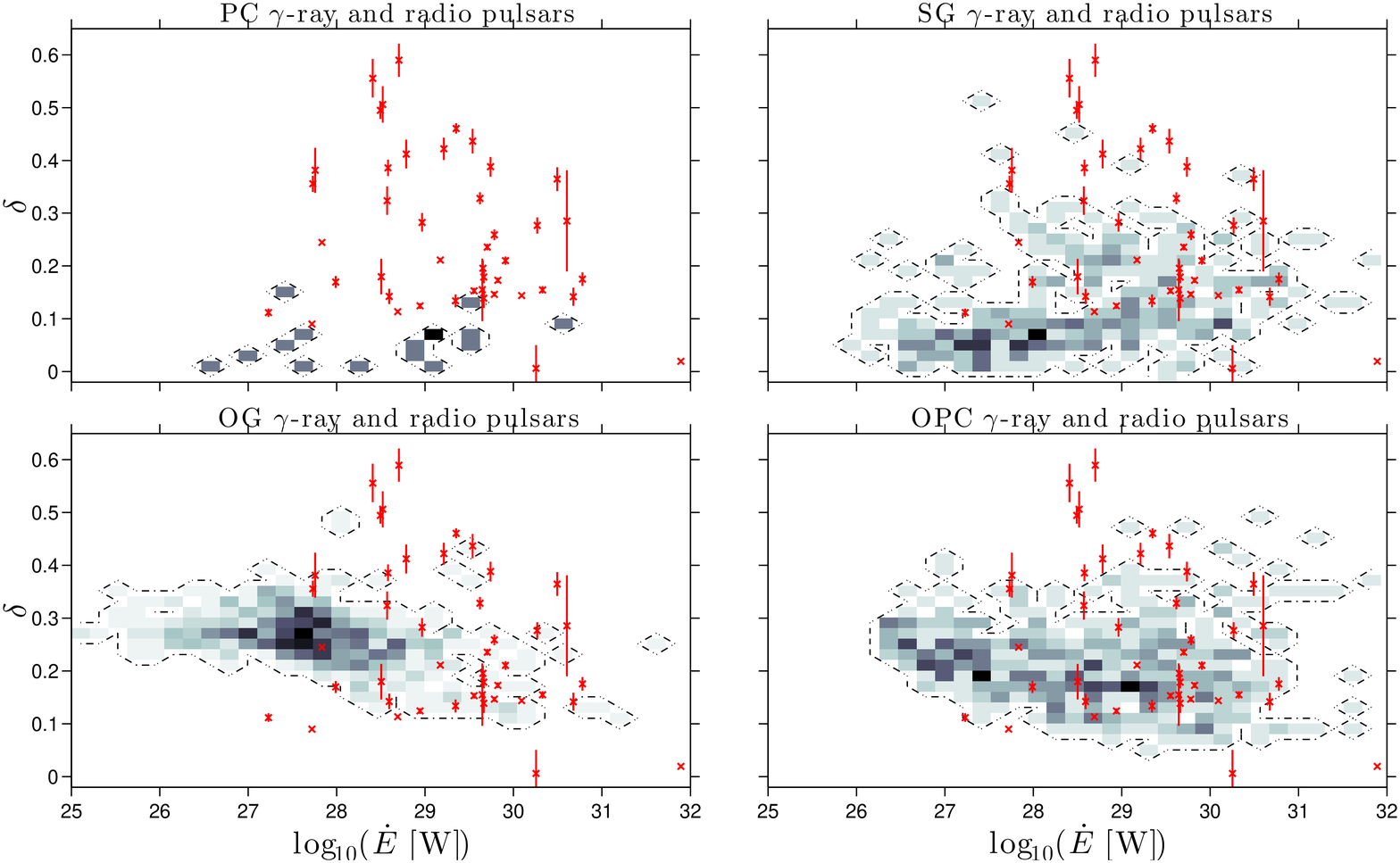} 
\caption{Distributions of the $\gamma$-ray peak separation, radio peak separation, and radio lag of simulated (grey) and observed (red) pulsars 
obtained, for each model, as a function of the spin-down power are shown in the top, central, and bottom panels, respectively. The model distributions 
have been obtained as two-dimensional number-density histograms. The dash-dotted line highlights the minimum number density contour. The LAT 
measurements are given with 1$\sigma$ error bar uncertainty.}
\label{G_PeakSep_d_Edot}
\end{center} 
\end{figure}

\subsection{Observable pulsar parameters: $\Delta_\gamma$, $\Delta_\mathrm{Radio}$, and $\delta$ as a function of spin-down power 
$\dot{E}$ and spin period P}
\label{Observable}

We compare the simulated and observed correlations between $\Delta_{\gamma}$, $\Delta_\mathrm{Radio}$, and $\delta$ and pulsar spin-down power 
$\dot{E}$ and spin period P. Simulated and observed $\dot{E}$ values were computed as described in PIERBA12\nocite{pghg12} while the  
LAT pulsar spin periods have been taken from PSRCAT2\nocite{aaa+13}. The statistical agreements between observed and simulated 
distributions are given in Table \ref{Tab5}. Figure \ref{G_PeakSep_d_Edot} compares observed and simulated 
trends of $\Delta_{\gamma}$, $\Delta_\mathrm{Radio}$, and $\delta$ as a function of $\dot{E}$ in the top, middle, and bottom panels, respectively. 
The comparisons of simulated and observed trends in the framework of all models reflect the same inconsistency shown in PIERBA12; none 
of the simulated emission geometries manage to explain the behaviour of the observed sample at high $\dot{E}$ showing a deficiency of 
high-$\dot{E}$ objects with respect to the observations. 
\begin{figure}
\begin{center}
\includegraphics[width=0.49\textwidth]{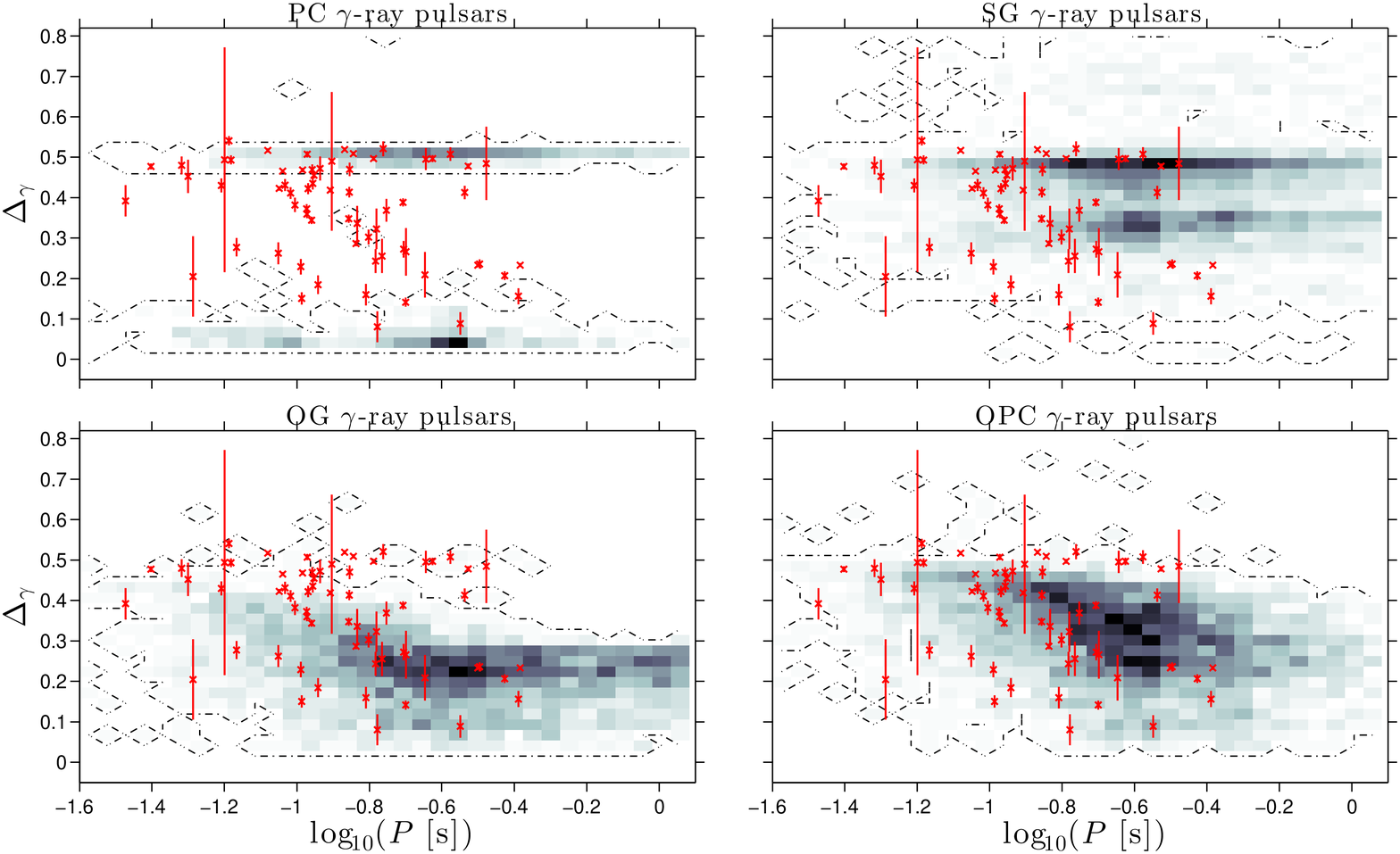} 
\includegraphics[width=0.49\textwidth]{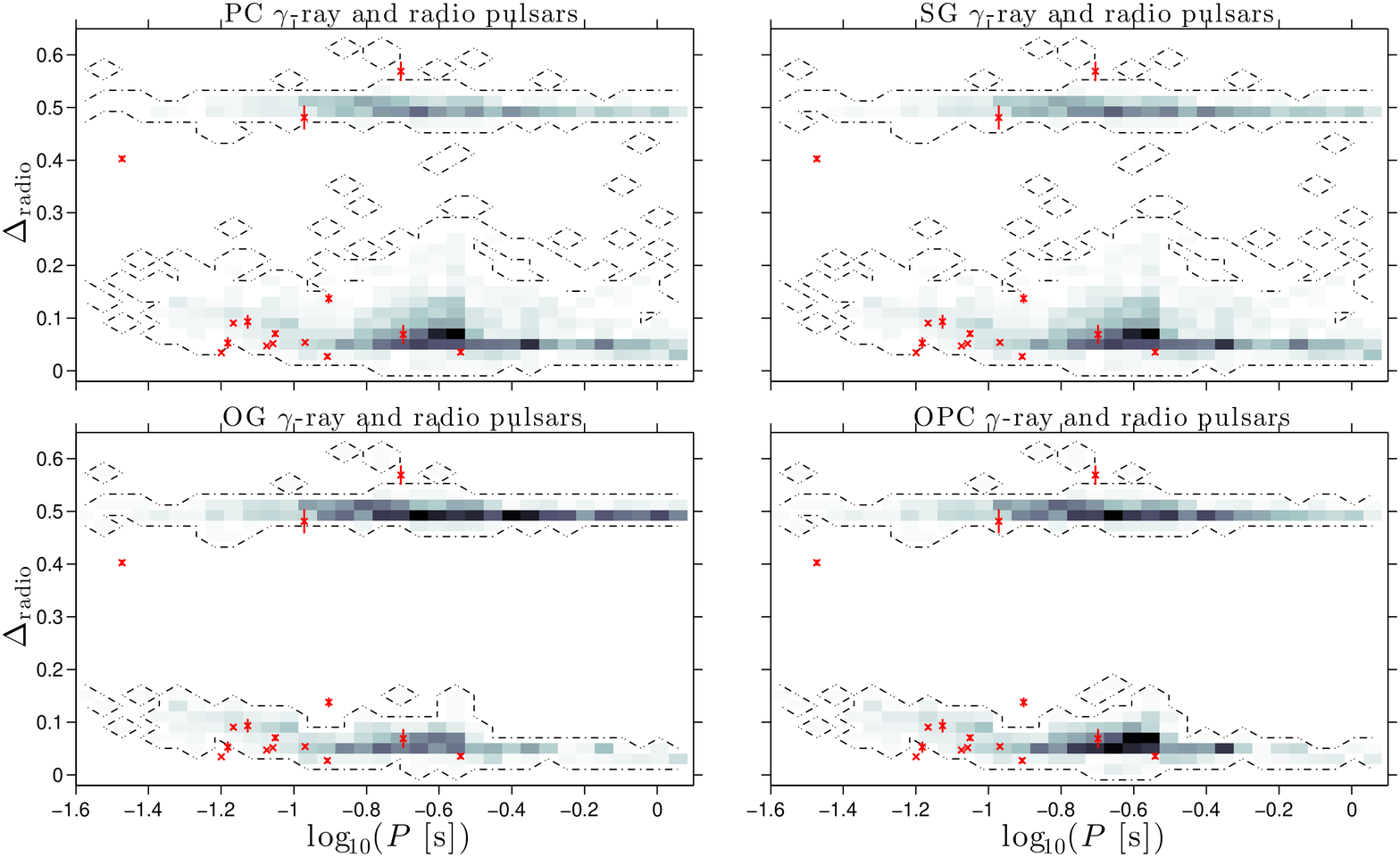} 
\includegraphics[width=0.49\textwidth]{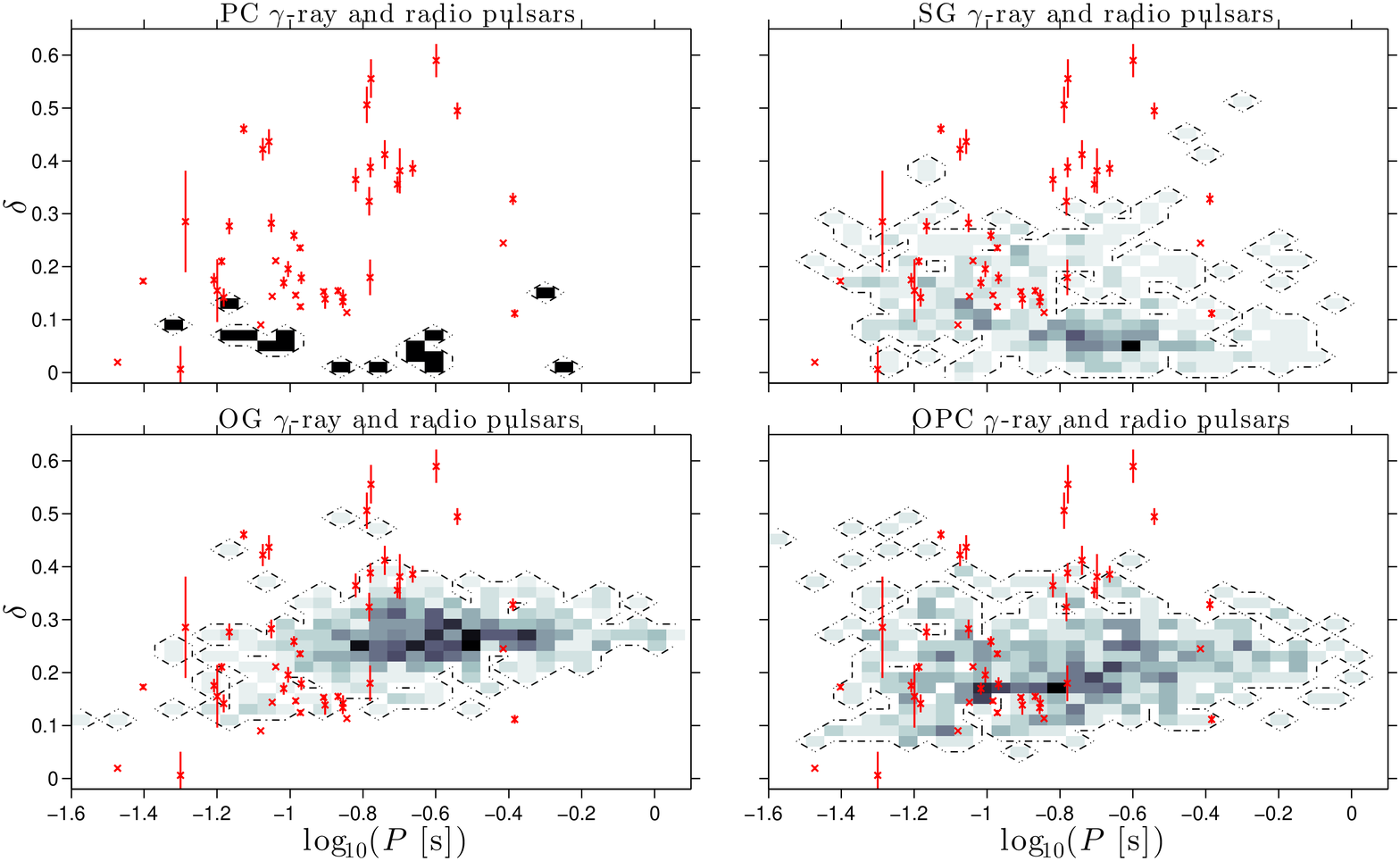} 
\caption{Distributions of the $\gamma$-ray peak separation, radio peak separation, and radio lag of simulated (grey) and observed (red) pulsars 
obtained, for each model, as a function of the spin period are shown in the top, central, and bottom panels, respectively. The model distributions 
have been obtained as two-dimensional number-density histograms. The dash-dotted line highlights the minimum number density contour. The LAT 
measurements are given with 1$\sigma$ error bar uncertainty.}
\label{G_PeakSep_d_Per}
\end{center}
\end{figure}
In the plane $\Delta_\gamma$-$\dot{E}$ the PC model completely fails to explain the observations. As already shown in Figure 
\ref{G_PeakSepHist}, the SG model predicts a double $\Delta_\gamma$ distribution. The first distribution is
centred at $\Delta_\gamma$=0.5 and is given by the separation of the peaks coming from each magnetic pole caustic at high $\alpha$ and 
does not change with increasing $\dot{E}$. The second distribution is centred at $\Delta_\gamma$=0.35, is given by the separation of the 
double peaks generated by the  same magnetic-pole caustic, and shows a mild decrease with increasing $\dot{E}$ due to the shrinking 
of the SG beam as $\dot{E}$ increases. 
The models SG and OPC best explain the observations with OPC best predicting the observed behaviour. The OG model over
predicts low $\dot{E}$ objects for $\dot{E}<10^{26.5}$W, while OPC does not predict any object in the same interval.
The $p_\mathrm{value}$ values listed in Table \ref{Tab5} shows that in most of the cases, OPC and OG describe the observed distributions at 
the higher CL with the SG predictions best explaining the observations on plane $\dot{E}$-$\Delta_\gamma$.

The $\dot{E}$-$\Delta_\gamma$ correlation was also studied in the framework of PSRCAT2\nocite{aaa+13}.  A comparison of the LAT pulsars 
$\dot{E}$-$\Delta_\gamma$ correlation obtained here and in the framework of PSRCAT2\nocite{aaa+13} shows consistency between the two 
$\Delta_\gamma$ computation techniques. As a result of more accurate $\Delta_\gamma$ measures, the $\dot{E}$-$\Delta_\gamma$ 
correlation obtained by PSRCAT2\nocite{aaa+13} shows clearer evidence of a double trend that is similar to our SG model predictions but centred at 
lower $\Delta_\gamma$ values of 0.43 and 0.22, with no apparent decrease of the $\Delta_\gamma$ as $\dot{E}$ increase. 
In the $\dot{E}$-$\Delta_\mathrm{Radio}$ plane all $\gamma$-ray-selected radio pulsars show the same double trend at constant $\dot{E}$ values, 
0.05 and 0.5, with the OPC-selected radio sample preferred to explain the observations. In the $\dot{E}$-$\delta$ plane, all implemented 
emission geometries fail to explain how the radio lag changes with increasing $\dot{E}$. Both PC and SG models do not show any 
trend in $\delta$ changing with increasing $\dot{E}$, while the one-pole caustic models, OG and OPC, show a mild trend in $\delta$ decreasing 
with increasing $\dot{E}$. 

Figure \ref{G_PeakSep_d_Per} compares observed and simulated trends of $\Delta_{\gamma}$, $\Delta_\mathrm{Radio}$, and $\delta$ as a function 
of spin period P in the top, middle, and bottom panels, respectively. In contrast tom the $\dot{E}$ computation, which requires assumptions on the 
pulsar mass, radius, and moment of inertia, the spin period is na assumption-independent of observed characteristics. An increase of P implies 
an increase of the light-cylinder radius corresponding to a decrease of the open magnetic field-line region. 
In the P-$\Delta_\gamma$ plane, the observed sample shows a 
two-component distribution: one with objects characterised by $\Delta_\gamma\sim$0.5 as $P$ increases and one showing 
$\Delta_\gamma$ decreasing as P increases for $\Delta_\gamma\lsimeq$0.45. Both OG and OPC predict the observed decreasing 
component as P increases but overestimate the high P objects and fail to explain the observed flat component at
$\Delta_\gamma\sim$0.5. The OPC best explains the observations since best describes the observed distribution over the entire $P$ interval. 
In the PC case, the model predictions completely fail to explain the observed decreasing trend. The increase of the light-cylinder 
radius implies a shrinking of the conical emission beams and a  decrease of $\Delta_{\gamma}$ with P. 
\begin{figure*}
\begin{center}
\includegraphics[width=0.495\textwidth]{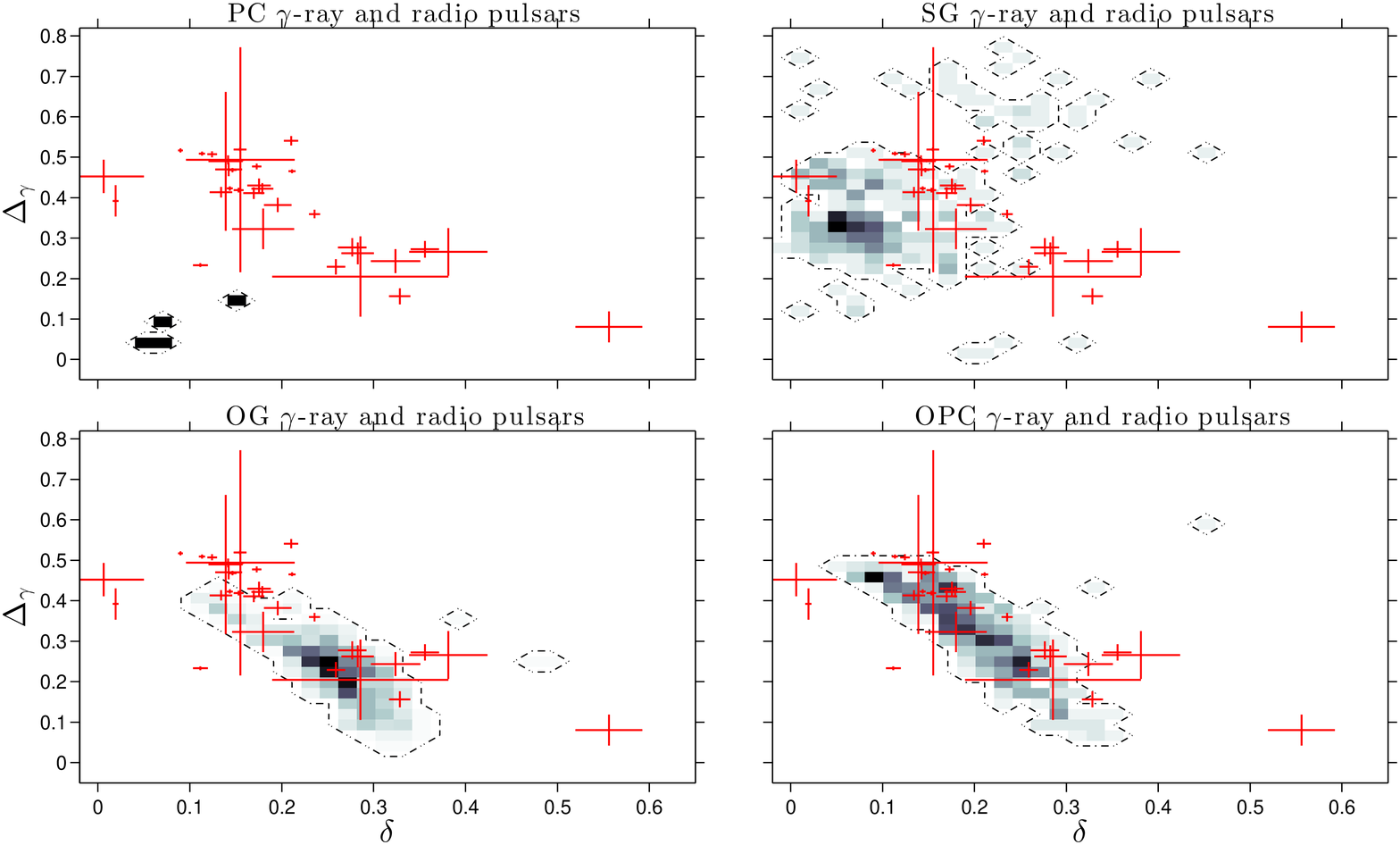}
\includegraphics[width=0.495\textwidth]{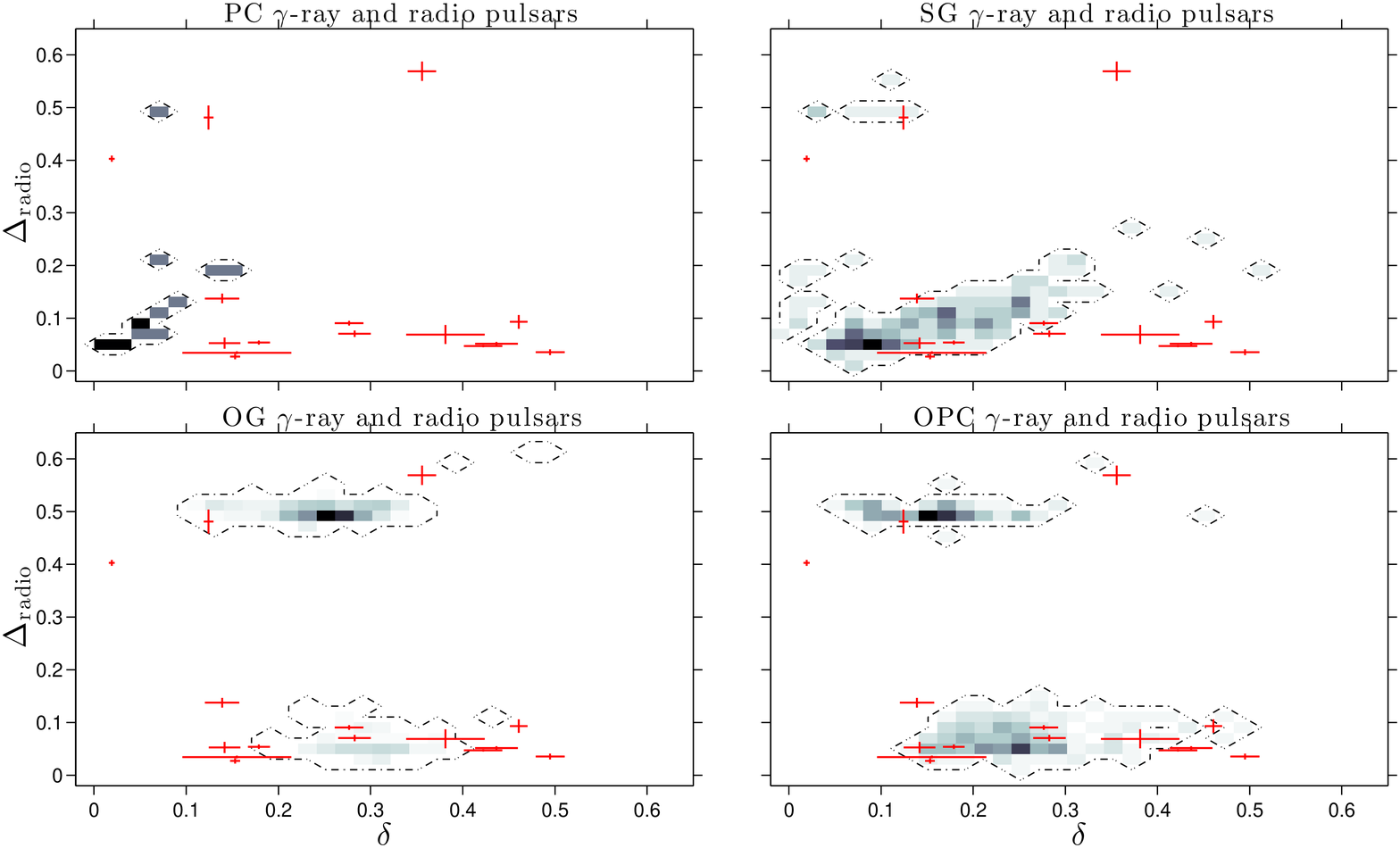}
\caption{Distributions of the $\gamma$-ray and radio peak separation of simulated (grey) and observed (red) pulsars 
obtained, for each model, as a function of radio lag are shown in the left and right panels, respectively. The model distributions 
have been obtained as two-dimensional number-density histograms. The dash-dotted line highlights the minimum number density contour. 
The LAT measurements are given with 1$\sigma$ error bar uncertainty.}
\label{RadLagPksepG}
\end{center}
\end{figure*}
Because of the 
paucity of PC simulated objects, a mild peak separation decrease is observed just for the $\gamma$-ray-selected radio simulated 
objects and for P$<$125ms. The SG model predicts a constant double distribution centred at $\Delta_{\gamma}$ values of 0.35 and 0.5. 
The model explains the constant $\Delta_{\gamma} \sim 0.5$ branch but fails 
to describe the observed decreasing trend. The fact that PS/SG and OG/OPC best predict the flat and decreasing component of the observed
distribution, respectively, is because of the two-pole and one-pole nature of their emission, respectively; the two-pole emission predicts 
more 0.5 separated peaks from 
both the poles, while in one-pole emission, 0.5 separated peaks only occur in the shorter P light curves.
In the P-$\Delta_\mathrm{Radio}$ plane the radio objects selected in framework of each $\gamma$-ray 
model give the very same explanation of the observed points and cannot be used to discriminate the model that best explains the observations.

In the plane P-$\delta$ none of the emission geometries manages to describe the observed trend but the OPC models seem to give  a 
better description of the observations. The shrinking of the open field-line
region as a consequence of the increase of P implies a simultaneous shrinking of both radio and $\gamma$-ray beams in the framework
of each model. In the SG case (and mildly in the PC), we observe a decreasing trend in $\delta$ as P increases, which could be explained 
by a faster shrinking of the radio emission beam with respect to the PC beam and as a consequence of the SG caustic approaching the radio 
beam as P increases, respectively. The OG and OPC model predictions are not able to explain the observations; both OG and OPC simulations 
are characterised by a mild increase of $\delta$ as P increases, which is not present in the observed distribution. The statistical 
significance with which the modelled distributions for $\Delta_{\gamma}$, $\Delta_\mathrm{Radio}$, and $\delta$ as a function of $P$ explain 
the observations partly mimic that obtained for the same parameters as a function of $\dot{E}$. The OPC model explains the observed distributions 
at the higher CL in the planes $\Delta_{\gamma}$-P and $\delta$-P, while the PC model predictions best explain the observations on the plane
$P$-$\Delta_\mathrm{Radio}$.


\subsection{$\gamma$-ray and radio peak separation as a function of the radio lag}
\label{RlagPsep}

The relation $\Delta_{\gamma}$ vs. $\delta$ is particularly important in the study of the pulsar emission geometry and 
magnetosphere structure since it correlates with information on the magnetospheric regions where $\gamma$-ray and radio 
emission are generated: $\Delta_{\gamma}$ is a function of the pulsar gap width structure while $\delta$ constrains the 
relative position of  $\gamma$-ray and radio emission regions. The $\Delta_{\gamma}$ vs. $\delta$ correlation was first
studied by Romani and Yadigaroglu in 1995 \nocite{ry95}. By analysing the light curves of six radio pulsars detected in 
$\gamma$-rays from the \emph{Compton gamma-ray observatory} (CGRO), those authors concluded that $\Delta_{\gamma}$
shows a decreasing trend of increasing $\delta$. With the sizeable increase of the number of $\gamma$-ray and radio active 
pulsars because of the advent of the \emph{Fermi} satellite, it was finally possible to give more precise estimates of the 
$\delta$-$\Delta_{\gamma}$relation. \cite{wrwj09} performed a first measurement of $\delta$ and $\Delta_{\gamma}$ for six 
RL LAT pulsars confirming the trend $\Delta_{\gamma}$ decreasing with increasing $\delta$. The trend was further confirmed 
in the framework of PSRCAT1\nocite{aaa+10} on a sample of 17
RL pulsars and by independent analyses on the same pulsar sample performed by \cite{pie10}. The last  and more accurate 
$\delta$ and $\Delta_{\gamma}$ measurements were obtained in the framework of PSRCAT2\nocite{aaa+13}; these measurements concern 32 
pulsars and confirm the previous findings of $\Delta_{\gamma}$ decreasing with increasing $\delta$. 
In all the above mentioned cases, the radio lag $\delta$ was computed as the distance between the fiducial phase computed 
according to PSRCAT2 criteria (estimated position of the magnetic pole) and the phase of the following$\gamma$-ray peak. 
Since radio and $\gamma$-ray peak refer to the same magnetic pole, no $\delta>0.5$ is allowed.

We compared simulated and observed correlations  $\Delta_{\gamma}$ vs. $\delta$ and $\Delta_\mathrm{Radio}$ vs. $\delta$. 
For both simulated and observed objects, $\Delta_{\gamma}$ and $\Delta_{\mathrm{Radio}}$ have been computed  as described 
in Section \ref{peak separation} while the computation technique of $\delta$ is described in Section \ref{Radio lag}. 
We studied how the radio peak separation changes with increasing $\delta$ to
constrain the structure of the radio emission region against the models.

Figure \ref{RadLagPksepG} compares observed and simulated trends for $\Delta_{\gamma}$ vs. $\delta$ and $\Delta_{\mathrm{Radio}}$ 
vs. $\delta$ in the left and right panels, respectively. In the $\delta$-$\Delta_{\gamma}$ plane, the LAT objects follow a clearly decreasing  
trend of $\Delta_{\gamma}$ when $\delta$ increases.
The PC model fails completely to explain the observations, predicting a distribution in total disagreement with the observations. 
Points are only predicted for small $\delta$ and $\Delta_{\gamma}$ values
and seem to show a trend of increasing $\Delta_{\gamma}$ with increasing $\delta$. The same increasing trend and inconsistency with
the observations was obtained by  \cite{pie10} comparing simulations and PSRCAT1 observations \nocite{aaa+10}. The SG 
model does not explain the observed decreasing trend, predicts observations mainly at low $\delta$ and intermediate $\Delta_{\gamma}$ 
and shows a mild increasing trend of $\Delta_{\gamma}$ with increasing $\delta$. Comparisons of SG simulations against observations were 
performed by \cite{wrwj09} and by \cite{pie10} on six $\gamma$-ray pulsars (see references in \cite{wrwj09}) and on the pulsars from
PSRCAT1, \nocite{aaa+10} respectively. The two studies found results consistent with each other and also conclude that the 
SG model does not explain the observed $\delta$-$\Delta_{\gamma}$ trend.
The one-pole emission geometry manages to explain the LAT findings successfully. Both the OG and OPC models manage to reproduce the 
shape of the observed distribution with the OPC showing higher agreement with observations. The $p_\mathrm{value}$ shown in Table 
\ref{Tab5} shows that, not including the PC model that shows a statistics that is too low to be correctly compared with observations, 
the OPC and OG are the models that explain the observed $\Delta_\gamma$ as a function of $\delta$ with the highest and 
second highest CL, respectively. Comparisons of OG simulations against observations were also performed by 
\cite{wrwj09} and \cite{pie10} on six $\gamma$-ray pulsars and on the pulsars from PSRCAT1, respectively (see \cite{wrwj09} 
and PSRCAT1\nocite{aaa+10} for references about the LAT pulsars). In both studies, the OG model predicts a decreasing trend of 
$\Delta_{\gamma}$ as $\delta$ increases consistent with the current findings.
It is proper to stress again (see Section \ref{Radio lag}) that the results obtained in this study are strongly dependent on the VRD 
magnetic field geometry assumed. The assumption of a more realistic DM geometry would imply larger radio to $\gamma$-ray lags predicted 
by all implemented $\gamma$-ray models \citep{khkc12}. This would shift all observed distributions towards higher $\delta$ and would probably 
imply an improvement of the SG model predictions in explaining the observations.

In the plane $\delta$-$\Delta_\mathrm{Radio}$ the LAT objects follow a clear flat trend with $\Delta_\mathrm{Radio}$ stable at 0.05 with 
increasing $\delta$. Both PC and SG models predict an increase of $\Delta_\mathrm{Radio}$ as $\delta$ increases, which is not observed in the
LAT data. The simultaneous increase of $\Delta_{\gamma}$ and $\Delta_\mathrm{Radio}$ as $\delta$ increases could be explained by the radio
emission beam size, which increases at a higher rate than the PC and SG beam sizes, thereby increasing the distance between the first radio 
peak and the first $\gamma$-ray peak. Both outer magnetosphere models, OG and OPC, correctly predict the observed behaviour with 
the OPC explaining the observed range of $\delta$ values. In the $\delta$-$\Delta_\mathrm{Radio}$ plane, the $p_\mathrm{value}$ 
shows that the OG and PC are the models that describe the observations with the highest and second highest CL, respectively.

\section{The $\alpha$-$\zeta$ plane}
\label{AZpl}
\begin{figure}
\begin{center}
\includegraphics[width=0.46\textwidth]{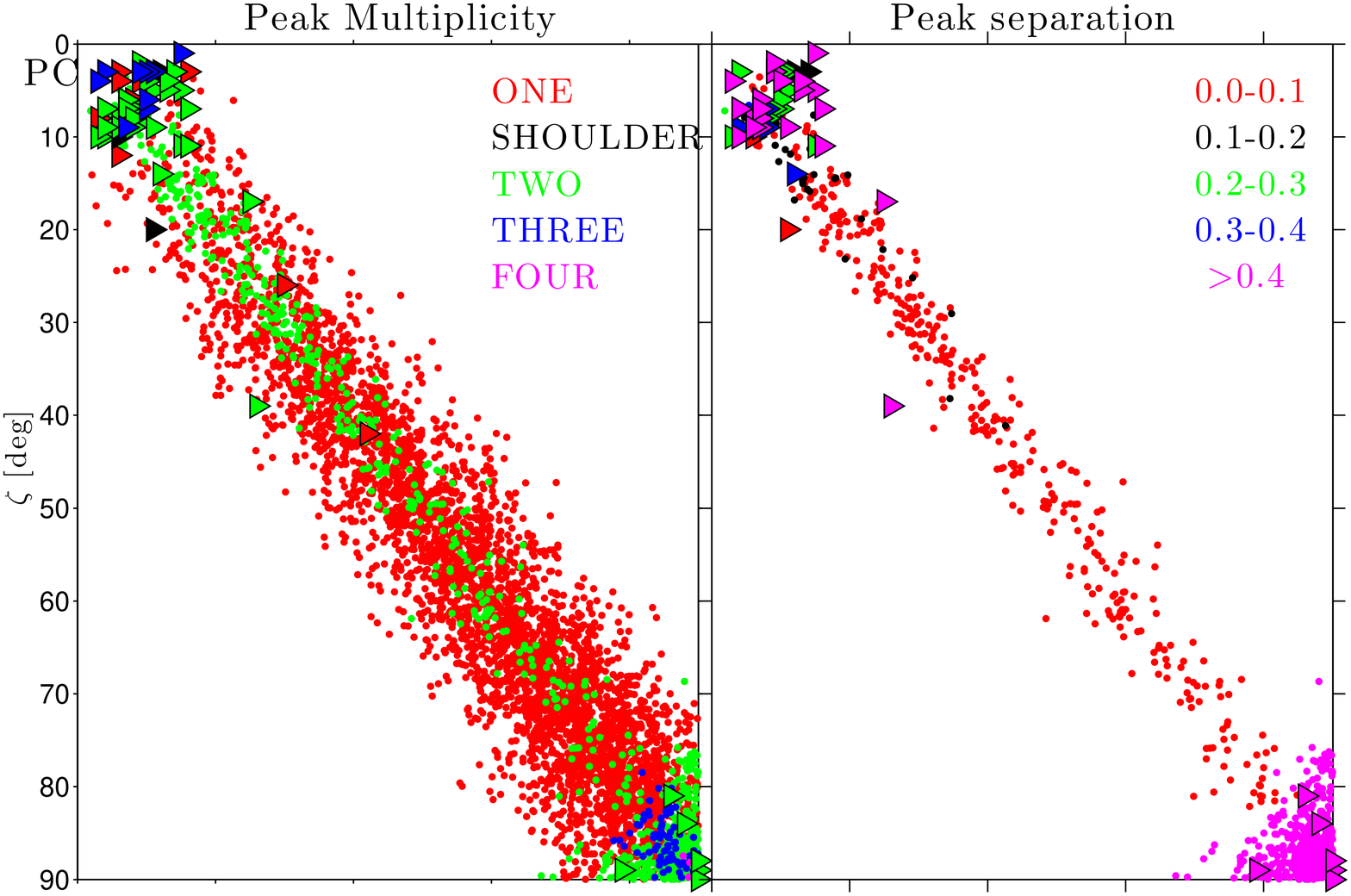}
\includegraphics[width=0.46\textwidth]{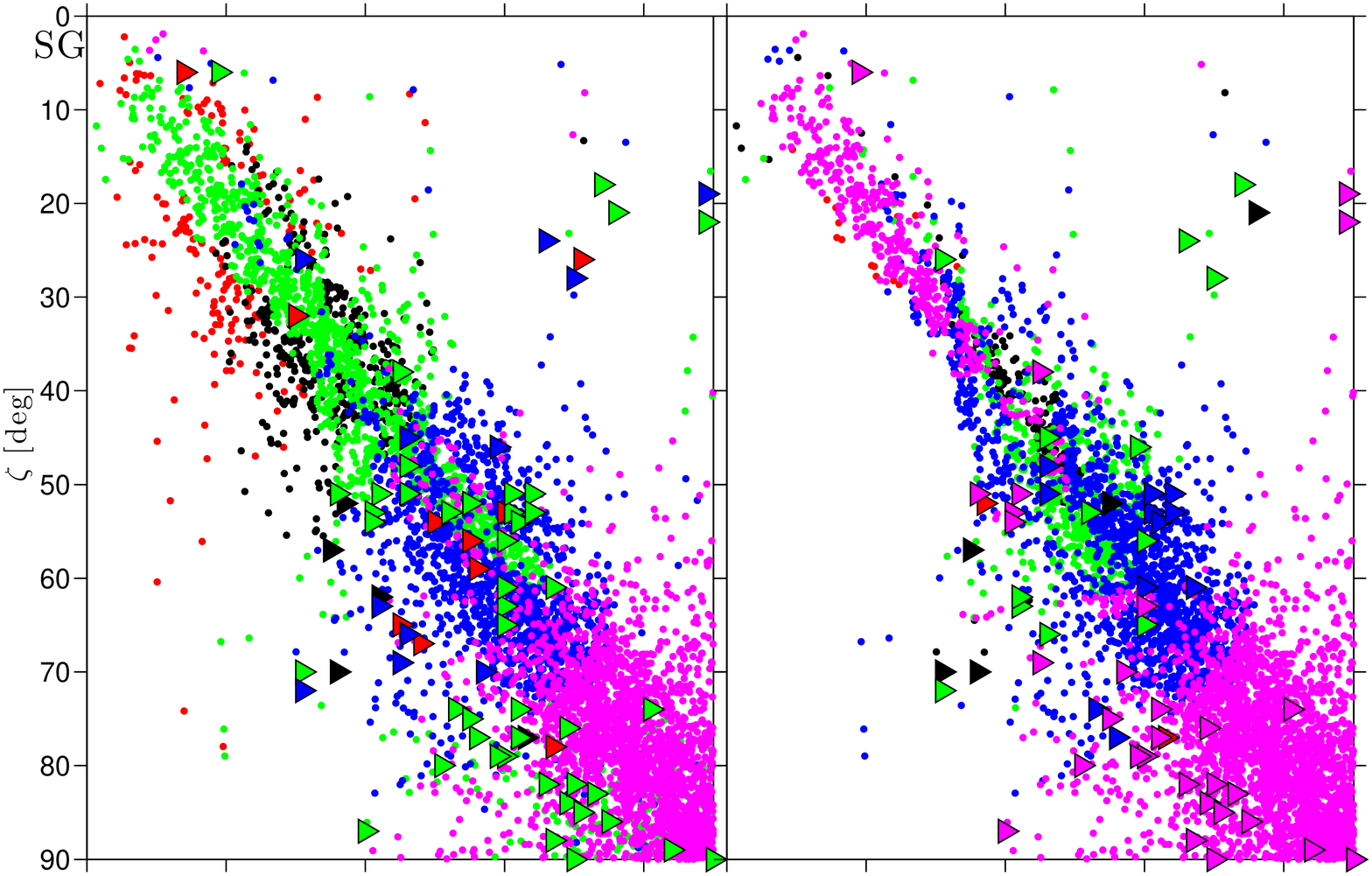}
\includegraphics[width=0.46\textwidth]{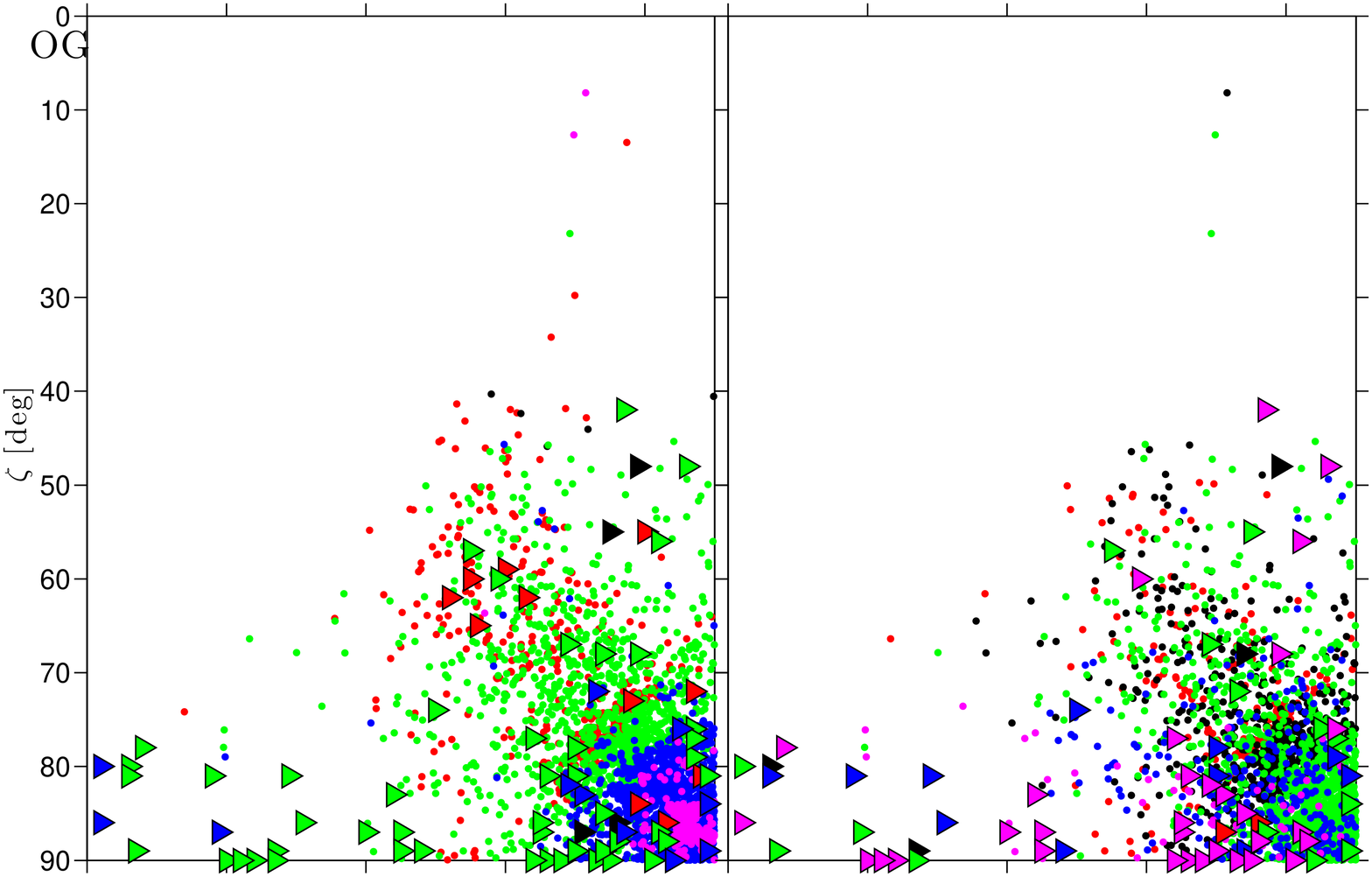}
\includegraphics[width=0.46\textwidth]{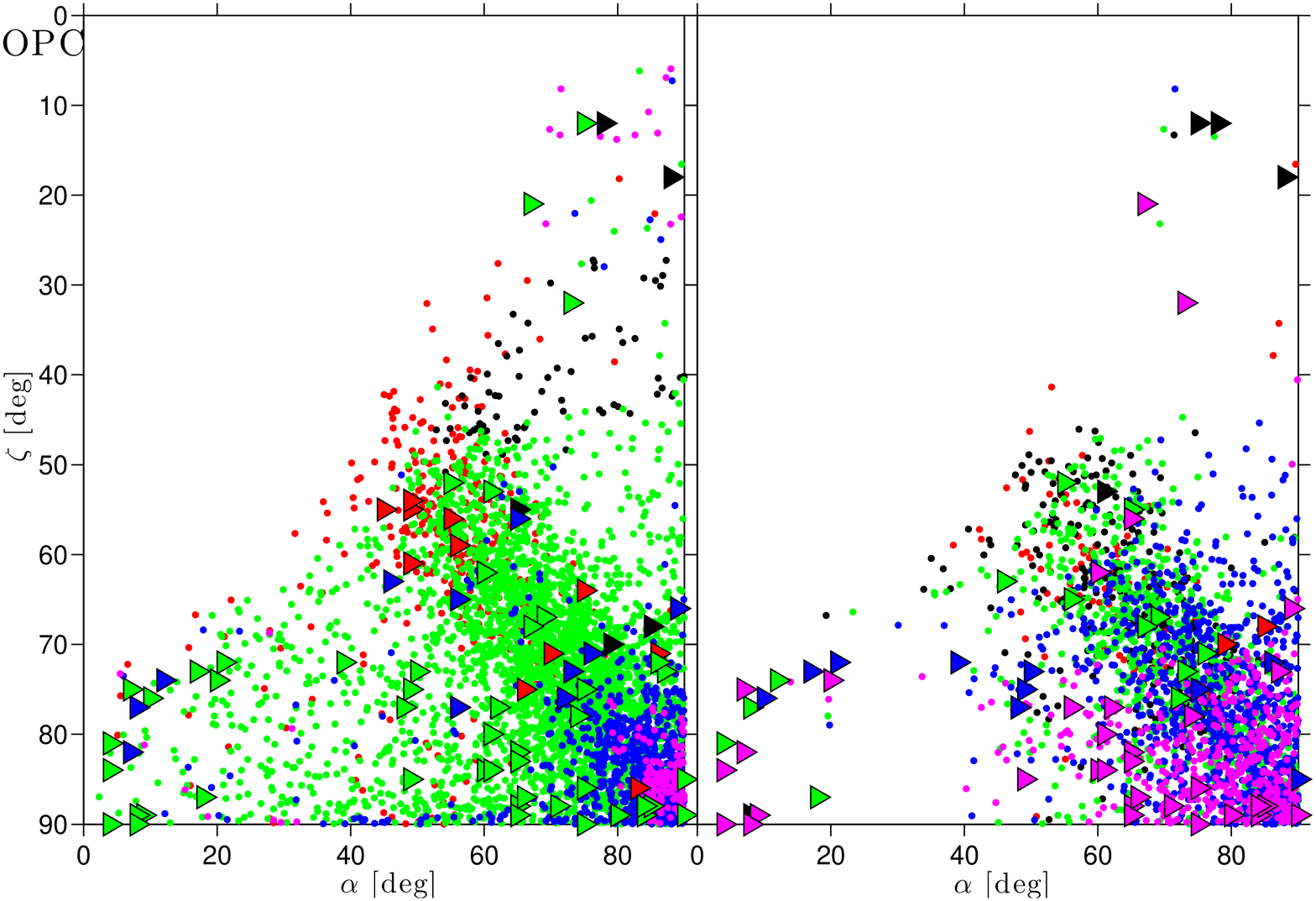}
\caption{From the first to the fourth row, the $\gamma$-ray light curve multiplicity (left column) and peak separation (right column)
as a function of $\alpha$ and $\zeta$ for LAT pulsars and PC, SG, OG, and OPC models are shown respectively. Big triangles and 
small points refer to LAT and simulated pulsars, respectively.}
\label{AZMulti}
\end{center}
\end{figure}

Figure \ref{AZMulti} compares light-curve multiplicity and peak separation as a function of $\alpha$ and $\zeta$ for modelled and observed $\gamma$-ray 
light curves. The $\alpha$ and $\zeta$ of the LAT pulsars are taken from PIERBA15\nocite{phg+15} while the $\alpha$ and $\zeta$ of the simulated objects
are those synthesised by PIERBA12\nocite{pghg12}. Parameters $\alpha$ and $\zeta$ are non-observable pulsar characteristics and, analogous to what we previously 
discussed in Section \ref{NonObservable}, a match between LAT pulsar estimates and model prediction suggests that that model explains the observed 
variety of $\gamma$-ray pulsar light curves.

The plane $\alpha$ vs. $\zeta$ as a function of the peak multiplicity and peak separation and in the 
framework of PC, SG (two pole emission geometries), and OG (one pole emission geometry) was also studied by \cite{wrwj09}. Our study shows that the 
geometrical characteristics as a function of the pulsar orientation of the light curves synthesised by PIERBA12\nocite{pghg12} are consistent with the 
simulations by \cite{wrwj09}. Moreover our analysis characterises the 
shoulder pulse shape that is recurrent among both synthesised and observed objects. Shoulder shapes are observed in the framework of the SG and OG/OPC
geometries; in the SG model, the shoulder shape is associate with the $\alpha$-$\zeta$ plane region between single- and double-peak profiles, while 
in the OG/OPC
geometry they do not represent the transition between single- and double-peak shapes but they are recurrent at high $\alpha$ and high $\zeta$. 
The best agreement between 
estimated parameters and the bulk of the model prediction is observed in the framework of the OPC model but there are inconsistencies in the peak separations plot (Figure \ref{AZMulti}), with the OG and OPC model predicting too few small $\alpha$ values than the fits to the LAT pulsars. 
The PC estimates completely fail to match the
major part of the PC simulations for all $\alpha$ and $\zeta$. The SG estimates show inconsistencies in the light-curve peak multiplicity and pulsar
orientation; many two peak pulsars are found at high $\alpha$ and low $\zeta$ where four peak profiles are found (see left side SG panel of Figure  \ref{AZMulti}). The OG/OPC model estimates are consistent with model predictions for what concern both multiplicity and peak separation. Other plots 
showing $\alpha$ and $\zeta$ as a function of spin period (just for the PC)/gap width (SG, OG, and OPC) and light-curve peak multiplicity, are shown in
Figures \ref{MultiAZ} and \ref{SepAZ}.
\begin{table*}
\begin{center}
\begin{tabular}{| l | c  c | c  c | c  c | c  c |}
\hline
& \multicolumn{2}{c|}{PC} & \multicolumn{2}{c|}{SG}& \multicolumn{2}{c|}{OG}& \multicolumn{2}{|c|}{OPC}\\
\hline
& $D$ & $p_\mathrm{value}$ [\%] & $D$ & $p_\mathrm{value}$ [\%] & $D$ & $p_\mathrm{value}$ [\%] & $D$ & $p_\mathrm{value}$ [\%]  \\
\hline
\hline
\multicolumn{9}{|l|}{Fig. \ref{G_PeakSepBeam}}\\
\hline
$\Delta_\gamma$ vs. $f_\Omega$ 		& 0.72&0.18 	& \cellcolor{gray!60}0.40&\cellcolor{gray!60}0.27 	& \cellcolor{gray!30}0.55&\cellcolor{gray!30}0.25 	& 0.35&24 \\
\hline
$\Delta_\mathrm{radio}$ vs. $f_\Omega$	& \cellcolor{gray!30}0.48&\cellcolor{gray!30}0.18	& 0.29&0.16	& \cellcolor{gray!60}0.35&\cellcolor{gray!60}0.26	& 0.29&0.17  \\
\hline
\hline
\multicolumn{9}{|l|}{Fig. \ref{G_PeakSepalpha}}\\
\hline
$\Delta_\gamma$ vs. $\alpha$ 			& \cellcolor{gray!60}0.69&\cellcolor{gray!60}0.26 	 & \cellcolor{gray!30}0.23&\cellcolor{gray!30}0.22 	 & 0.57&0.21 	 & 0.43&0.20  \\
\hline
$\Delta_\mathrm{radio}$ vs. $\alpha$	& \cellcolor{gray!60}0.54&\cellcolor{gray!60}0.23 	 & \cellcolor{gray!30}0.24&\cellcolor{gray!30}0.20 	 & 0.39&0.19	& 0.39&0.16 \\
\hline
\hline
\multicolumn{9}{|l|}{Fig. \ref{G_PeakSepzeta}}\\
\hline
$\Delta_\gamma$ vs. $\zeta$ 			& 0.73&0.23 	 & \cellcolor{gray!30}0.23&\cellcolor{gray!30}0.22 	 & \cellcolor{gray!60}0.45&\cellcolor{gray!60}0.23 	 & 0.34&0.22 \\
\hline
$\Delta_\mathrm{radio}$ vs. $\zeta$		& 0.51&0.14 	 & 0.24&0.15 	 & \cellcolor{gray!60}0.38&\cellcolor{gray!60}0.22	 & \cellcolor{gray!30}0.37&\cellcolor{gray!30}0.18 \\
\hline
\hline
\multicolumn{9}{|l|}{Fig. \ref{RadLagBAZ}}\\
\hline
$\delta$ vs. $f_\Omega$ 				&  0.50&0.23	 & 0.48&0.24 	 & \cellcolor{gray!60}0.32&\cellcolor{gray!60}0.27  	&  \cellcolor{gray!30}0.34&\cellcolor{gray!30}25\\
\hline
$\delta$ vs. $\alpha$					&  0.52&0.24	 & \cellcolor{gray!30}0.49&\cellcolor{gray!30}0.25	 & 0.38&0.24 	&  \cellcolor{gray!60}0.33&\cellcolor{gray!60}0.28\\
\hline
$\delta$ vs. $\zeta$					&  0.51&0.22	 & \cellcolor{gray!30}0.49&\cellcolor{gray!30}0.24	 & 0.32&0.21 	& \cellcolor{gray!60}0.35&\cellcolor{gray!60}0.27 \\
\hline
\hline
\multicolumn{9}{|l|}{Fig. \ref{G_PeakSep_d_Edot}}\\
\hline
$\Delta_\gamma$ vs. $\dot{E}$ 		& 0.74&0.20 	 & \cellcolor{gray!60}0.53&\cellcolor{gray!60}0.25 	 & 0.52&0.23 	 & \cellcolor{gray!30}0.49&\cellcolor{gray!30}0.23 	 \\
\hline
$\Delta_\mathrm{radio}$ vs. $\dot{E}$	&  \cellcolor{gray!30}0.44&\cellcolor{gray!30}0.23	 &  0.40&0.20	 & 0.43&0.19 	 & \cellcolor{gray!60}0.40&\cellcolor{gray!60}0.29 	\\
\hline
$\delta$ vs. $\dot{E}$				&  0.56&0.24	 & \cellcolor{gray!30}0.49&\cellcolor{gray!30}0.26 	 & \cellcolor{gray!60}0.42&\cellcolor{gray!60}0.28 	 & 0.39&0.24  	\\
\hline
\hline
\multicolumn{9}{|l|}{Fig. \ref{G_PeakSep_d_Per}}\\
\hline
$\Delta_\gamma$ vs. $P$ 			&  0.71&0.22	 & \cellcolor{gray!30}0.53&\cellcolor{gray!30}0.24 	 & 0.47&0.23 	 & \cellcolor{gray!60}0.43&\cellcolor{gray!60}0.27 	 \\
\hline
$\Delta_\mathrm{radio}$ vs. $P$		&  \cellcolor{gray!60}0.48&\cellcolor{gray!60}0.23	 & \cellcolor{gray!30}0.43&\cellcolor{gray!30}0.19 	 & 0.49&0.16 	 & 0.45&0.17  \\
\hline
$\delta$ vs. $P$					&  0.54&0.25	 & 0.50&0.24 	 & \cellcolor{gray!30}0.39&\cellcolor{gray!30}0.25 	 & \cellcolor{gray!60}0.38&\cellcolor{gray!60}0.25 \\
\hline
\hline
\multicolumn{9}{|l|}{Fig. \ref{RadLagPksepG}}\\
\hline
$\Delta_\gamma$ vs. $\delta$ 		 	&  \cellcolor{gray!30}0.43&\cellcolor{gray!30}0.25 	& 0.38&0.23  	& 0.28&0.24 	& \cellcolor{gray!60}0.25&\cellcolor{gray!60}0.25 	\\
\hline
$\Delta_\mathrm{radio}$ vs. $\delta$		&  \cellcolor{gray!30}0.16&\cellcolor{gray!30}0.32	& 0.13&0.23  	& \cellcolor{gray!60}0.12&\cellcolor{gray!60}0.41  	&  0.08&0.21  	\\
\hline
\end{tabular}
\end{center}
\caption{Estimates of the consistency between observed/estimated and simulated two-dimensional distributions shown 
in Figures \ref{G_PeakSepBeam} to \ref{RadLagPksepG}. The 2D-2KS statistics $D$ ranges between 0 and 1 for distributions showing total 
agreement and total disagreement, respectively. The $p_\mathrm{value}$ is the probability to obtain the observed $D$ value under 
the assumption that the two distributions are obtained from the same distribution (null hypothesis). This is equivalent to rejecting the null 
hypothesis at a confidence level of 100-($p_\mathrm{value}$)\%. The 2D-2KS statistics and distributions are described in Section 
\ref{Met2KS2}. The $D$ and $p_\mathrm{value}$ parameters relative to the first and second most consistent distributions are highlighted 
in dark grey cells and light grey cells, respectively.}
\label{Tab5}
\end{table*}

\section{Summary}
\label{summary}

We compared the morphological characteristics of observed $\gamma$-ray and radio light curves with the 
same characteristics computed on a synthesised pulsar light-curve population in the framework of different 
$\gamma$-ray and radio geometrical models. The observed $\gamma$-ray and radio light curves are the 
young or middle-aged ordinary pulsars published in \cite{aaa+13}, while the simulated pulsar 
light curves are those synthesised by \cite{pghg12} assuming the magnetospheric emission 
geometry from \cite{dhr04} and in the framework of four $\gamma$-ray models and a radio model: Polar 
Cap \citep[PC;][]{mh03}, Slot Gap \citep[SG;][]{mh04a}, Outer Gap \citep[OG;][]{crz00}, One Pole Caustic 
\citep[OPC;][]{wrwj09,rw10}, and radio core plus cone models \citep{gvh04,sgh07,hgg07}.

For observed and simulated light curves and for each model we defined a series of morphological characteristics, 
namely peak number, light-curve minima, width of the peaks, among others, and built a $\gamma$-ray and 
radio shape classification according to the recurrence of these characteristics in the light curve. We evaluated 
the precise peak phases by fitting the light curve with a number of Gaussian and/or Lorentzian functions 
equal to light-curve peak number and computed $\gamma$-ray and radio peak separation distributions 
($\Delta_{\gamma}$ and $\Delta_\mathrm{Radio}$, respectively) and the radio loud pulsars, radio-radio lag distribution 
($\delta$) for observed objects and for the objects simulated in the framework of each model.

We studied how $\Delta_{\gamma}$, $\Delta_\mathrm{Radio}$, and $\delta$ changes as a function of observable 
pulsar characteristics, namely spin period (P) and spin-down power ($\dot{E}$), and as a function of non-observable 
pulsar characteristics like magnetic obliquity ($\alpha$), observer line of sight ($\zeta$), and $\gamma$-ray 
beaming factor ($f_{\Omega}$). The observable pulsar parameters are taken from the second catalogue of $\gamma$-ray
LAT pulsars, \citep{aaa+13} while the non-observable LAT pulsar parameters were estimated in the framework of 
PC, SG, OG, and OPC models by \cite{phg+15}. We compared the observed distributions of $\Delta_{\gamma}$, 
$\Delta_\mathrm{Radio}$, and $\delta$  with the same distributions obtained in the framework of the implemented  
$\gamma$-ray emission geometries. We also compared observed and simulated trends for $\Delta_{\gamma}$, 
$\Delta_\mathrm{Radio}$, and $\delta$ as a function of the observable parameters P and $\dot{E}$, and as a function 
of the non-observable/estimated pulsar parameters $\alpha$, $\zeta$, and $f_{\Omega}$. The comparison of the 
$\Delta_\mathrm{Radio}$ distribution and trends within each $\gamma$-ray models is possible because the RL pulsar 
population changes as a function of the different $\gamma$-ray visibility of each model and shows how the characteristics 
of a unique radio population change in the framework of its $\gamma$ ray-selected subsamples.

We studied how the recurrence of the shape classes changes between the whole simulated sample and its visible
subsample and we obtained that no selection effects due to the light-curve shapes affect the pulsar visibility. This
allowed us to compare the observed morphological characteristics with the same characteristics obtained on the 
whole simulated sample and within each model without applying the LAT $\gamma$-ray pulsar visibility to the 
simulated pulsar sample.

The agreement between the observed and simulated one-dimensional distribution in the framework of each model has
been quantified by computing the 2KS test. Each  2KS statistics $D$, has been 
computed jointly with its $p_{value}$ expressing the probability of obtaining the $D$ value assuming 
that the two one-dimensional distribution are obtained from the same underlying distribution (null hypothesis). The agreement between 
the observed/estimated and simulated two-dimensional distributions has been quantified by computing the two-dimensional 2KS statistics 
$D$ with relative $p_\mathrm{value}$ through bootstrap resampling, as described in Section \ref{Met2KS2}.

We obtain that none of the proposed emission geometries explain the LAT pulsar morphology. The OPC model
provides, overall, a best explanation of the observed morphological characteristics and trends but the two-pole caustic 
emission from the SG is necessary to explain some important morphological characteristics. Comparisons of observed 
and simulated $\gamma$-ray peak multiplicity (number of light curve peaks) show that the OPC model manages 
to explain the observed light-curve complexity. We studied how the $\gamma$-ray peak multiplicity changes
from RQ to RL pulsars and we observed an increase of the single-peak, $\gamma$-ray light curves among RL objects
that is not explained in the framework of the implemented emission geometries.

The comparison of observed and simulated $\gamma$-ray and radio peak separations show that the SG and OPC 
models explain the observations in a complementary way. The SG best explain the observed wide separated peaks 
but do not manage to explain the observation at low peak separation while the OPC gives a good explanation of the 
low peak-separation region but completely fails to explain wide peak separation. This suggest that the OPC model
best explains the structure of the  $\gamma$-ray peak but the two-pole caustic SG model is required to explain the 
0.5 separated features generated by the emission from the two magnetic poles appearing in the light curves at high
$\alpha$ and $\zeta$. Among the $\gamma$ ray-selected radio pulsars, all the models explain the radio peak-separation 
distribution but the SG model gives the best description of the observed trend.
Overall, none of the assumed emission geometries explain the observed radio-lag distribution with all models 
underestimating wide radio lag, $\delta>0.4$. The OPC model best explains the observed distribution 
since it predicts the lack of measures for $\delta\sim0.09$ and partly explains the peak observed at $\delta\sim0.15$. 

The comparison of simulated and observed trends for $\Delta_{\gamma}$, $\Delta_\mathrm{Radio}$, and $\delta$
as a function of the non-observable pulsar parameters suggests a higher agreement between observation and 
estimations in the framework of the OPC. In all studied cases, the OPC and SG 
estimated trends show a best match with the modelled trends with the OPC estimates best matching the model 
predictions for $\Delta_{\gamma}$ vs. $f_{\Omega}$, $\Delta_\mathrm{Radio}$ vs. $\alpha$, $\Delta_{\gamma}$ vs. 
$\zeta$, and $\Delta_\mathrm{Radio}$ vs. $\alpha$, and particularly for $\delta$ vs. $\alpha$, $\zeta$, and $f_{\Omega}$. 
This confirms that the OPC model predicts the ranges of light-curve shapes that best explains the LAT findings.
The comparison of simulated and observed trends for $\Delta_{\gamma}$, $\Delta_\mathrm{Radio}$, and $\delta$ 
as a function of P and $\dot{E}$ confirms the lack of high $\dot{E}$ objects found, for all models, by \cite{pghg12}. 
The observed lack affects all simulated distributions that in no case explains the LAT findings.

In the plane  $\delta$-$\Delta_{\gamma}$ and $\delta$-$\Delta_\mathrm{Radio}$, the only model that correctly explains
the observations is the OPC. The $\delta$-$\Delta_{\gamma}$  plane is a tracker of the magnetospheric structure of the
$\gamma$-ray gap emission region and of the relative position of $\gamma$-ray and radio emission regions. This allows 
us to conclude that, under the assumption that our radio core plus cone emission geometry is correct for the major part of
the LAT pulsars, the outer magnetosphere of a pulsar is the most likely location for $\gamma$-ray photon production. 
The $\delta$-$\Delta_\mathrm{Radio}$ plane is a tracker of the radio emission beam structure and of its position with respect
to the $\gamma$-ray emission region and the OPC model gives the best explanation of the observed trend.
  
We obtained a map of peak multiplicity and $\Delta_{\gamma}$ as a function of $\alpha$ and $\zeta$ and studied for which
model, the $\alpha$, $\zeta$ estimated for LAT pulsars match the model prediction for the same peak multiplicity and 
peak separation. We obtain that the OPC model prediction for LAT pulsars $\alpha$ and $\zeta$ best match the simulations
for the same peak multiplicity and $\Delta_{\gamma}$.

Despite the larger agreement between the OPC model prediction and observation, one has to note that $\delta$ is strongly dependent
on the assumed magnetospheric structure that in this study is a VRD. \cite{khkc12} has shown that $\delta$ 
is larger in DMs and increases with increasing conductivity from 0 (VRD magnetosphere) 
to $\infty$ (FFE magnetosphere). We can expect that the same model predictions obtained in the framework of a non-VRD 
magnetosphere (conductivity larger than 0) would imply larger values of $\delta$ that would modify the $\delta$ distribution 
and the $\Delta_{\gamma}$ as a function of $\delta$ in the framework of the implemented $\gamma$-ray models. A $\delta$ 
increase has already been shown in the OG model prediction by \cite{khkc12} and would 
surely increase the agreement between the SG prediction and observations. Particularly interesting is the model recently proposed by \cite{khk14}, 
which assumes a FFE magnetosphere
with an infinite conductivity, inside the light cylinder and a DM with finite but high conductivity, outside 
the light cylinder (FFE Inside, Dissipative Outside, FIDO). The FIDO model best explains the observed $\delta$ and $\Delta_{\gamma}$
distributions and predict the correct variation of $\Delta_{\gamma}$ as a function of $\delta$ merging characteristics of both 
SG and OG models. 

The results obtained by \cite{khkc12} and by \cite{khk14}, in conjunction with the results of the current study, suggest a complementary 
nature of SG and OPC in explaining 
different aspects of the observed phenomenology, support the existence of a hybrid SG/OG emission mechanism and a non-VRD 
magnetosphere geometry, and calls for a more physical explanation of the OPC model to better understand the pulsar magnetosphere.


\begin{acknowledgements}
 
 The \textit{Fermi} LAT Collaboration acknowledges generous ongoing 
support from a number of agencies and institutes that have supported 
both the development and operation of the LAT as well as scientific 
data analysis. These include the National Aeronautics and Space 
Administration and the Department of Energy in the United States, the 
Commissariat \`a l'Energie Atomique and the Centre National de la 
Recherche Scientifique / Institut National de Physique Nucl\'eaire et de 
Physique des Particules in France, the Agenzia Spaziale Italiana and the 
Istituto Nazionale di Fisica Nucleare in Italy, the Ministry of Education, 
Culture, Sports, Science and Technology (MEXT), High Energy Accelerator 
Research Organization (KEK) and Japan Aerospace Exploration Agency 
(JAXA) in Japan, and the K.~A.~Wallenberg Foundation, the Swedish 
Research Council and the Swedish National Space Board in Sweden.

Additional support for science analysis during the operations phase is 
gratefully acknowledged from the Istituto Nazionale di Astrofisica in Italy 
and the Centre National d'\'Etudes Spatiales in France.

MP acknowledges the Nicolaus Copernicus Astronomical Center, grant 
DEC-2011/02/A/ST9/00256, for providing software and 
computer facilities needed for the development of this work. MP
gratefully acknowledges Eric Feigelson for useful discussions and
suggestions. 
PLG thanks the National Science Foundation through Grant No. AST-1009731 
and the NASA Astrophysics Theory Program through Grant No. NNX09AQ71G 
for their generous support.

The authors gratefully acknowledge the Pulsar Search and Timing Consortia,
all the radio scientists who contributed in providing the radio light curves used 
in this paper, and the radio observatories that generated the radio profiles used
in this paper:
the Parkes Radio Telescope is part of the Australia Telescope which is funded 
by the Commonwealth Government for operation as a National Facility managed 
by CSIRO; the Green Bank Telescope is operated by the National Radio Astronomy 
Observatory, a facility of the National Science Foundation operated under cooperative 
agreement by Associated Universities, Inc; the Arecibo Observatory is part of the National 
Astronomy and Ionosphere Center (NAIC), a national research center operated by Cornell 
University under a cooperative agreement with the National Science Foundation; the 
Nan\c cay Radio Observatory is operated by the Paris Observatory, associated with the 
French Centre National de la Recherche Scientifique (CNRS); the Lovell Telescope is 
owned and operated by the University of Manchester as part of the Jodrell Bank Centre 
for Astrophysics with support from the Science and Technology Facilities Council of the 
United Kingdom; the Westerbork Synthesis Radio Telescope is operated by Netherlands 
Foundation for Radio Astronomy, ASTRON.

 \end{acknowledgements}

\bibliographystyle{aa}
\bibliography{/home/mpierba/Pierba/Bib_Ref_Style/journals,/home/mpierba/Pierba/Bib_Ref_Style/psrrefs,/home/mpierba/Pierba/Bib_Ref_Style/crossrefs,/home/mpierba/Pierba/Bib_Ref_Style/modrefs,/home/mpierba/Pierba/Bib_Ref_Style/pierba.bib}


\appendix

\section{The $\gamma$-ray and radio light-curve shape classification}
\label{ShapeCLS}
\begin{table*}
\centering
\begin{tabular}{| l | c | m{2.5cm} |  m{2.5cm} |  m{2.5cm} |  m{2.5cm} |}
\hline
\multicolumn{1}{|c|}{Shape Class}		&  Multiplicity 	&  \multicolumn{1}{c|}{PC Light curve} 								& \multicolumn{1}{c|}{SG Light curve} 								& \multicolumn{1}{c|}{OG Light curve}  								& \multicolumn{1}{c|}{OPC Light curve}  	   							\\
\hline
\hline
 1- Bump							&   1 			&\includegraphics[height=2cm]{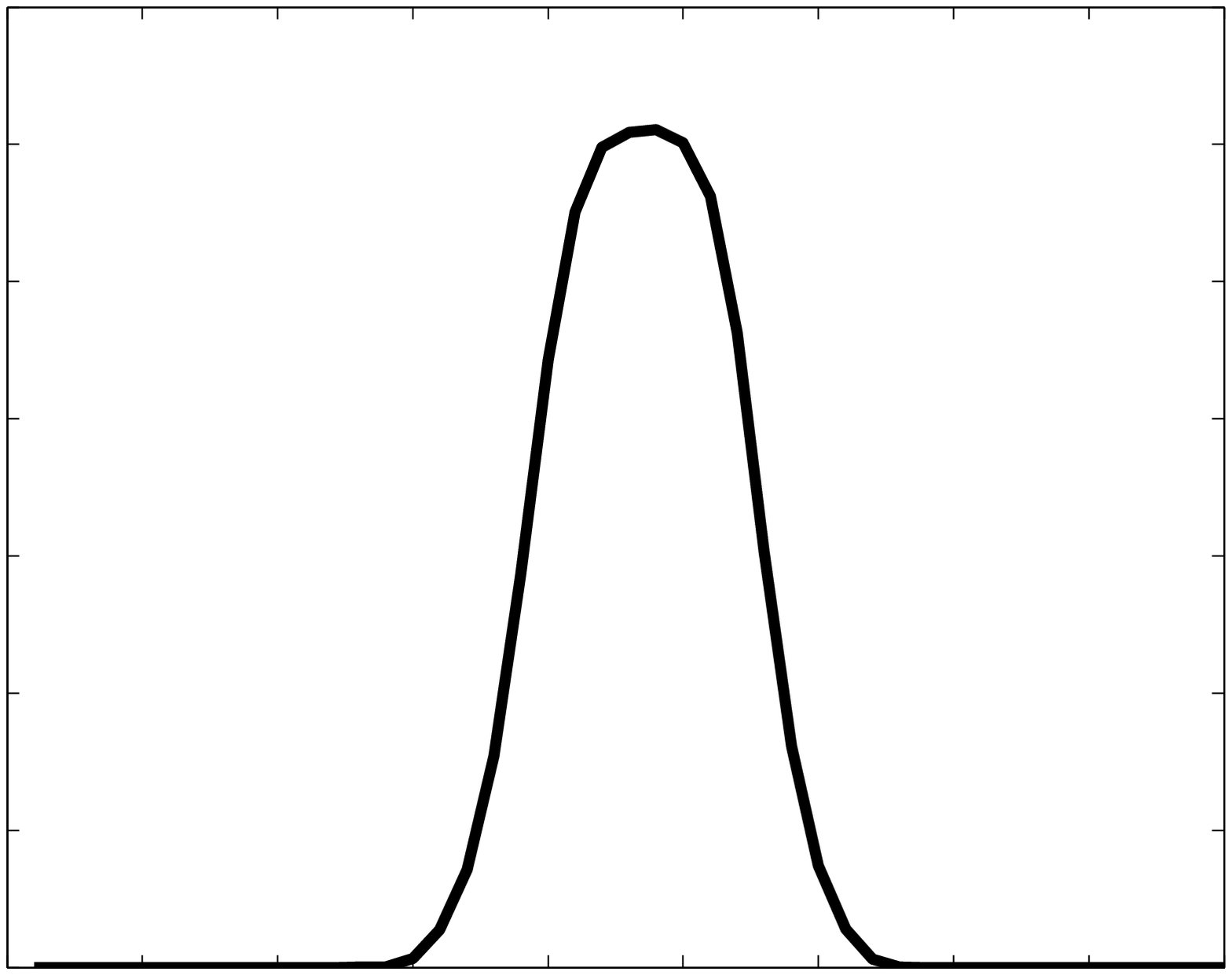}	&\includegraphics[height=2cm]{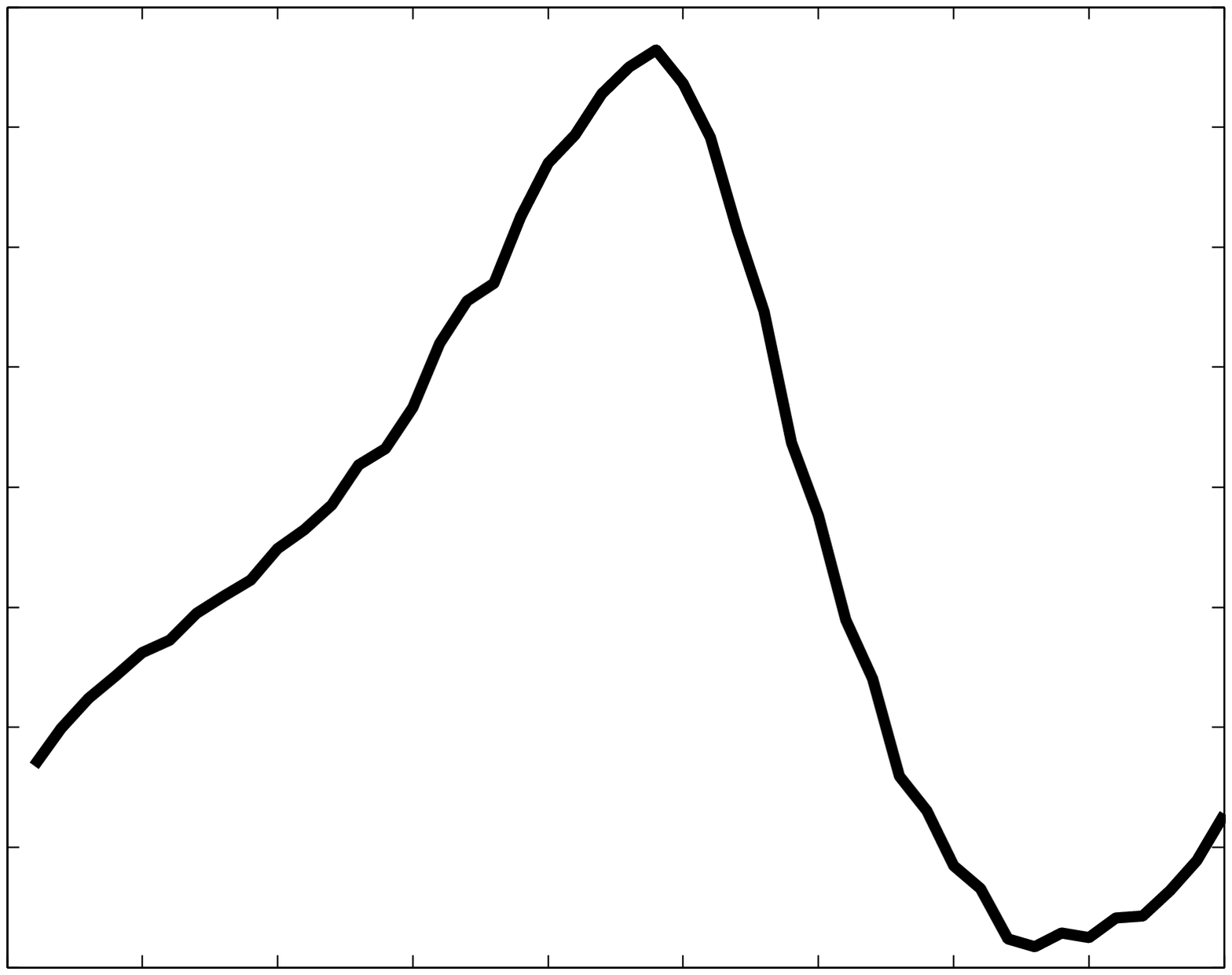} 	& 															&\includegraphics[height=2cm]{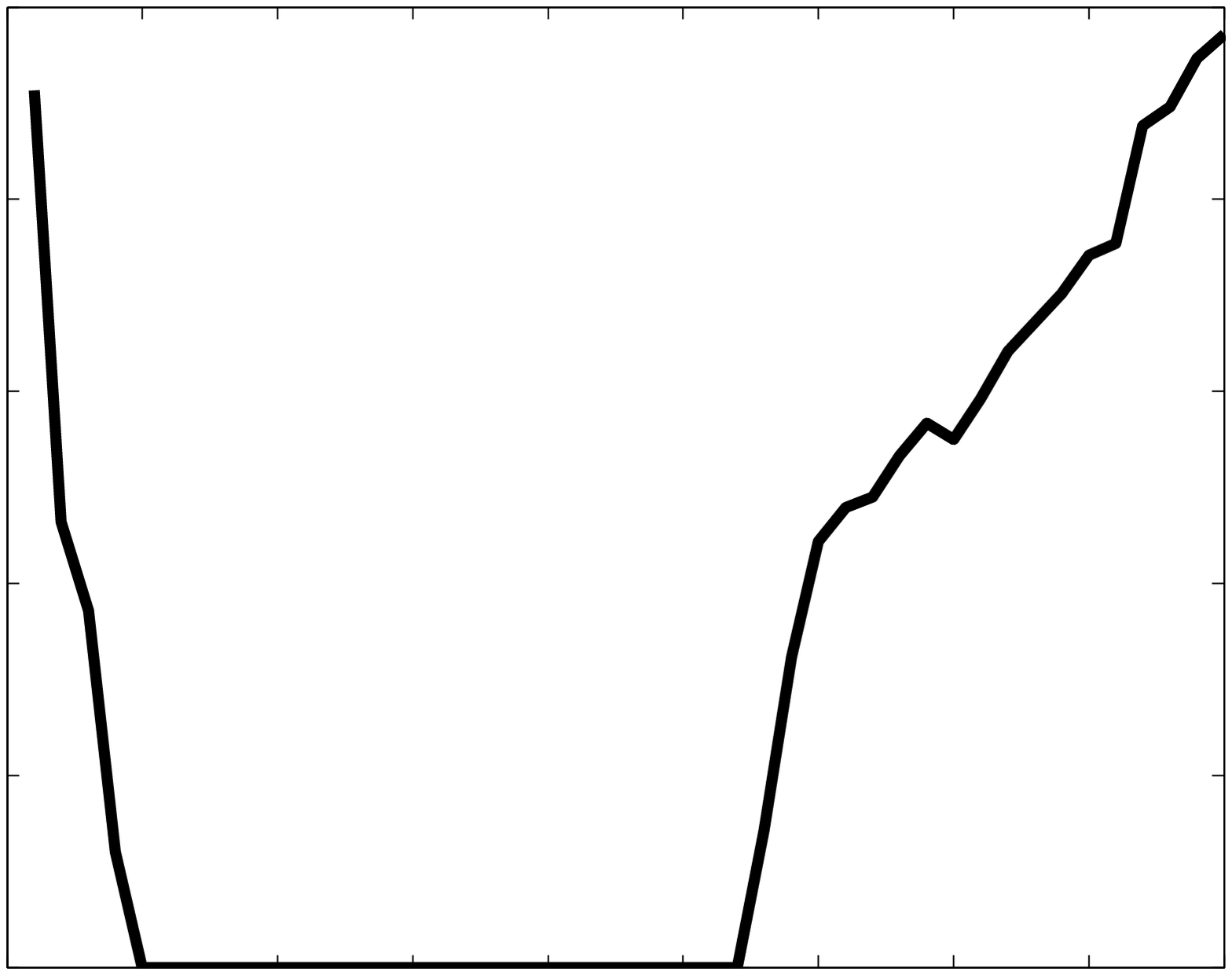}   	\\
\hline
 2- Sharp 							&   1 			&\includegraphics[height=2cm]{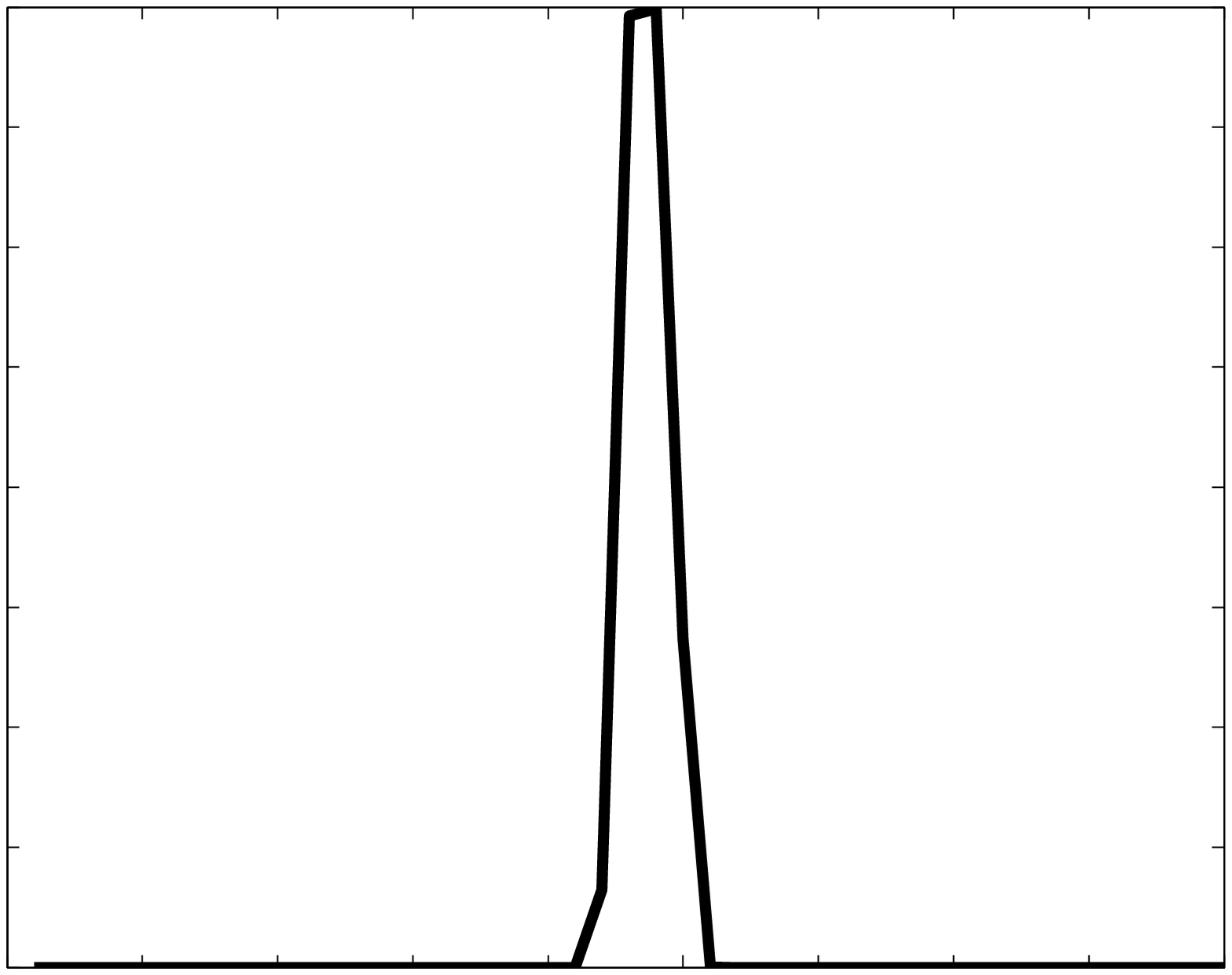} 	&\includegraphics[height=2cm]{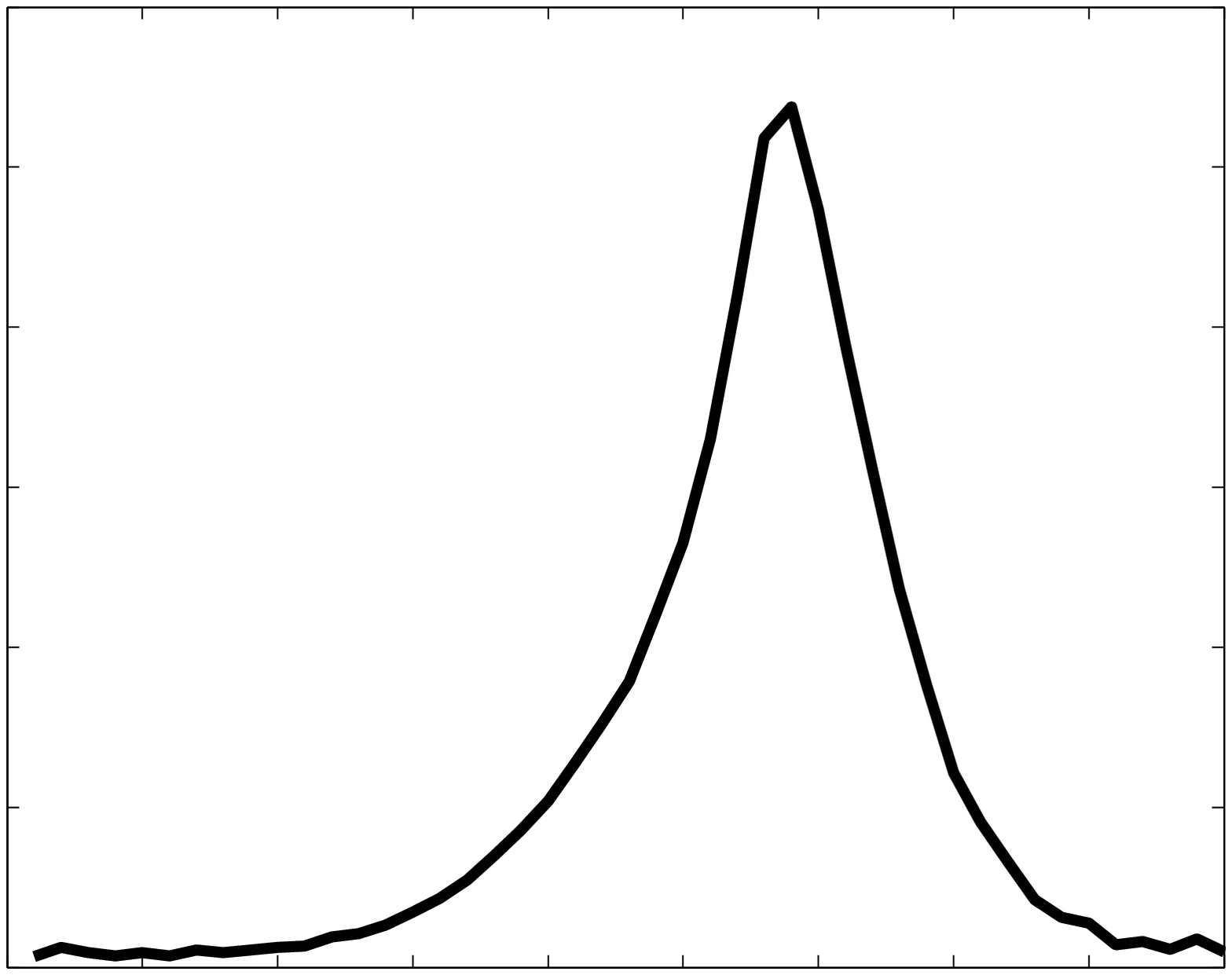} 	&\includegraphics[height=2cm]{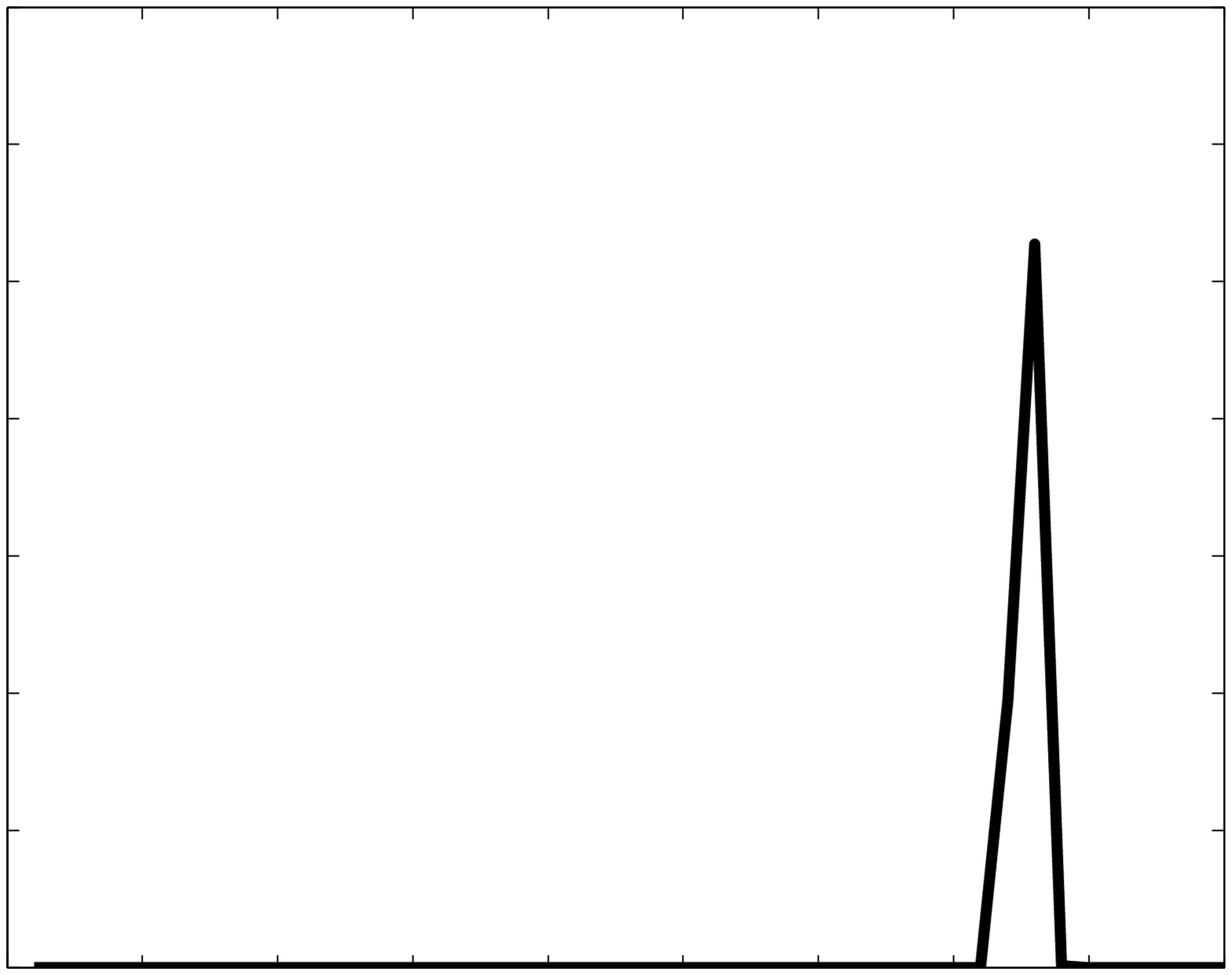} 	&\includegraphics[height=2cm]{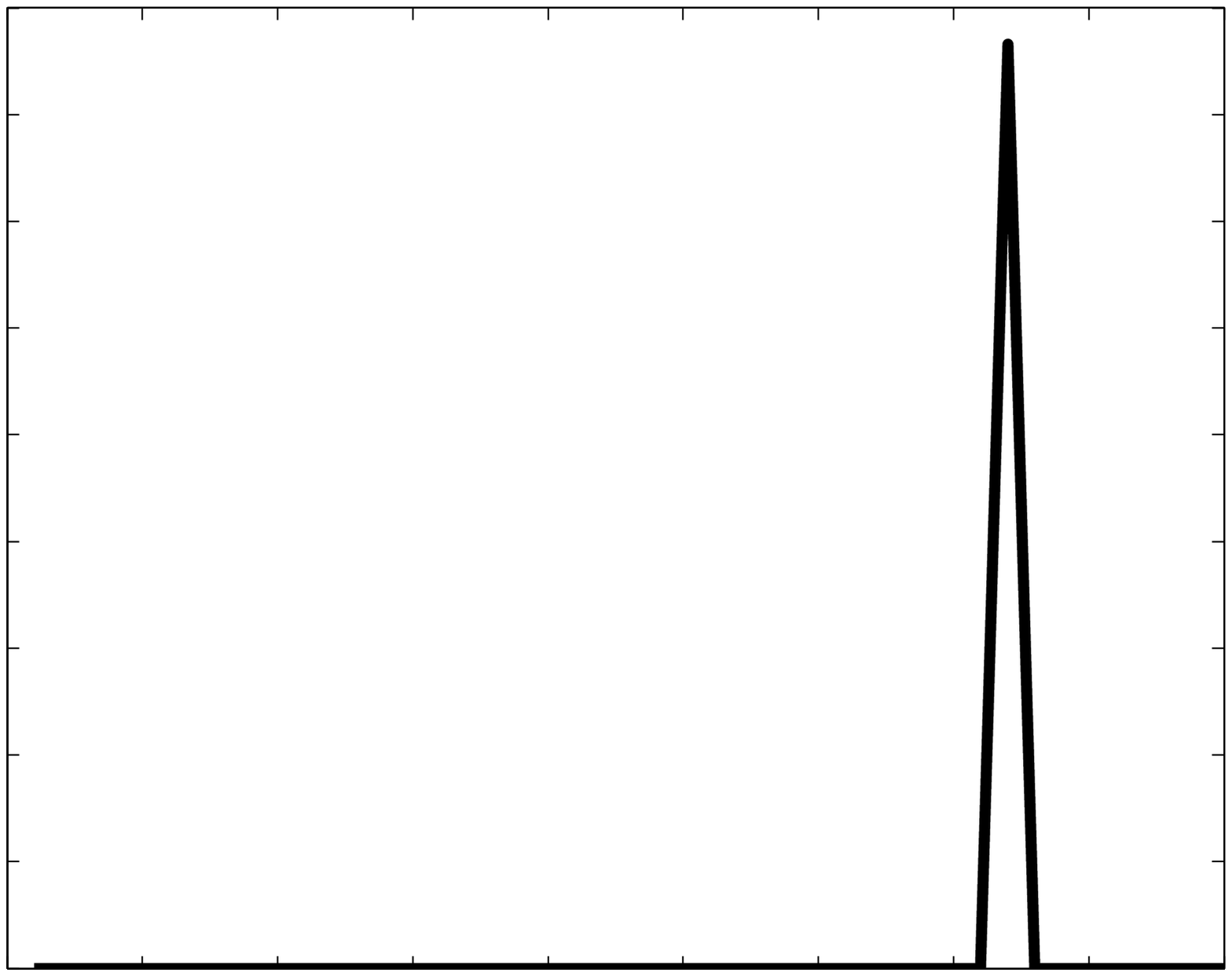}	\\
\hline
 3- Shoulder 						&   1 			&														 	&\includegraphics[height=2cm]{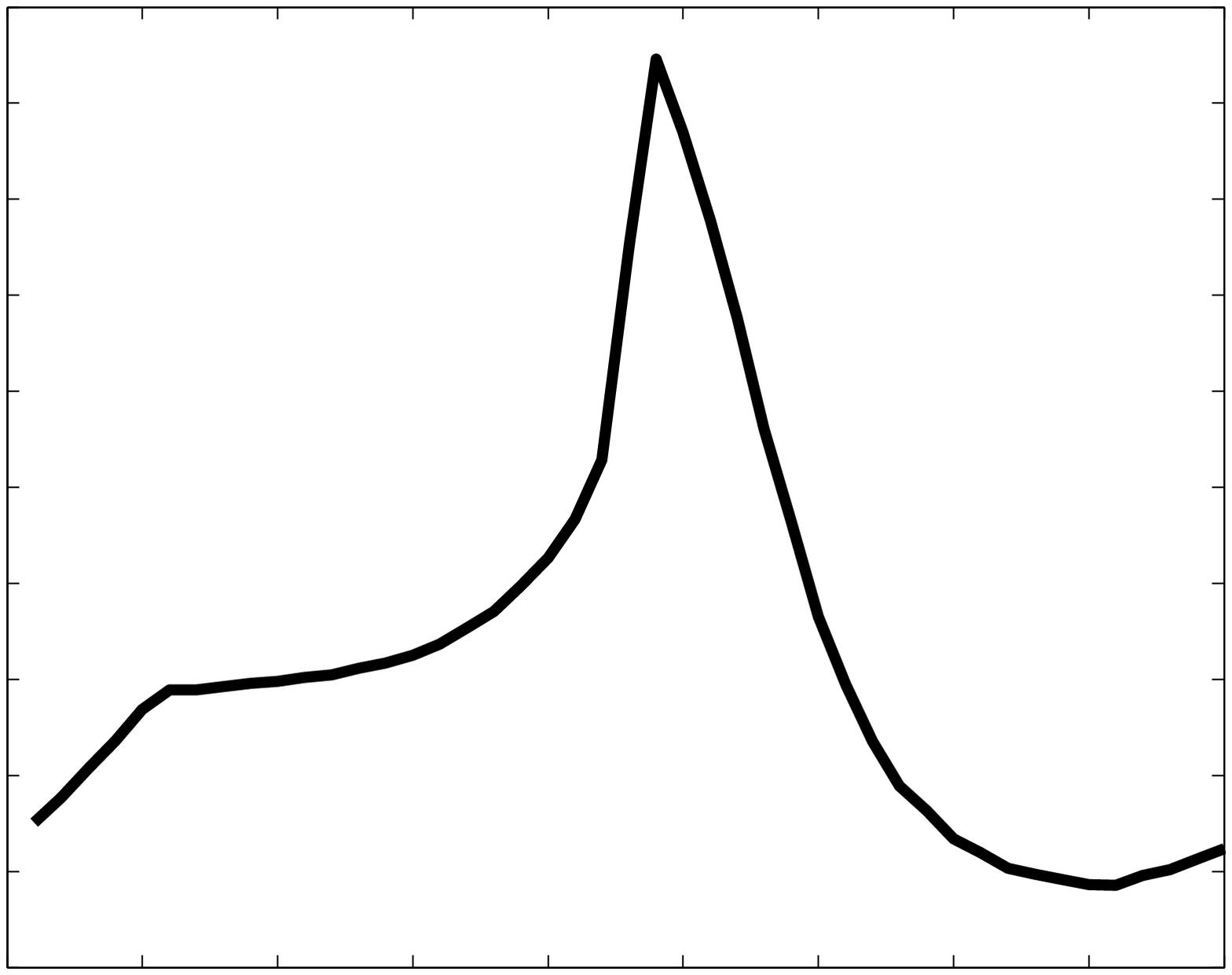} 	&\includegraphics[height=2cm]{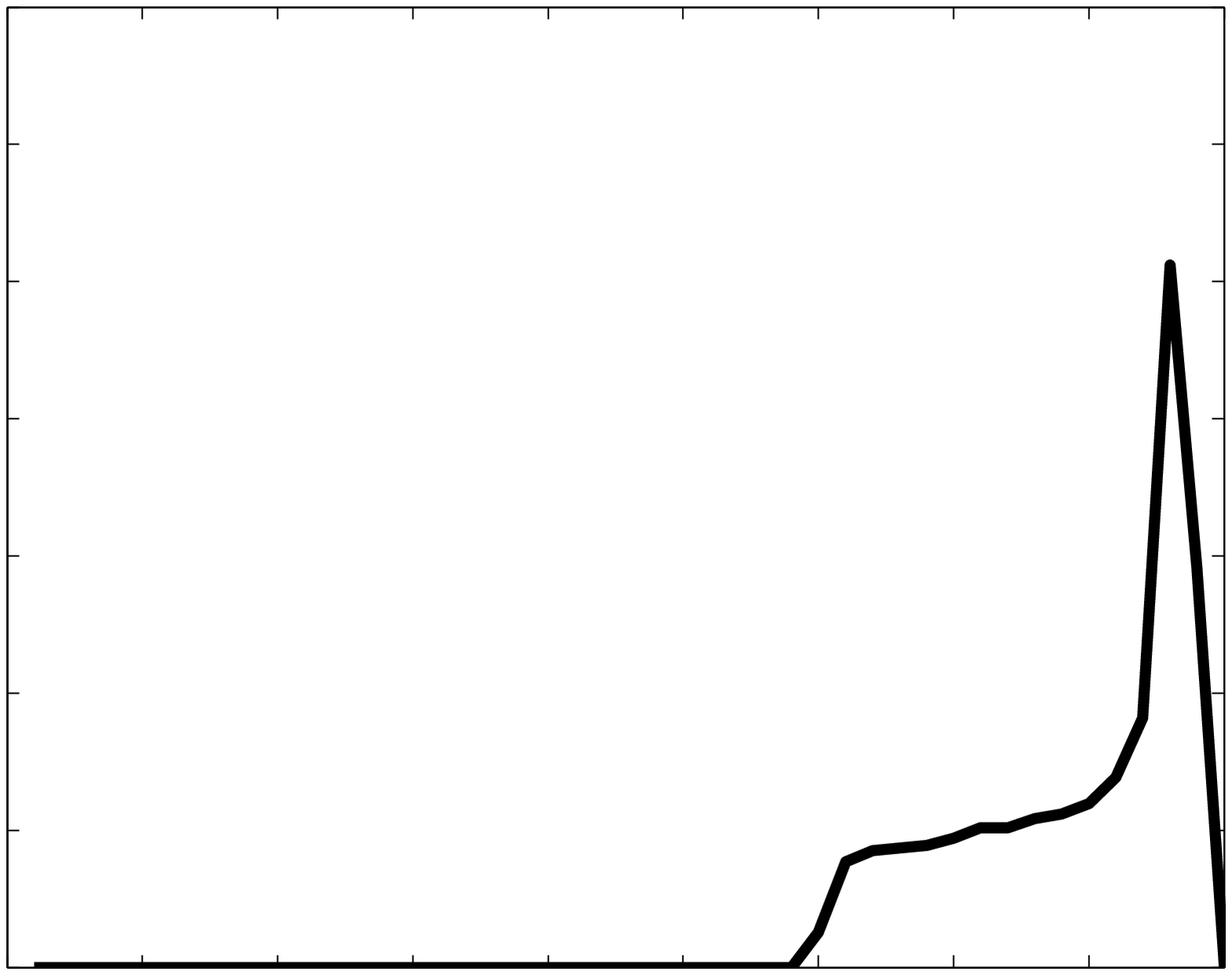} 	&\includegraphics[height=2cm]{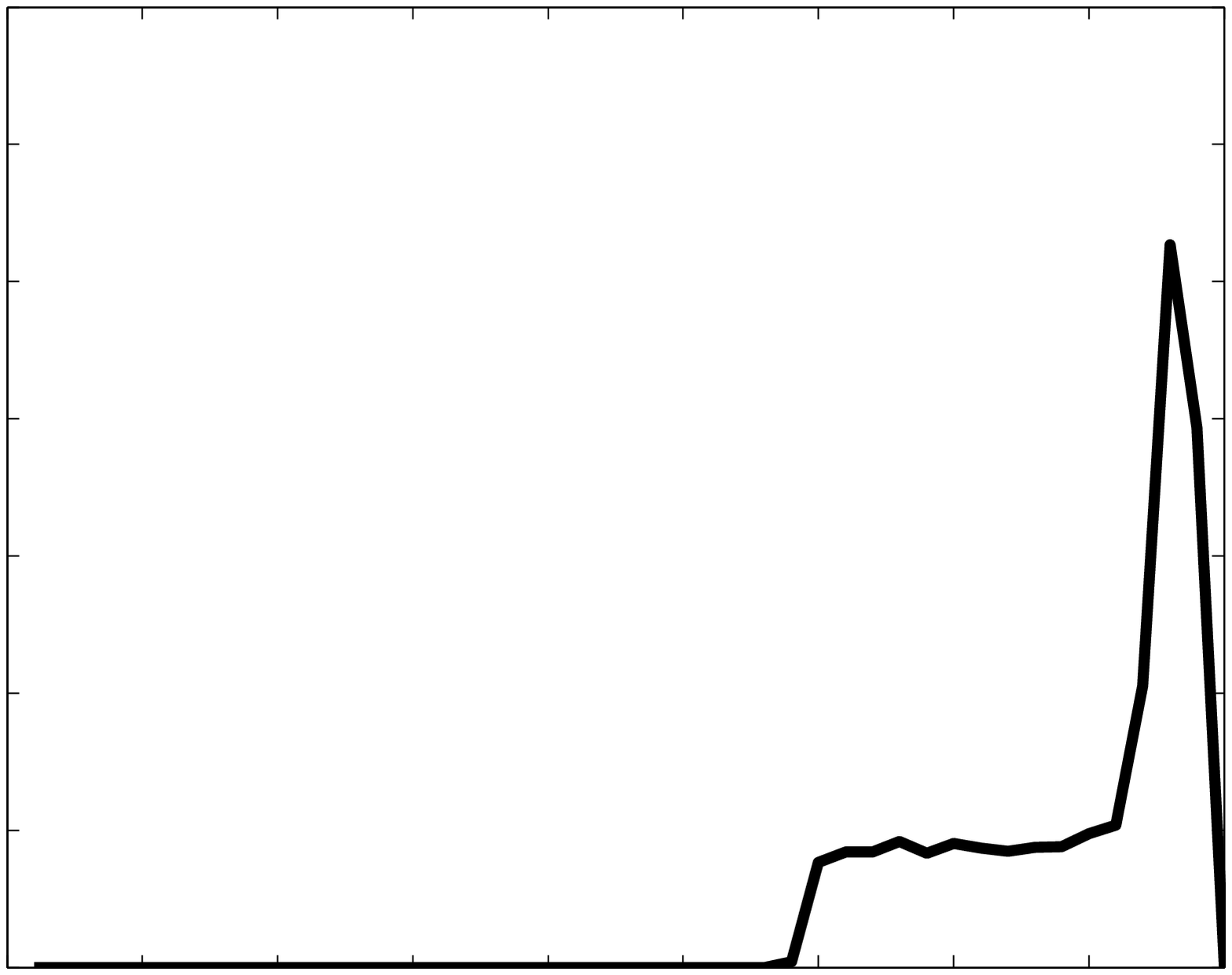}	\\
\hline
 4- Two 							&   2 			&\includegraphics[height=2cm]{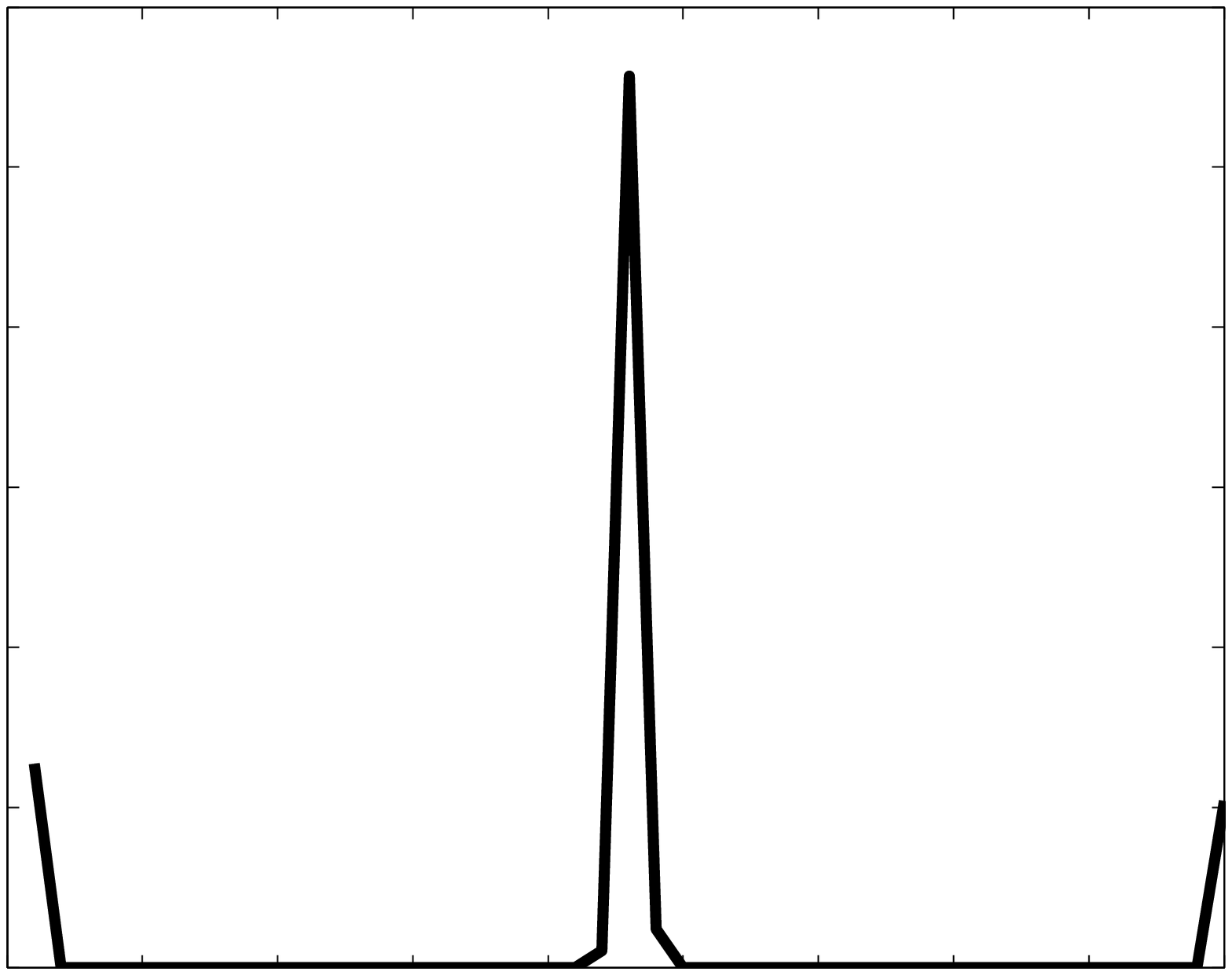} 	&													             	&\includegraphics[height=2cm]{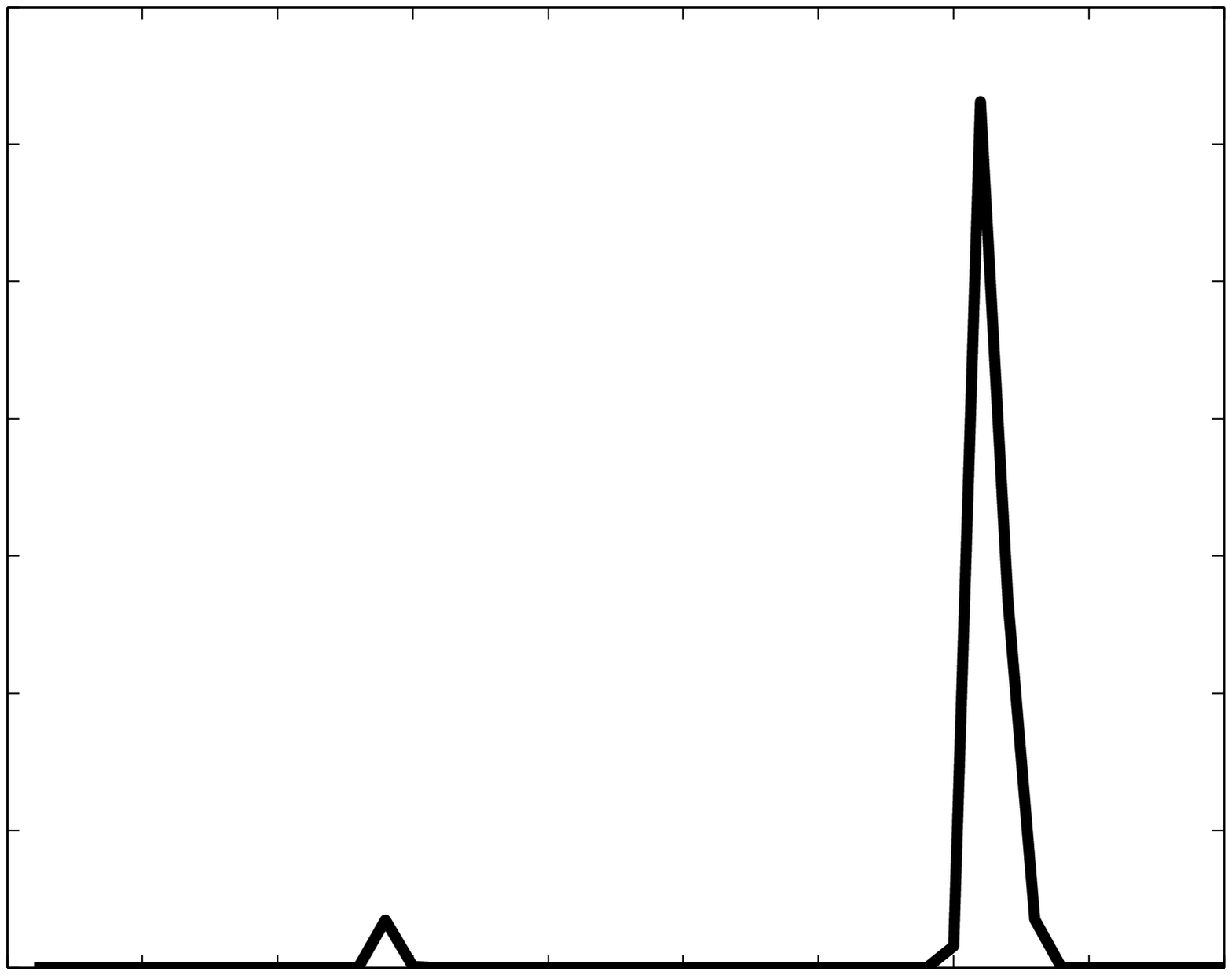} 	&\includegraphics[height=2cm]{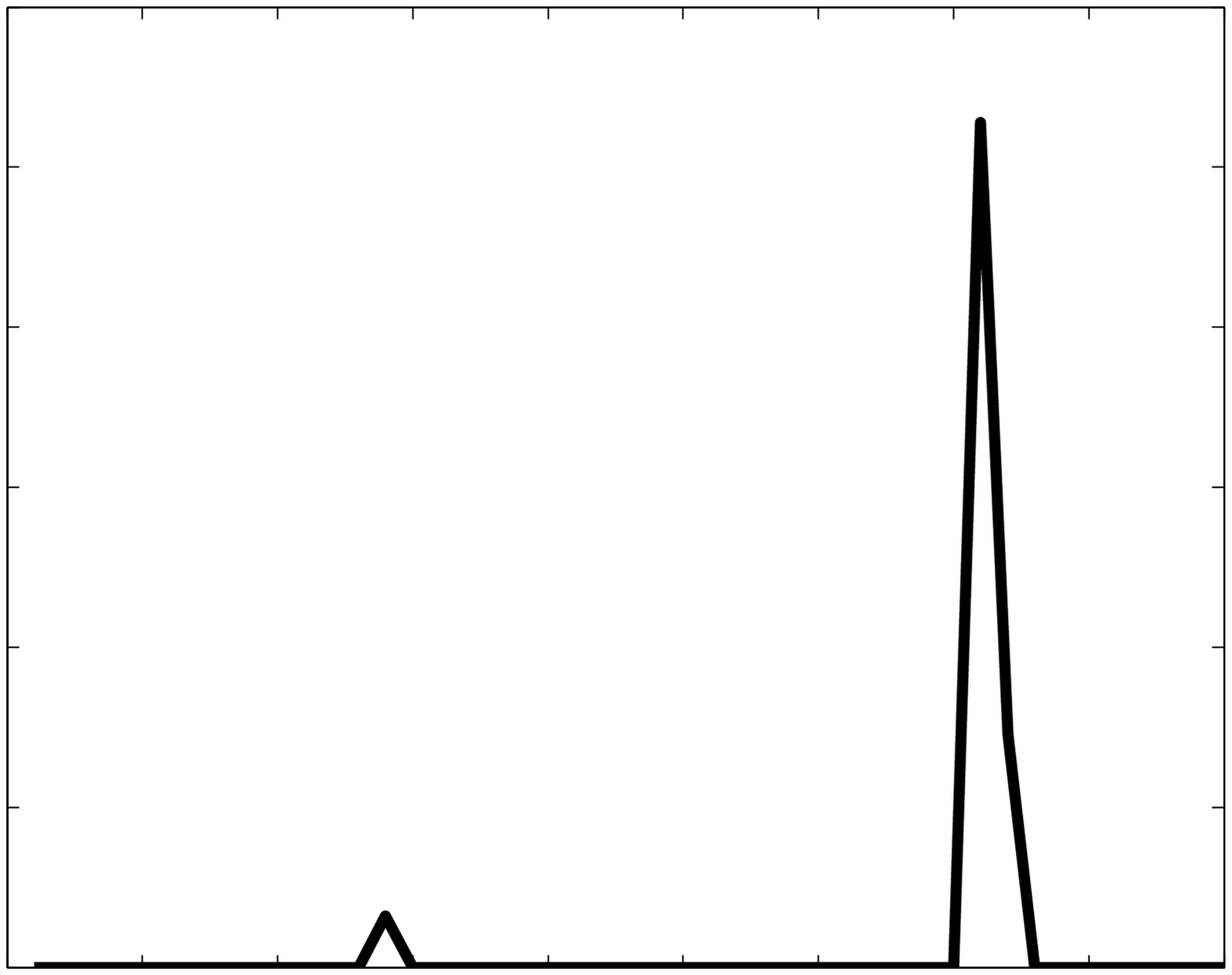}	\\
\hline
 5- Double 						&   2 			&\includegraphics[height=2cm]{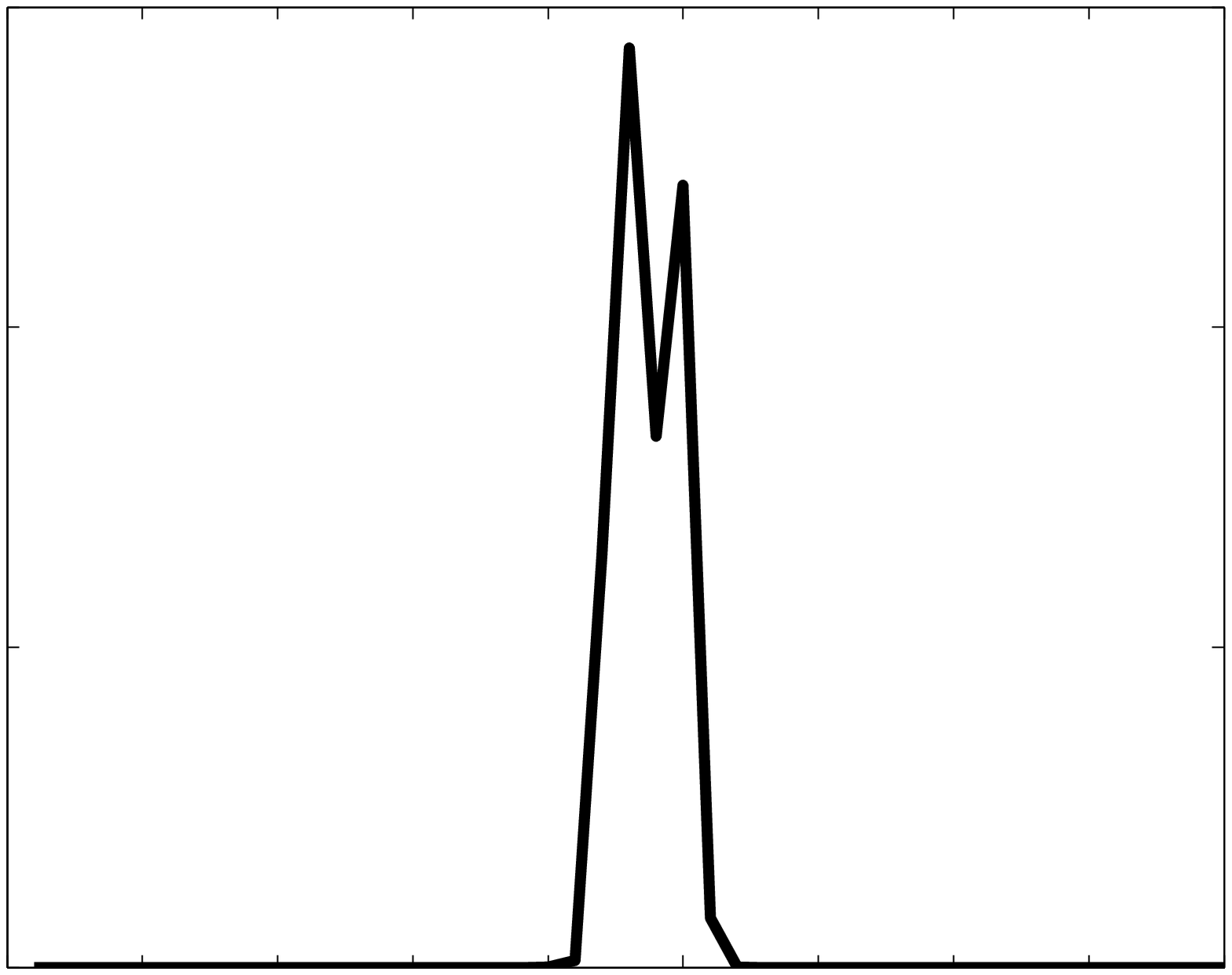} 	&\includegraphics[height=2cm]{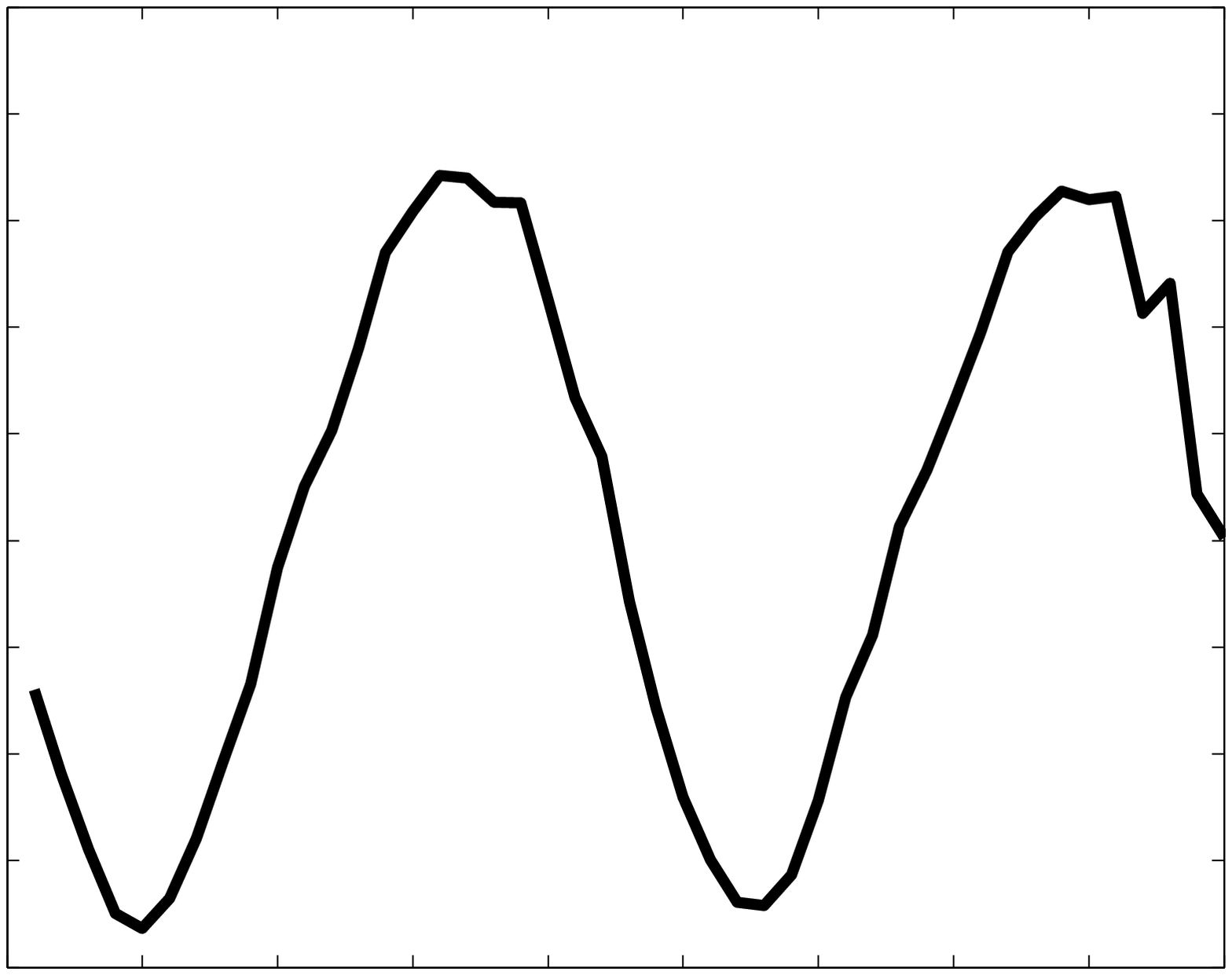} 	&\includegraphics[height=2cm]{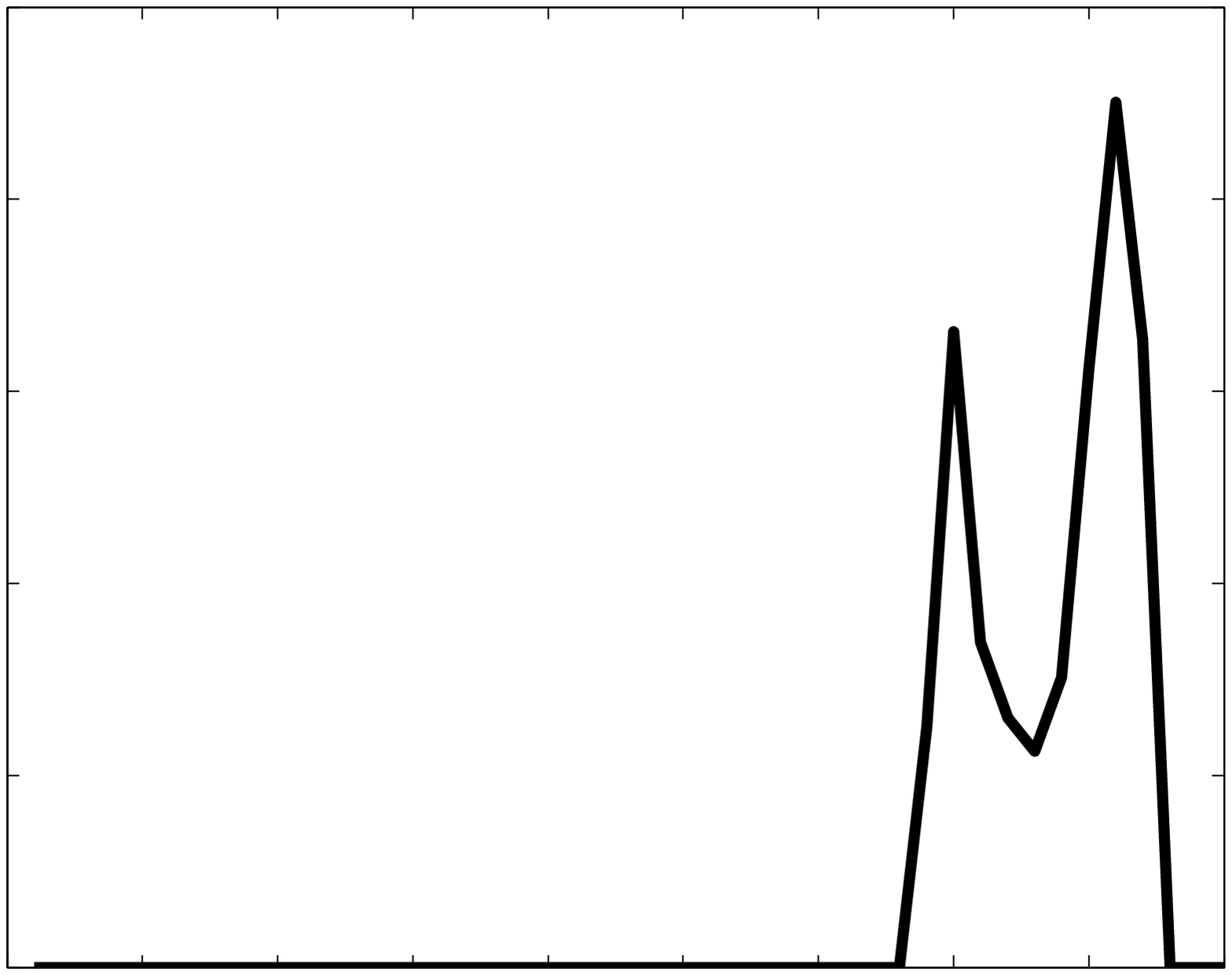} 	&\includegraphics[height=2cm]{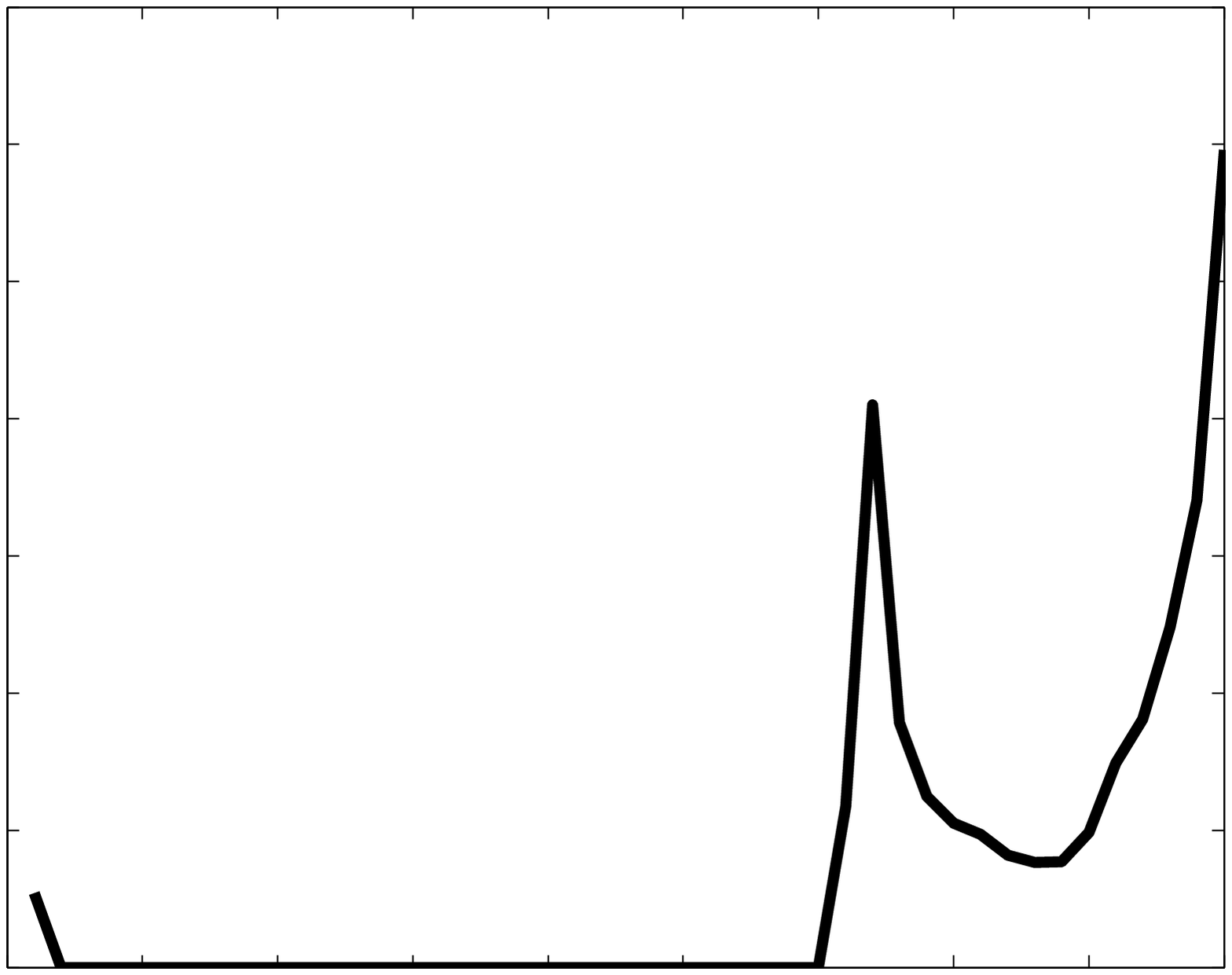}	\\
\hline
 6- Double+Single					&   3 			&\includegraphics[height=2cm]{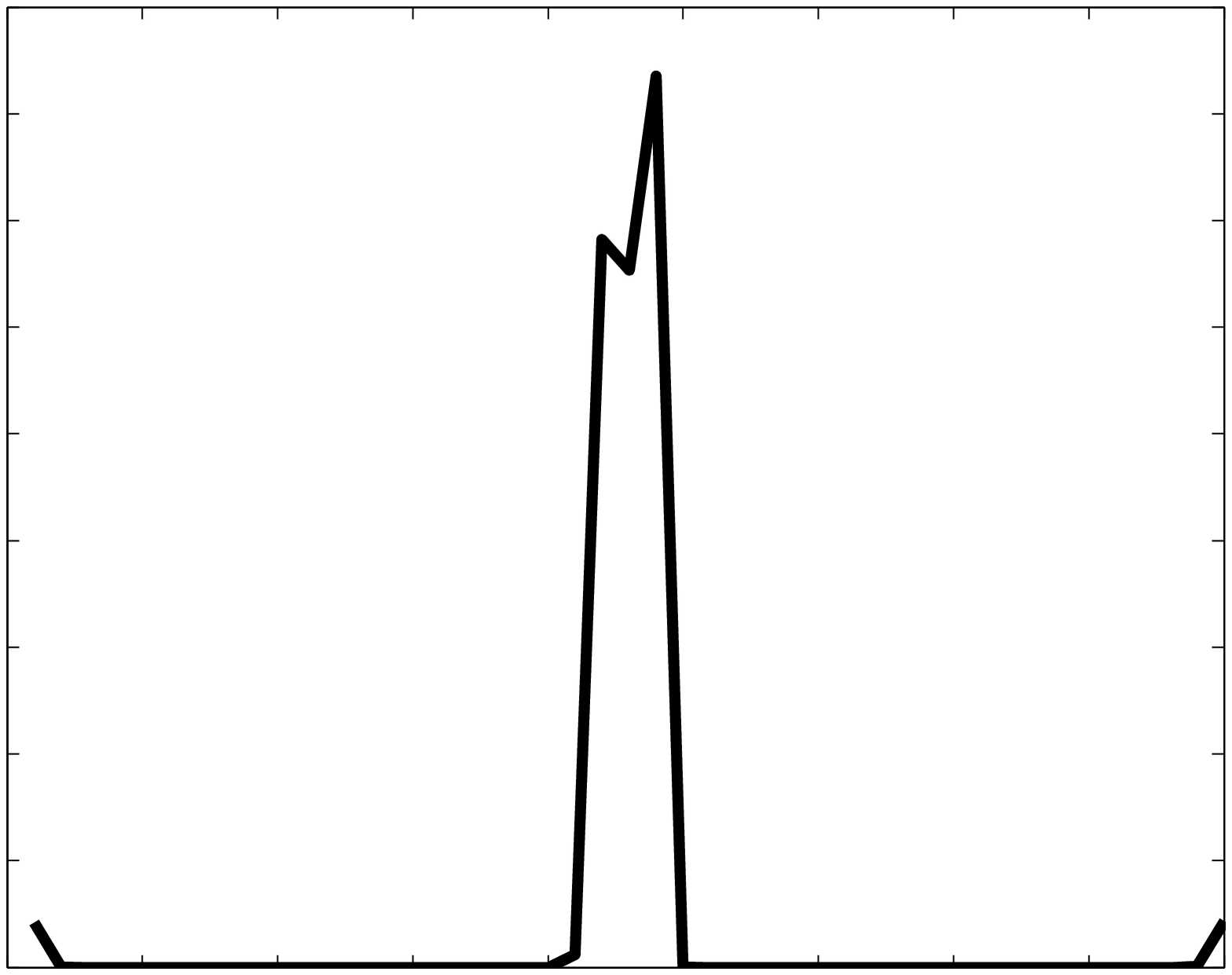} 	&\includegraphics[height=2cm]{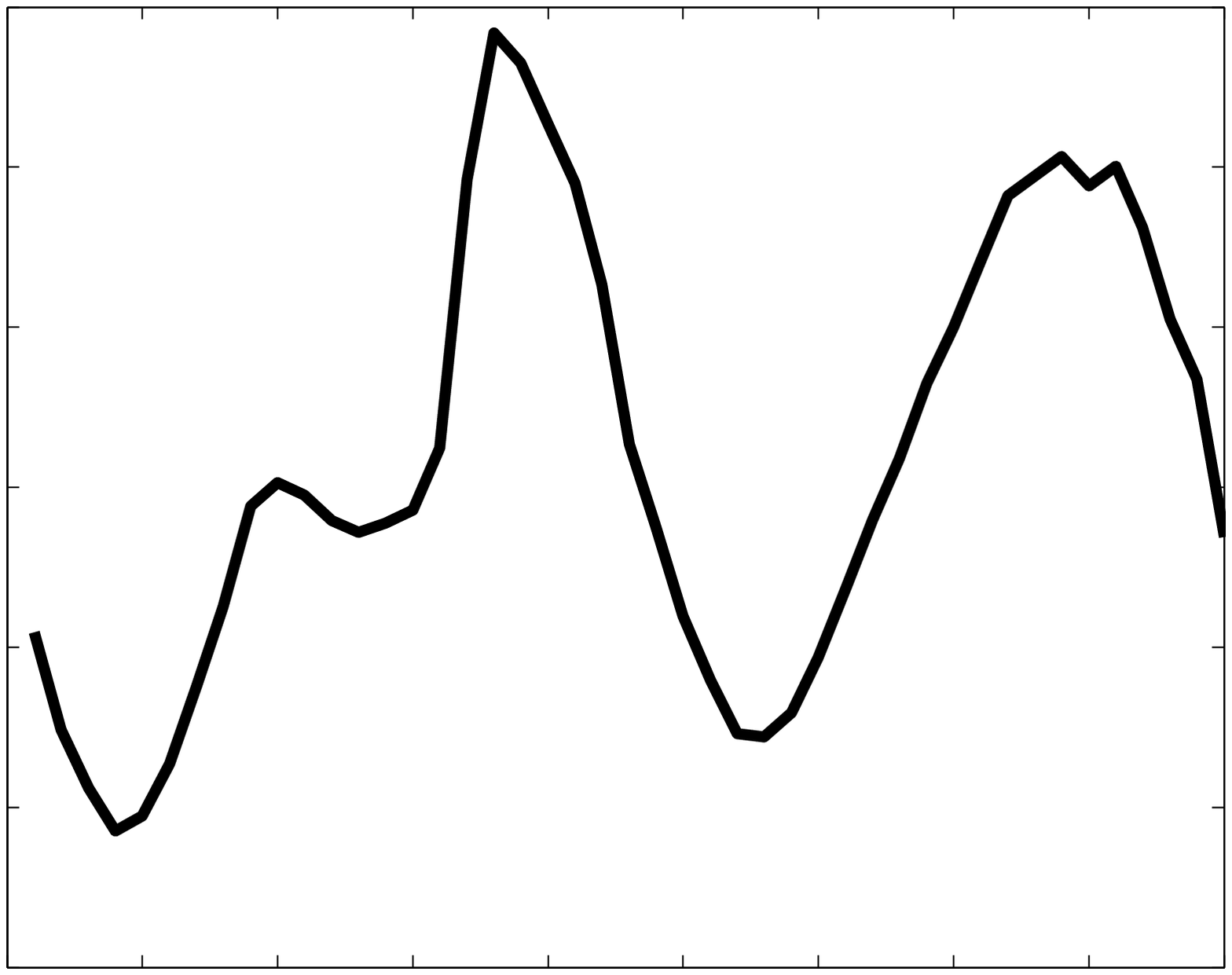} 	&															&															\\
\hline
 7- Triple/Three 					&   3 			&															&													             	&\includegraphics[height=2cm]{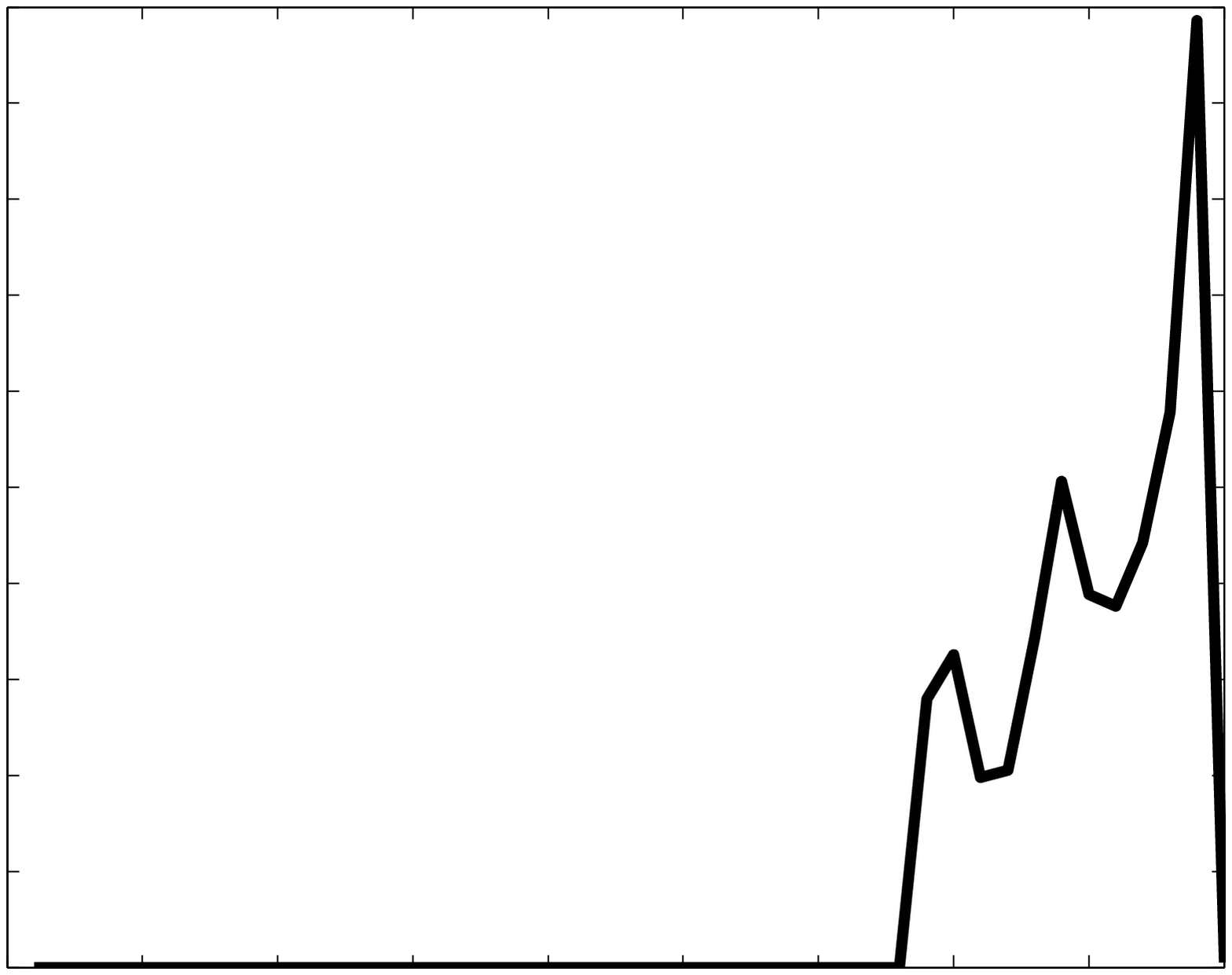} 	&\includegraphics[height=2cm]{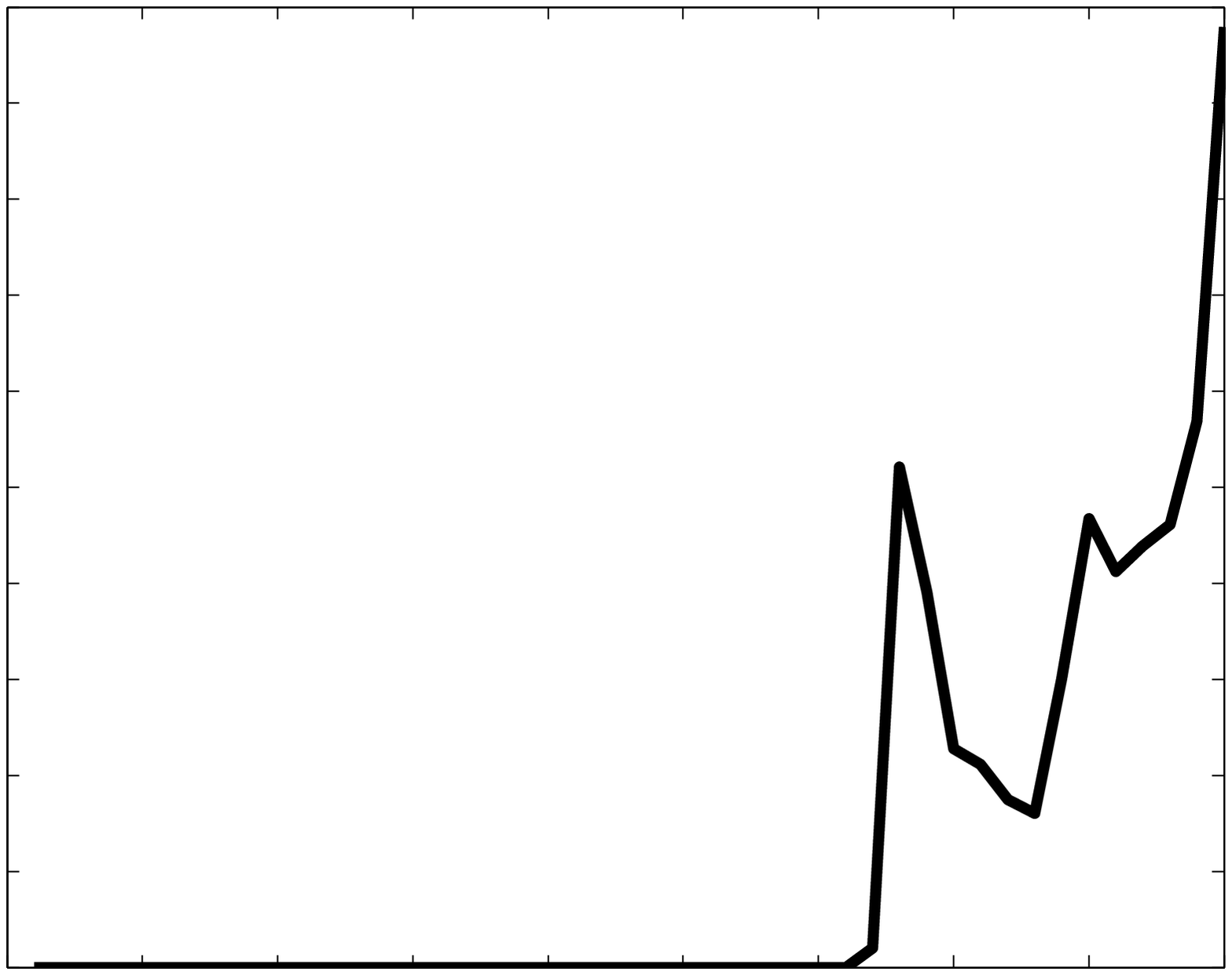}	\\
\hline
 8- Two double 						&   4 			&\includegraphics[height=2cm]{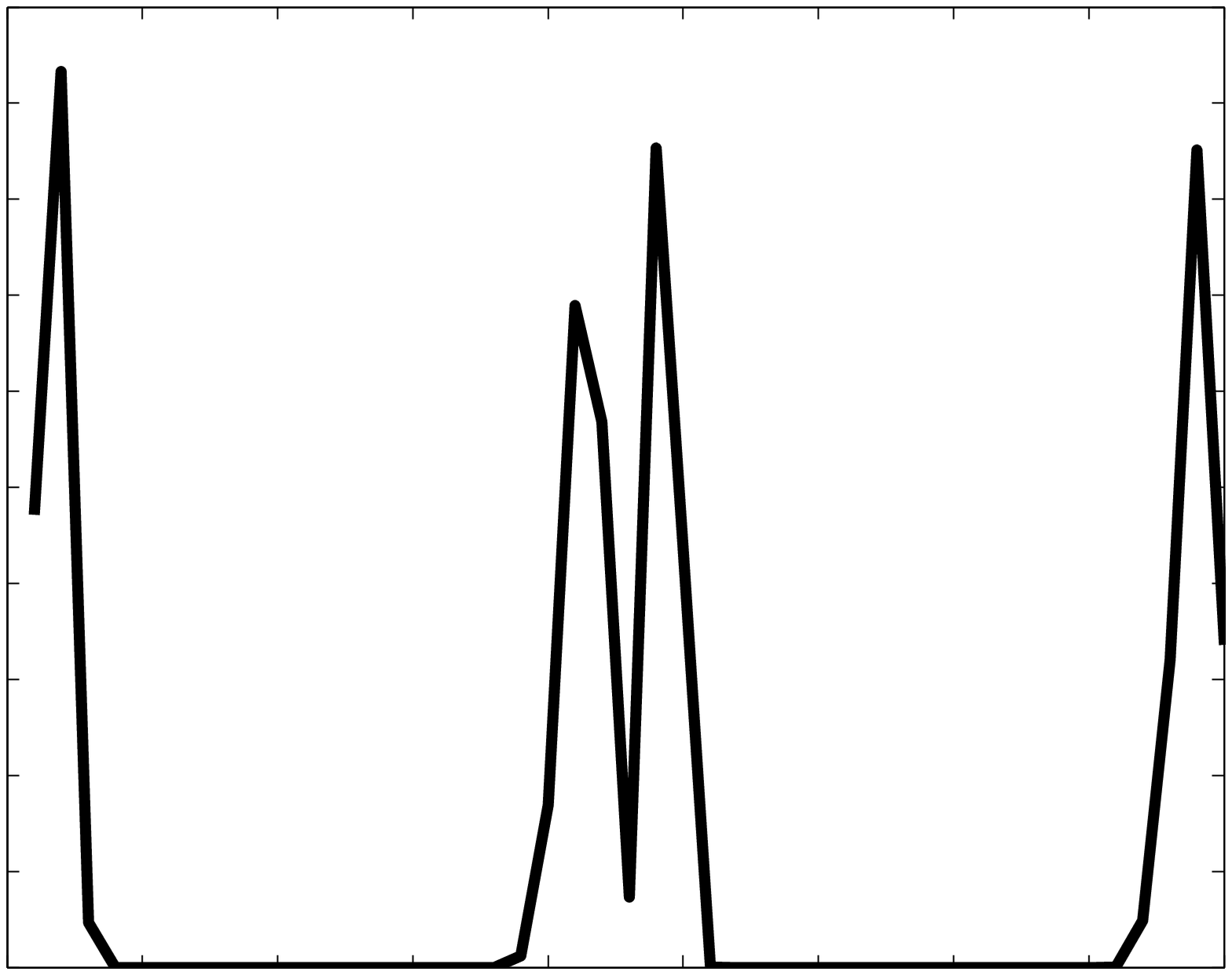} 	&\includegraphics[height=2cm]{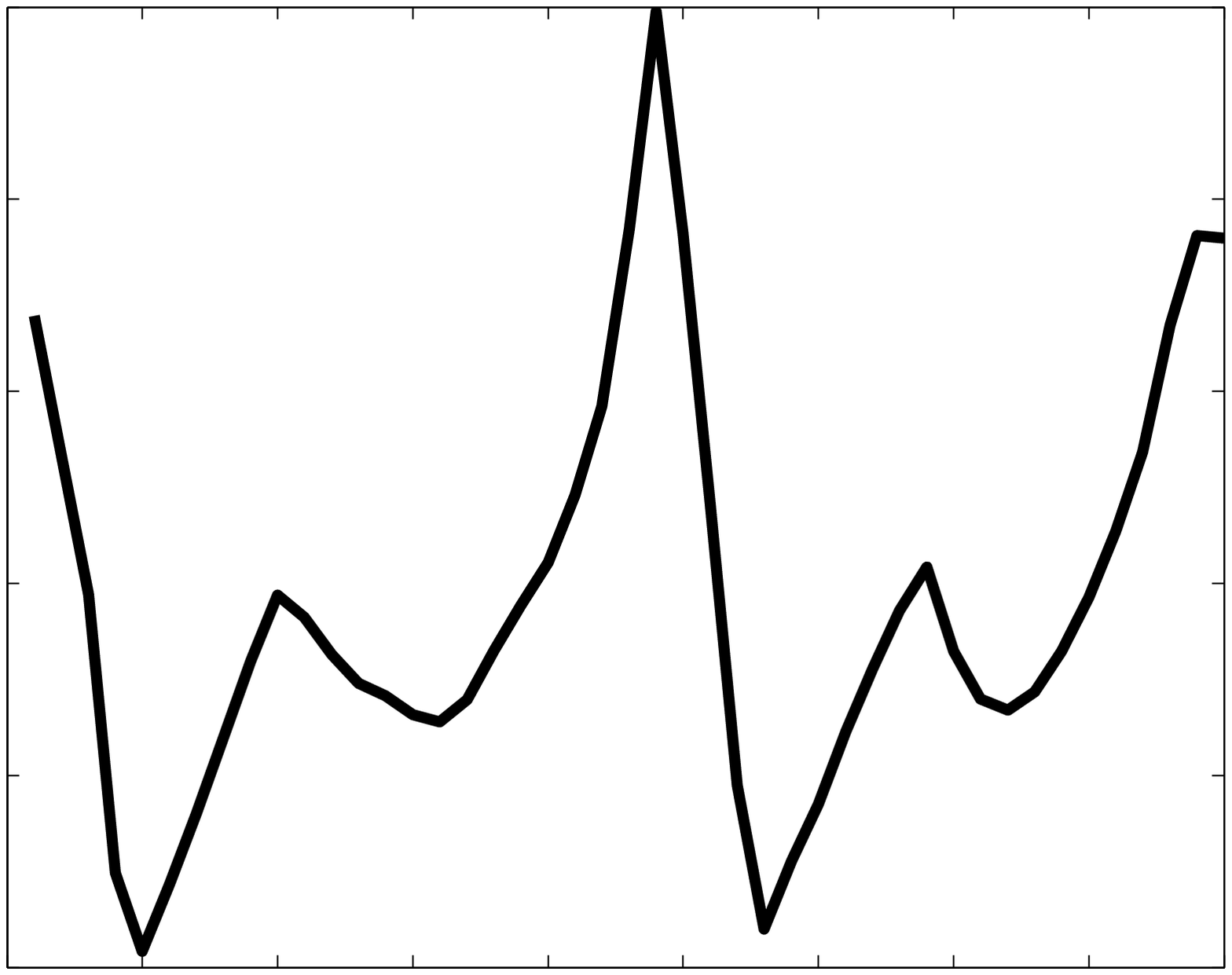}	&\includegraphics[height=2cm]{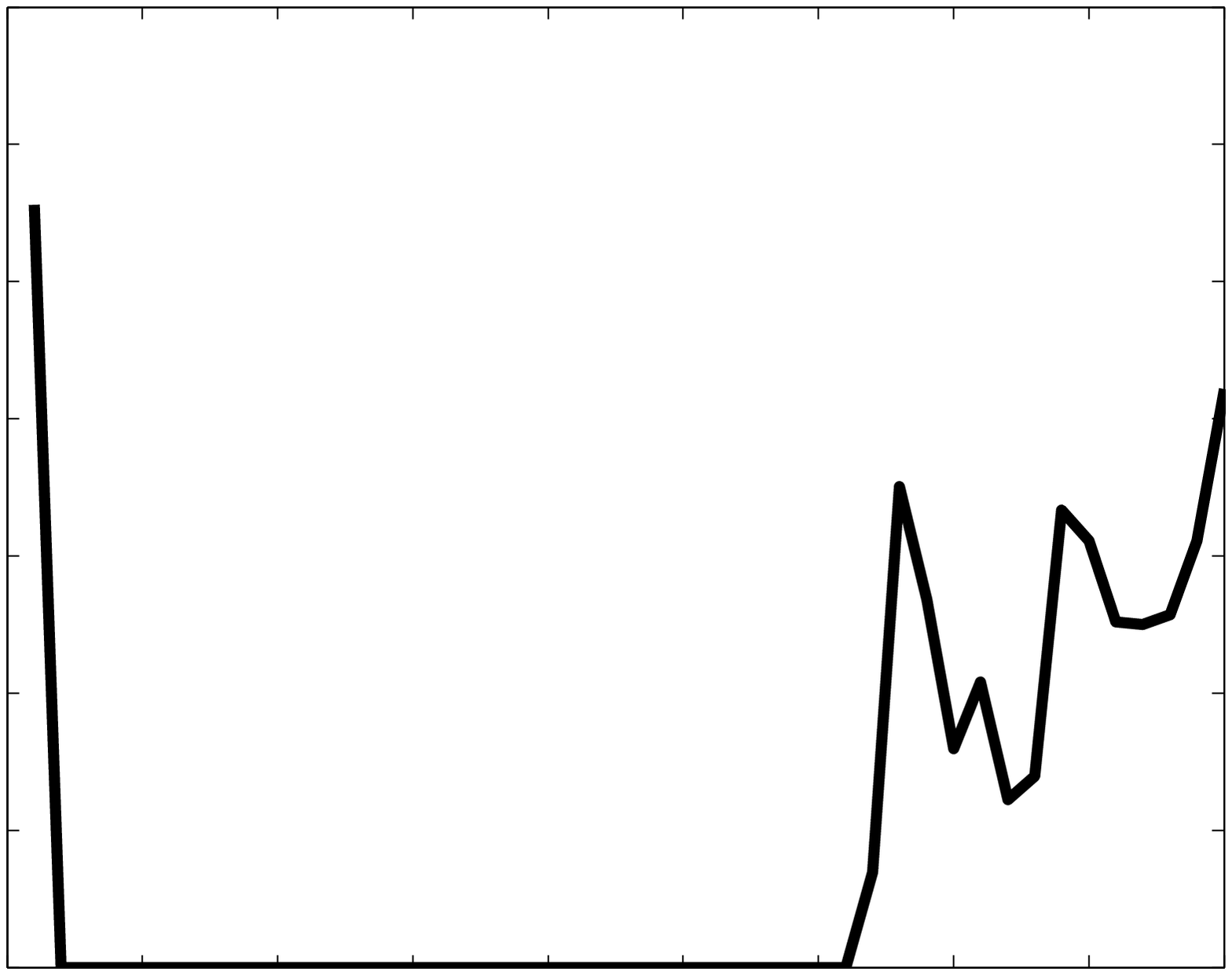} 	&\includegraphics[height=2cm]{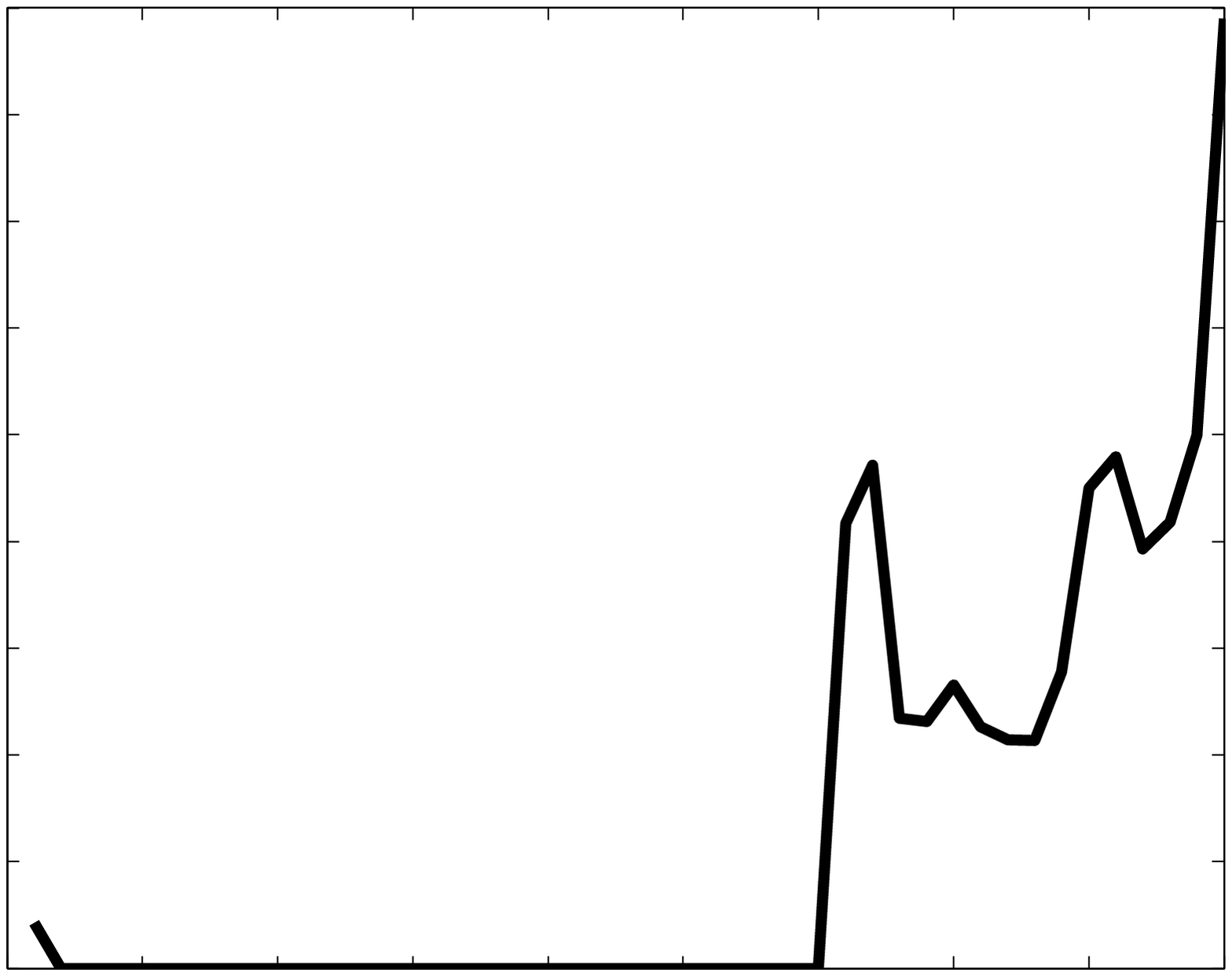}	\\
\hline
\end{tabular}
\caption{Shape classification and peak multiplicity for simulated and observed $\gamma$-ray light curves as defined in Table \ref{TabAssG}. 
The empty cell in correspondence of a shape class for a particular model indicates that this light-curve shape is not observed in the framework of this model.}
\label{TabShapeG}
\end{table*}

\begin{table*}
\centering
\begin{tabular}{| l | c | m{2.5cm} |}
\hline
\multicolumn{1}{|c|}{Shape Class}		&  Multiplicity 	&  \multicolumn{1}{c|}{Radio Core plus cone} 		       \\
\multicolumn{1}{|c|}{ 			  }		&              	 	&  \multicolumn{1}{c|}{and Radio LAT} 		       \\
\hline
\hline
 1- Bump							&   1 			&\includegraphics[height=2cm]{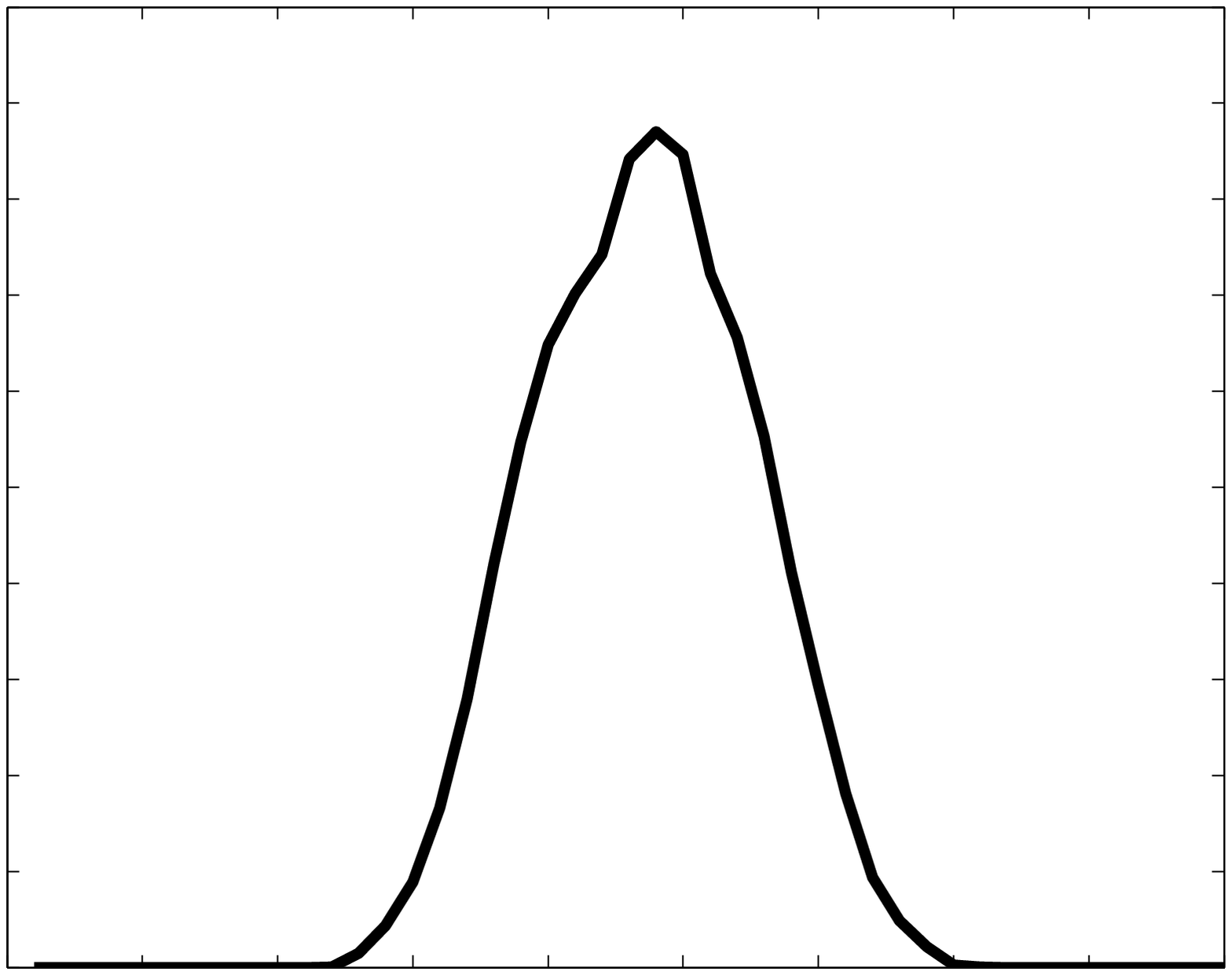}	\\
\hline
 2- Sharp 							&   1 			&\includegraphics[height=2cm]{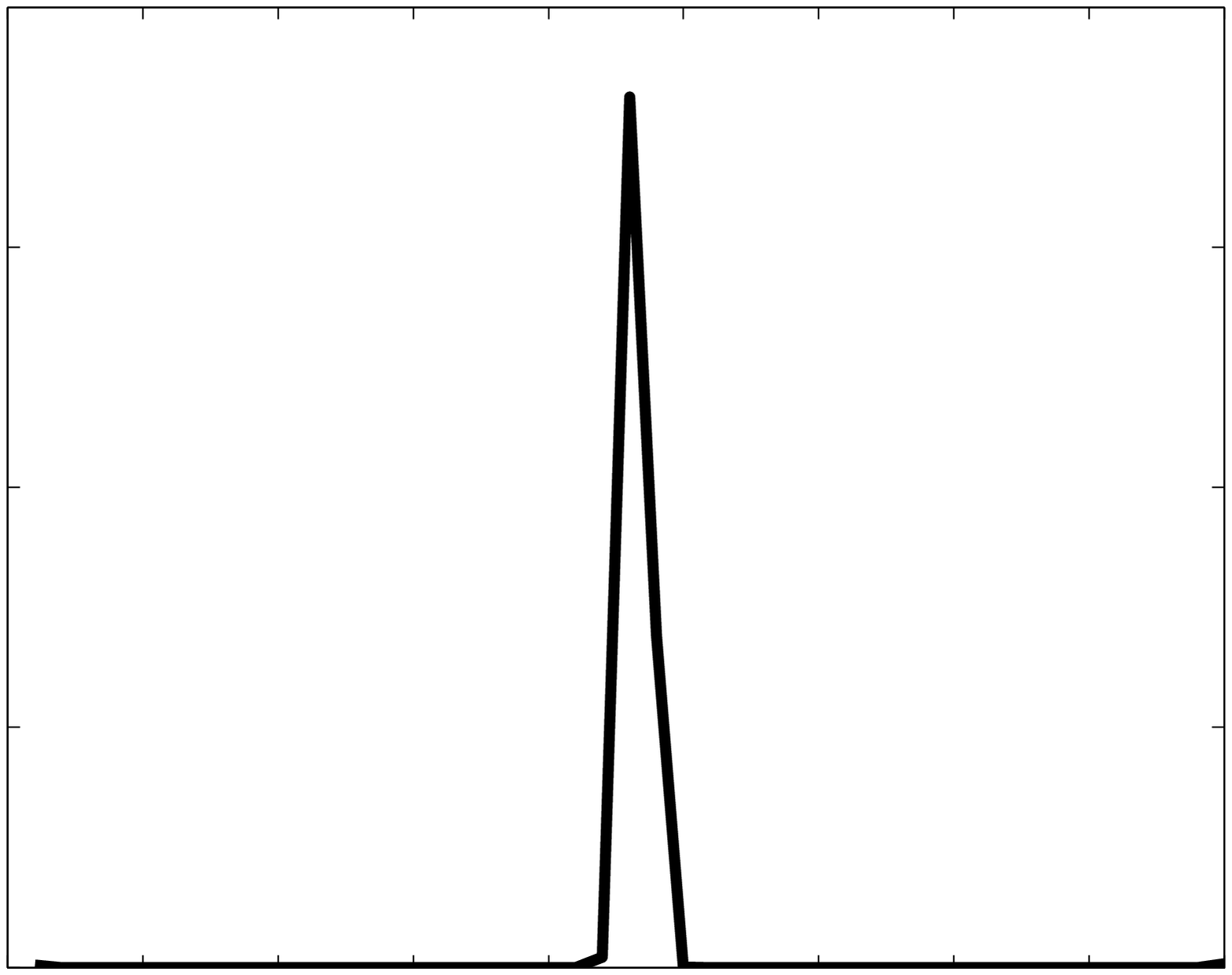} 	\\
\hline
 3- Two 							&   2 			&\includegraphics[height=2cm]{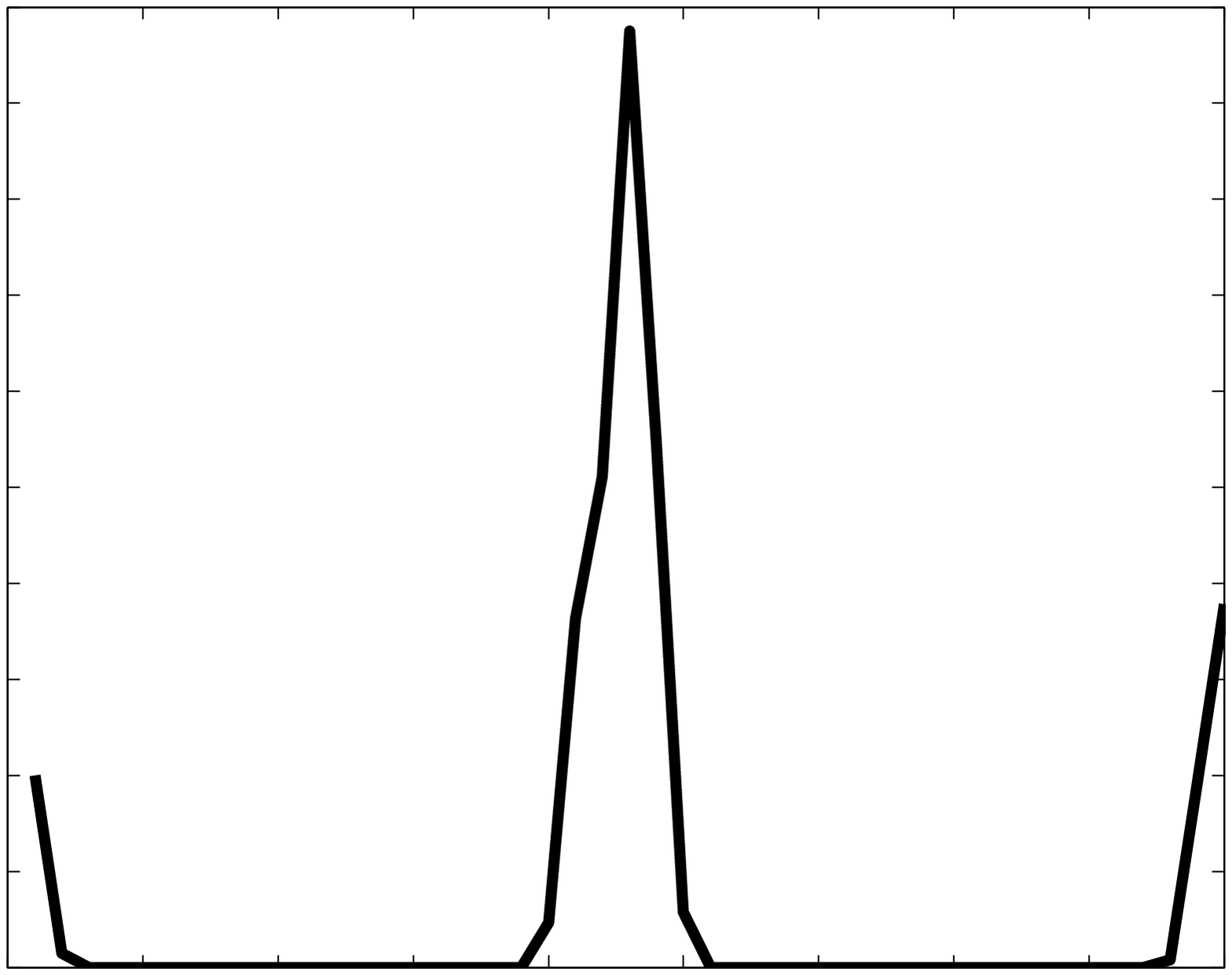} 	\\
\hline
 4- Double 						&   2 			&\includegraphics[height=2cm]{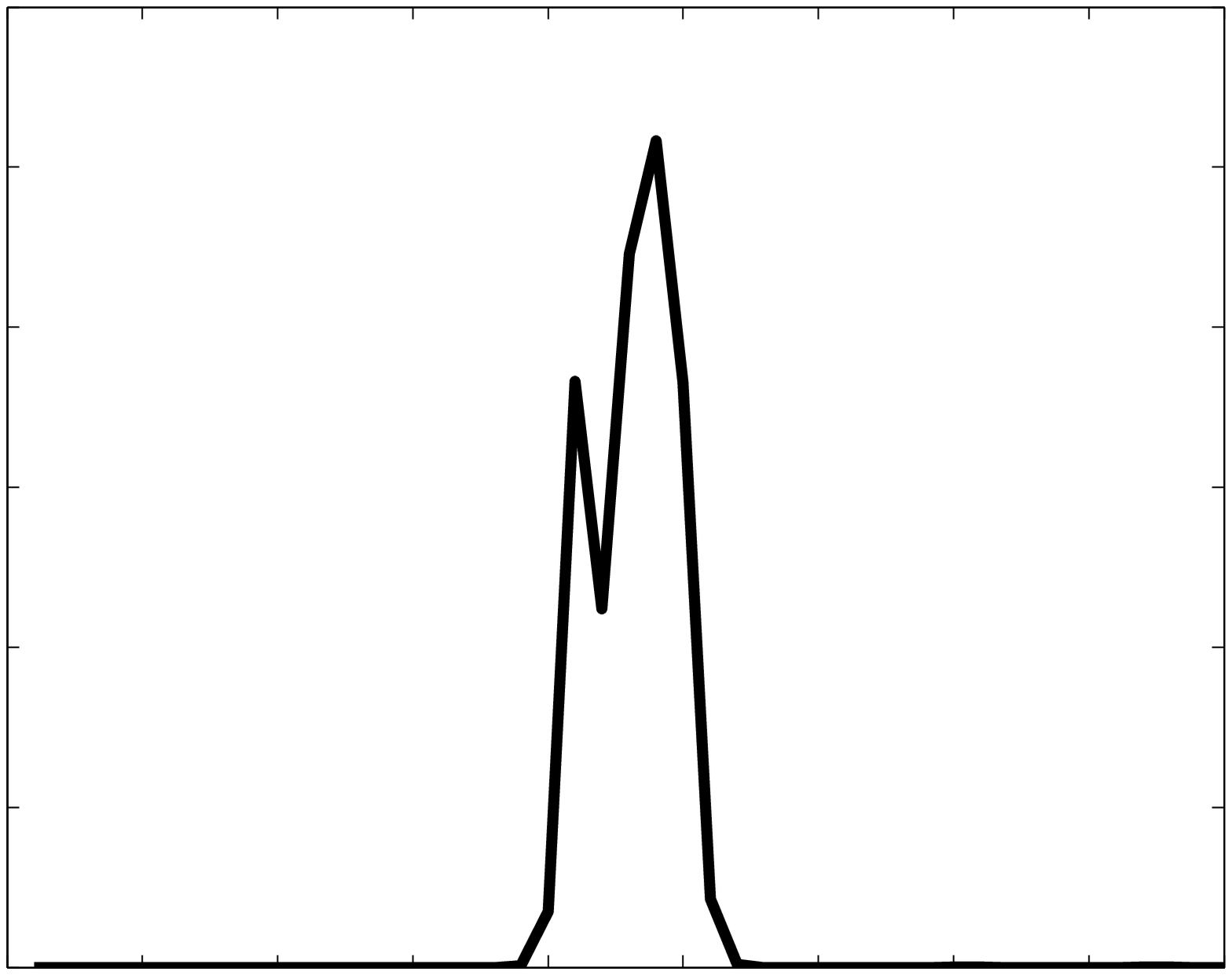} 	\\
\hline
 5- Double+Single					&   3 			&\includegraphics[height=2cm]{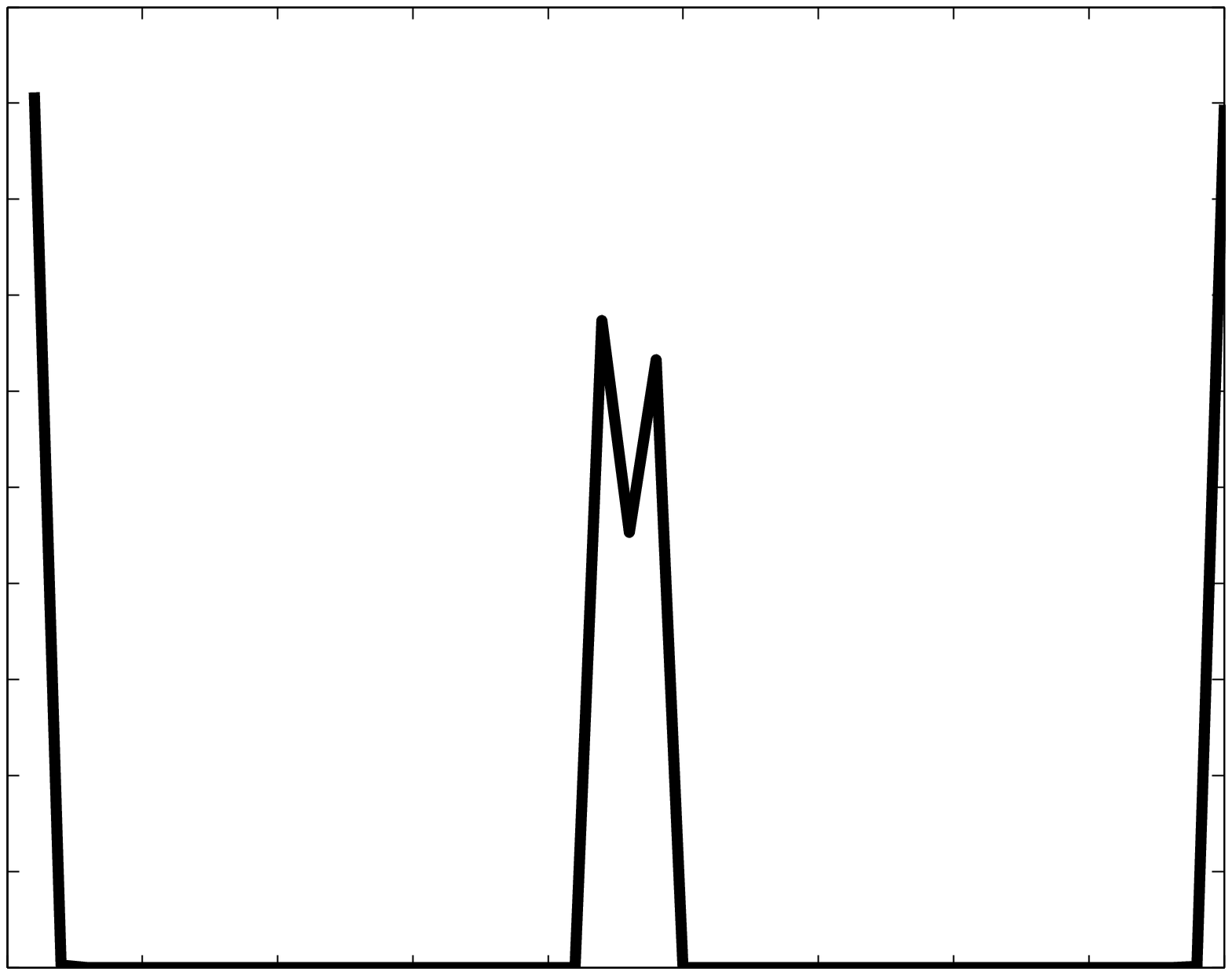} 	\\
\hline
 6- Two Double						&   4 			&\includegraphics[height=2cm]{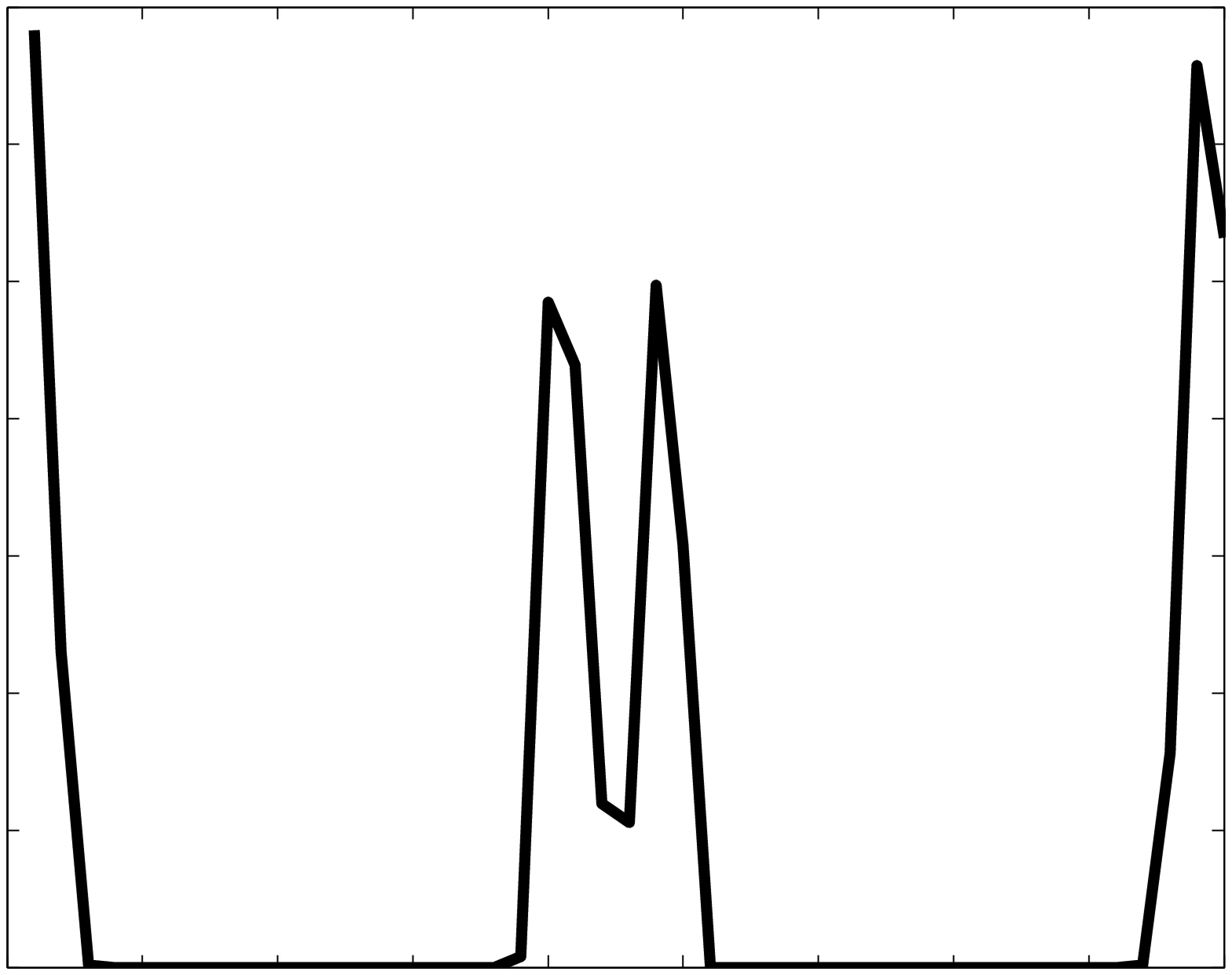} 	\\
\hline
 7- Triple 							&   3 			&\includegraphics[height=2cm]{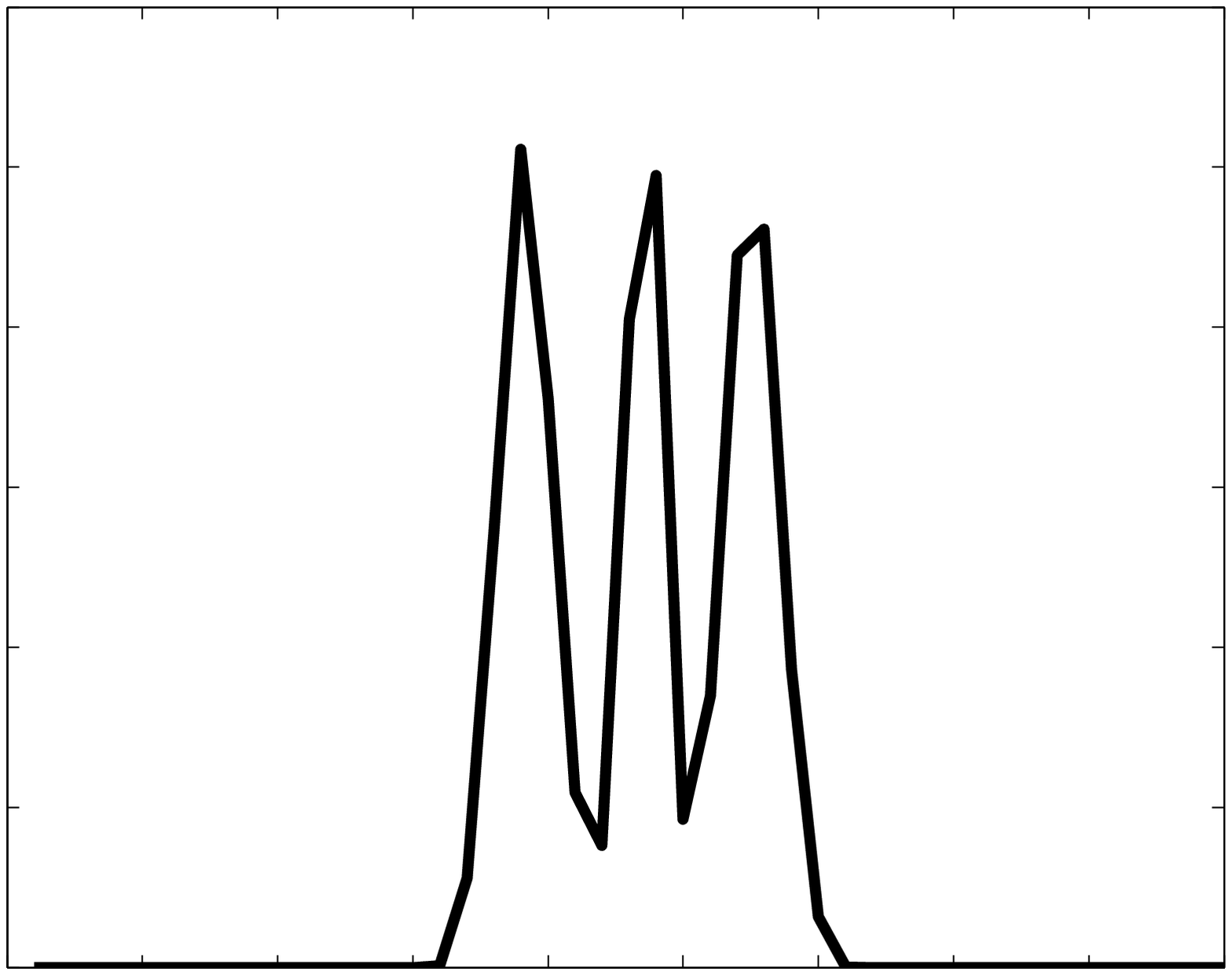} 	\\
\hline
 8- Three 							&   3 			&\includegraphics[height=2cm]{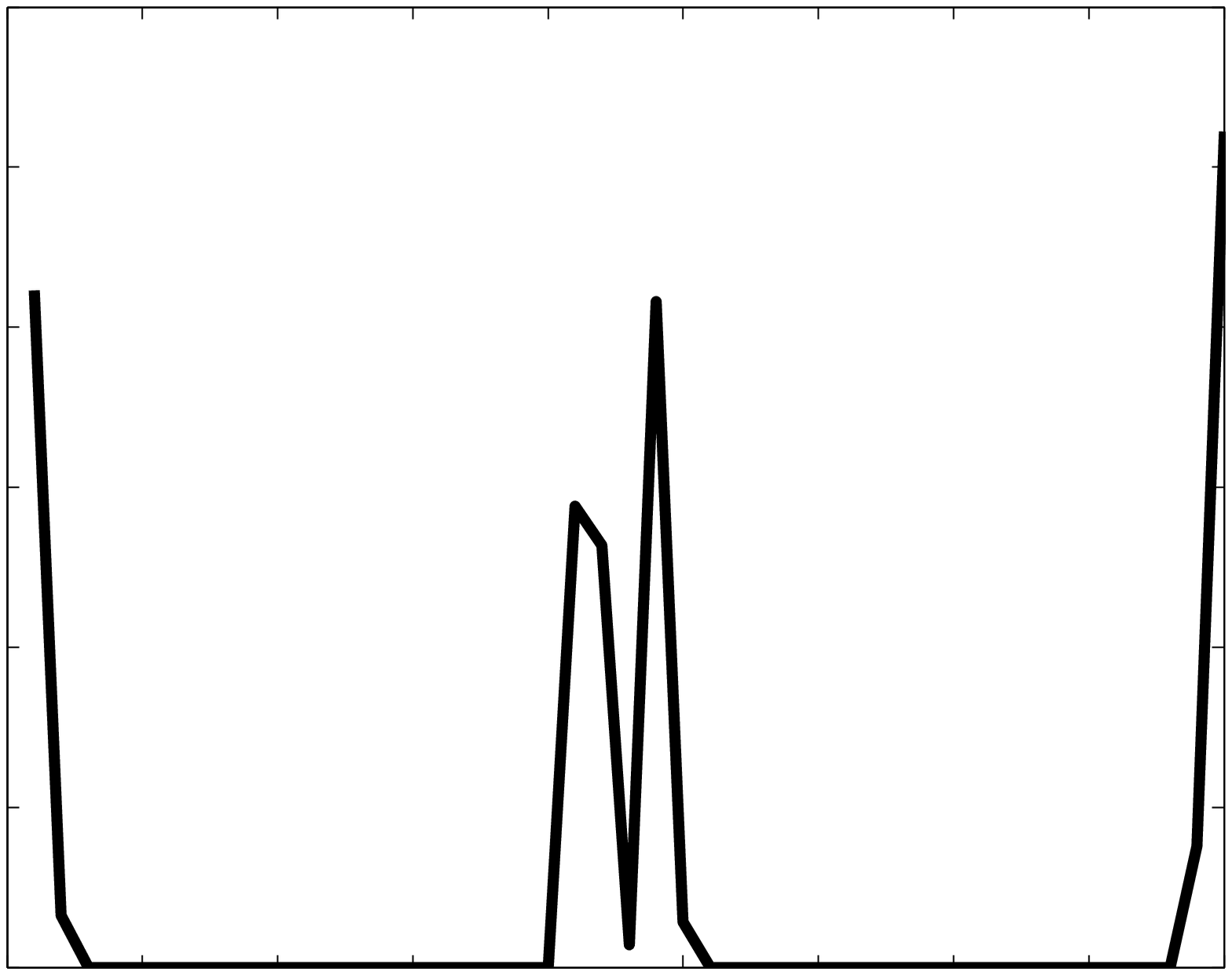} 	\\
\hline
 9- Two Triple 						&   4 			&\includegraphics[height=2cm]{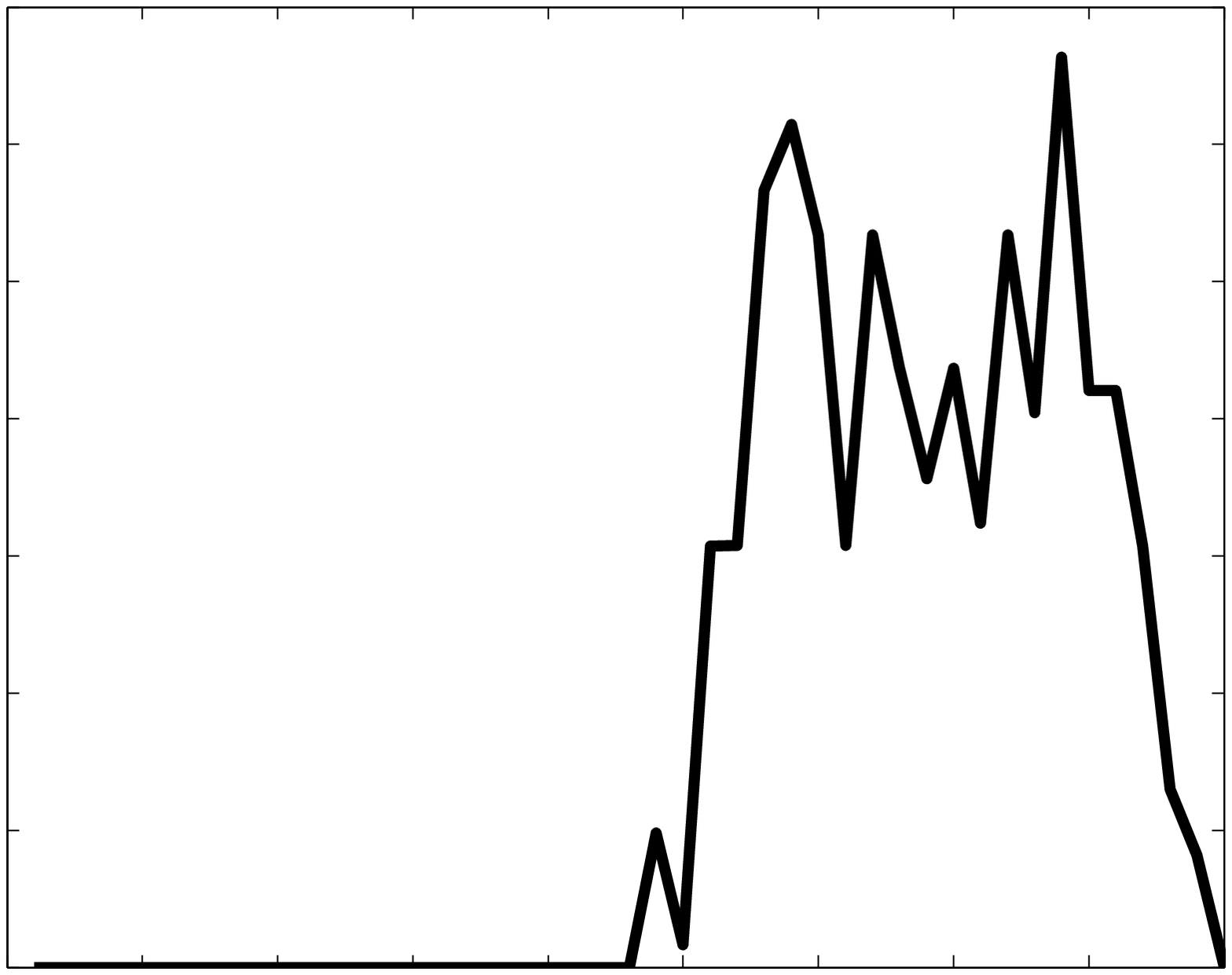} 	\\
\hline
\end{tabular}
\caption{Shape classification and peak multiplicity for simulated and observed radio light curves as defined in Table \ref{TabAssR}.}
\label{TabShapeR}
\end{table*}

\clearpage


\section{The pulsar $\gamma$-ray and radio emission pattern}
\label{PSREmispat}
\begin{figure*}[h!]
\begin{center}
\includegraphics[width=1.0\textwidth]{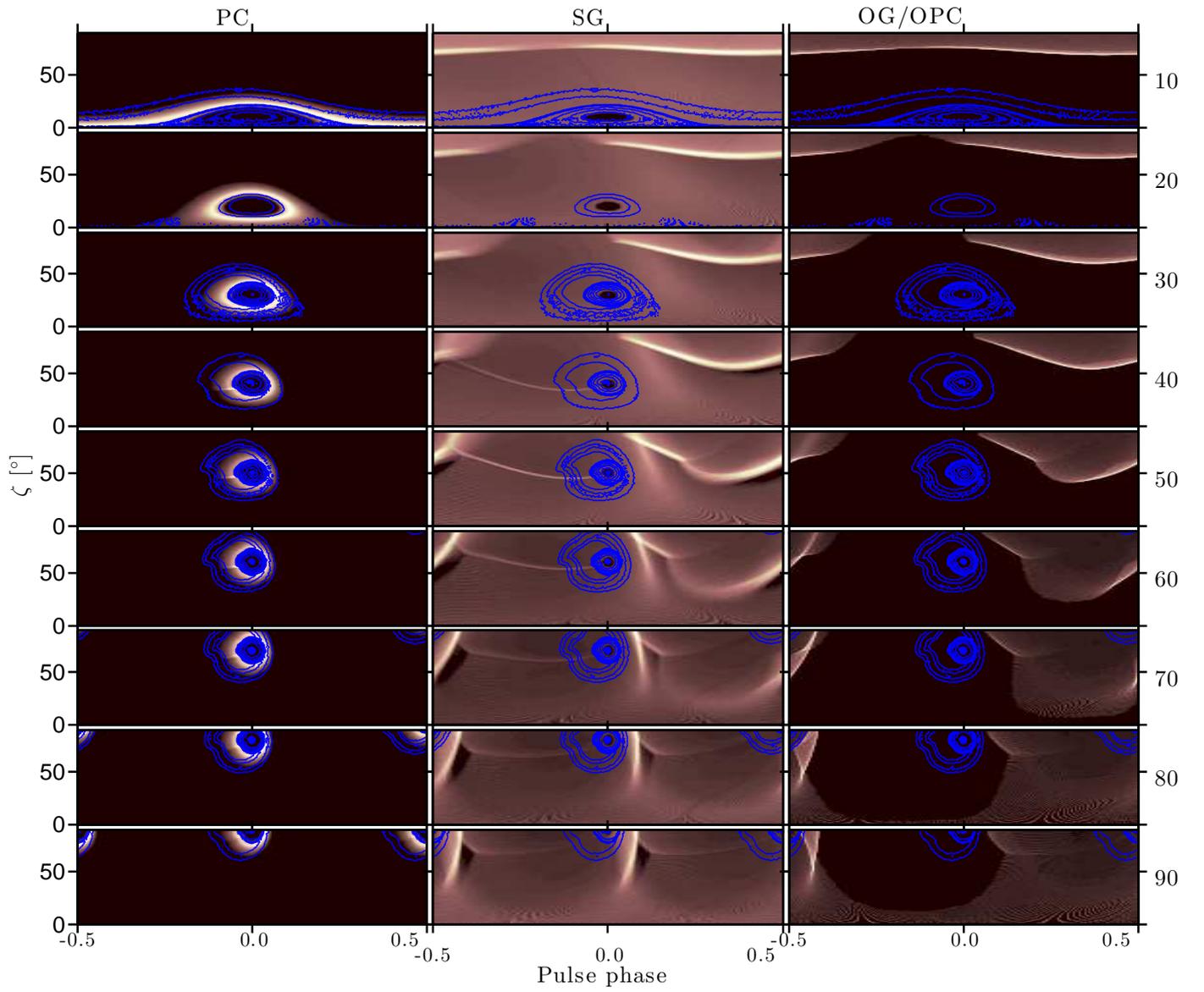} 
\caption{For each $\gamma$-ray emission model, the pulsar $\gamma$-ray (shaded surface) and radio (blue contours) emission patterns (phase plots) 
as a function of magnetic obliquity $\alpha$, are shown. Each phase-plot panel gives the pulsar light curve as a function of the observer line-of-sight 
$\zeta$ for each $\alpha$ value stepped every 10$^\circ$ in the interval $0^\circ<\alpha<90^\circ$. The $\gamma$-ray and radio phase plots have been
obtained for a magnetic field strength of 10$^8$ Tesla and spin period of 30 ms for the PC and radio cases, and gap widths of 0.04 and 0.01 for the 
SG and OG/OPC cases, respectively.}
\label{Em_Pat}
\end{center}
\end{figure*}

\clearpage


\section{Recurrence of $\gamma$-ray and radio shape classes}
\label{AppC}

\begin{table*}
\begin{center}
\begin{tabular}{| l || c  c || c  c || c  c || c  c |}
\hline
& \multicolumn{2}{c||}{PC} & \multicolumn{2}{c||}{SG}& \multicolumn{2}{c||}{OG}& \multicolumn{2}{|c|}{OPC}\\
\hline
& $D$ & $p_\mathrm{value}$ [\%] & $D$ & $p_\mathrm{value}$ [\%] & $D$ & $p_\mathrm{value}$ [\%] & $D$ & $p_\mathrm{value}$ [\%]  \\
\hline
\hline
Fig. \ref{RMulti}: Radio peak multiplicity  distribution							& 0.42 				& 3e-8				&\cellcolor{gray!60}0.37	& \cellcolor{gray!60}5e-6	& 0.45				& 1e-9				& \cellcolor{gray!30}0.417 	& \cellcolor{gray!30}4e-8 \\
\hline
Fig. \ref{G_HistoMultRLRQ}-left: RQ $\gamma$-ray peak  multiplicity distribution	& 0.72 				& 2.4 				& 0.38 				& 2e-2				& \cellcolor{gray!30}0.17	& \cellcolor{gray!30}33	& \cellcolor{gray!60}0.05 	& \cellcolor{gray!60}99.9\\
\hline
Fig. \ref{G_HistoMultRLRQ}-right: RL  $\gamma$-ray peak  multiplicity distribution	& \cellcolor{gray!30}0.36    				& \cellcolor{gray!30}1.8 	& 0.35 				& 6e-3				& 0.41 				& 1e-4				& \cellcolor{gray!60}0.22 	& \cellcolor{gray!60}2.6\\
\hline
Fig. \ref{G_HistoClas}-left: RQ shape-classes distribution  					& 0.72 				& 2.4 				& 0.38 				& 2e-2				& \cellcolor{gray!60}0.17	& \cellcolor{gray!60}33	& \cellcolor{gray!30}0.18 	& \cellcolor{gray!30}24\\
\hline
Fig. \ref{G_HistoClas}-right: RL shape-classes distribution 					& \cellcolor{gray!60}0.36    & \cellcolor{gray!60}1.8 	& \cellcolor{gray!30}0.40 	& \cellcolor{gray!30}3e-4	& 0.59 				& 2e-11				& 0.56 				& 2e-10\\\hline
\end{tabular}
\end{center}
\caption{Two-sample Kolmogorov-Smirnov statistics (2KS) and relative $p_\mathrm{value}$ between observed and simulated one-dimensional distributions 
shown in Figures \ref{RMulti} to \ref{G_HistoClas} and each model. The 2KS statistics ranges between 0 and 1 for distributions showing total 
agreement and total disagreement, respectively. The $p_\mathrm{value}$ is the probability to obtain observed 2KS value under 
the assumption that the two distributions are obtained from the same distribution (null hypothesis). This is equivalent to reject the null hypothesis at a 
confidence level of 100-($p_\mathrm{value}$). The 2KS test is described in Section \ref{Met2KS2}. The $D$ parameters relative to the first and second most 
consistent distributions are highlighted in dark grey and light grey cells, respectively}
\label{C2ksTab}
\end{table*}

\begin{figure}
\begin{center}
\includegraphics[width=0.49\textwidth]{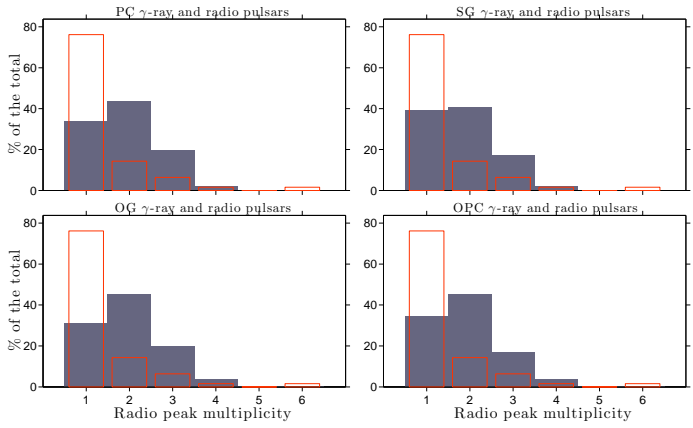} 
\caption{Recurrence of peak multiplicities for the simulated radio pulsar populations (grey) and LAT population (red) 
and each model. The radio peak multiplicities are defined in Table \ref{TabShapeG}.}
\label{RMulti}
\end{center}
\end{figure}

\begin{figure*}
\begin{center}
\includegraphics[width=0.49\textwidth]{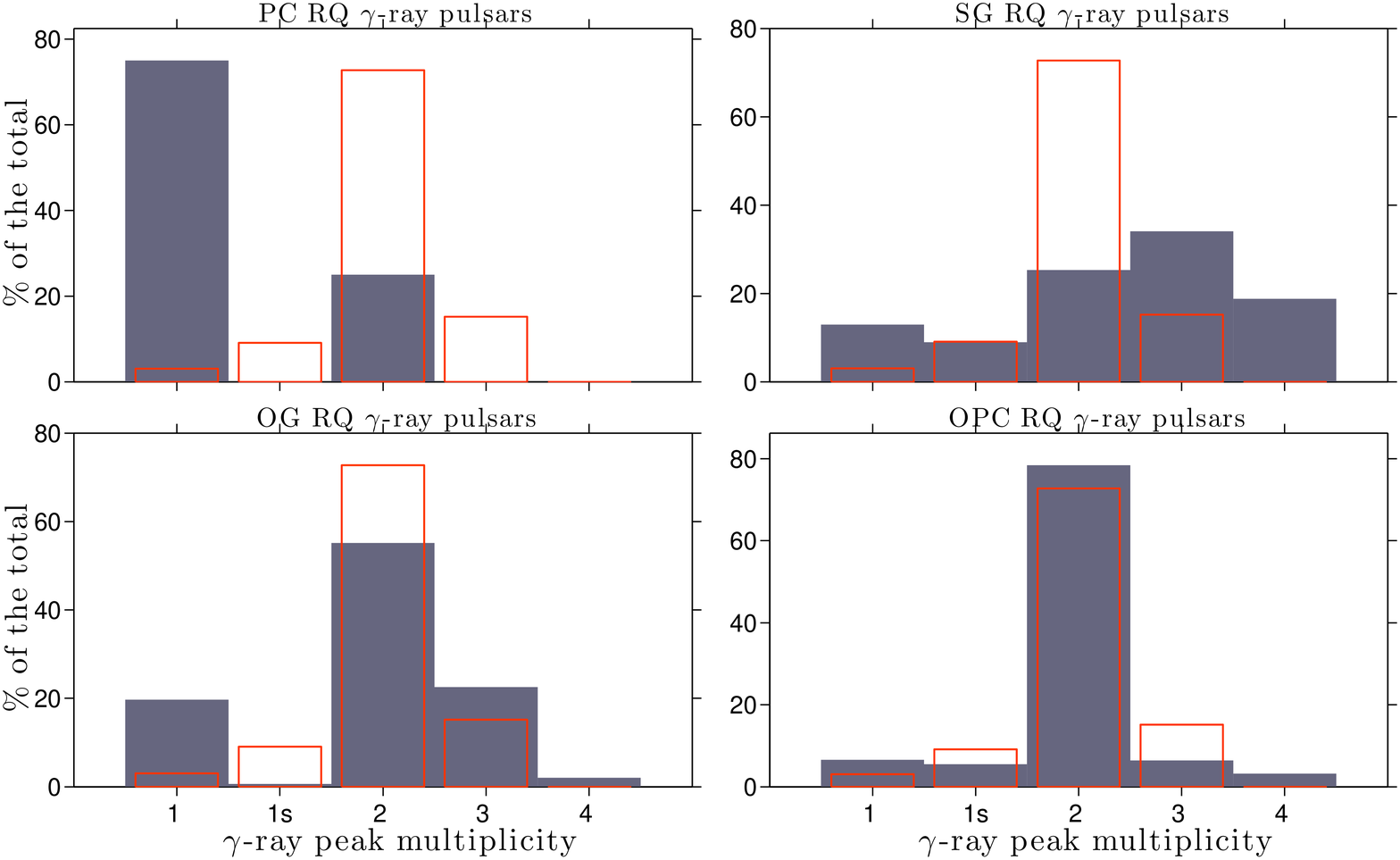} 
\includegraphics[width=0.49\textwidth]{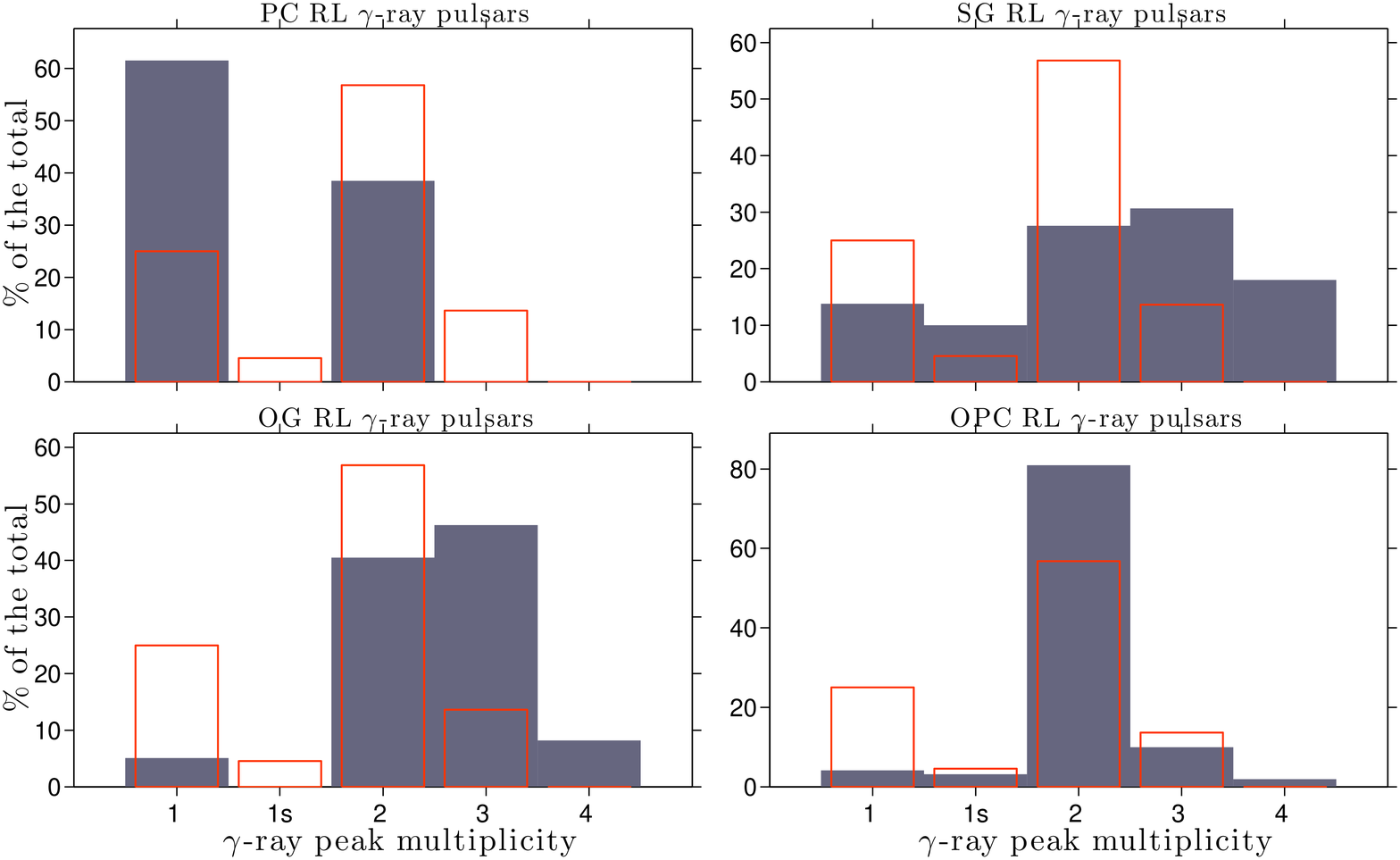} 
\caption{Recurrence of the $\gamma$-ray peak multiplicity for the RQ and RL pulsars of simulated (grey) and observed (red) 
populations and each model are shown in the top and bottom panels, respectively.}
\label{G_HistoMultRLRQ}
\end{center}
\end{figure*}

\begin{figure*}
\begin{center}
\includegraphics[width=0.49\textwidth]{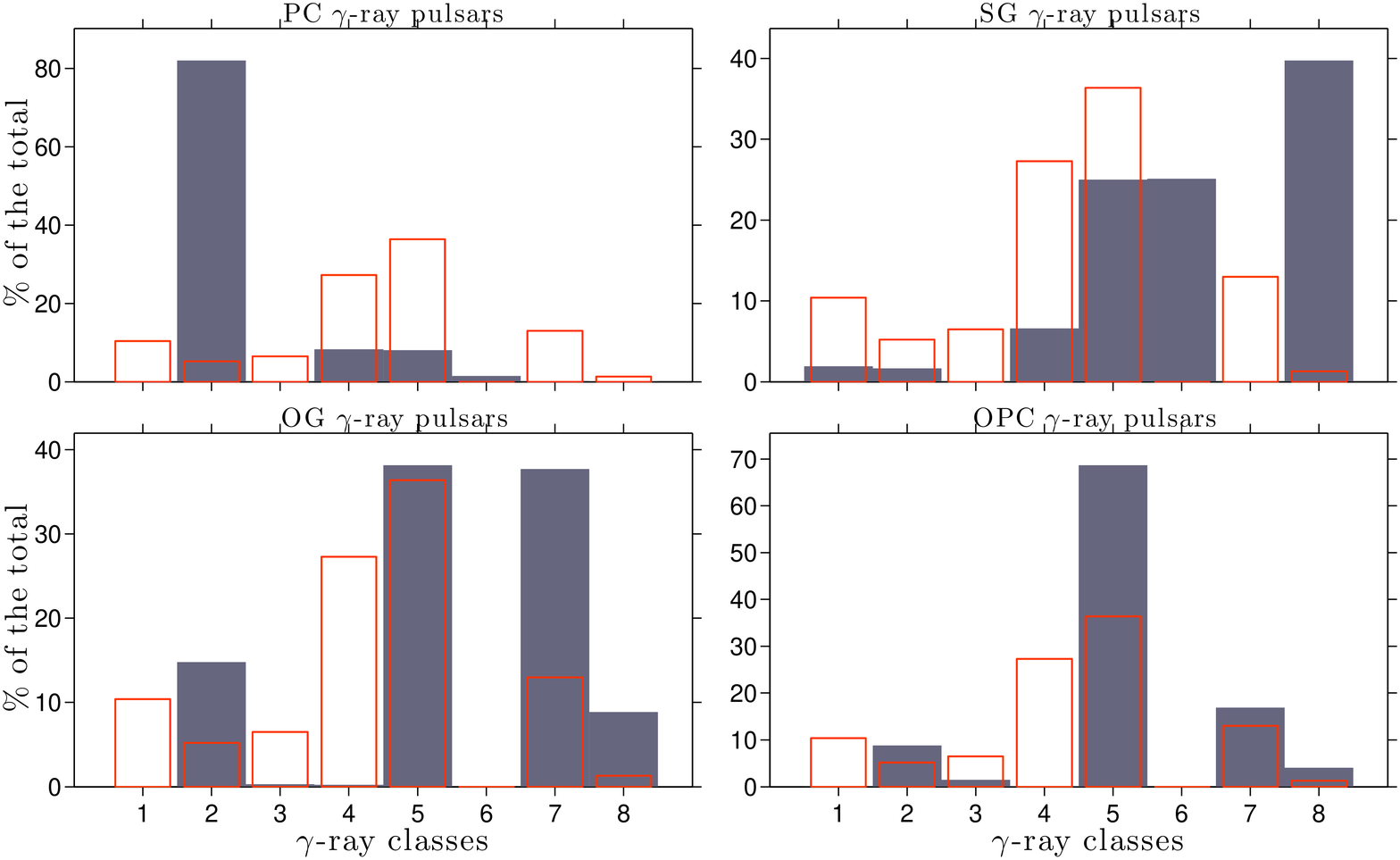} 
\includegraphics[width=0.49\textwidth]{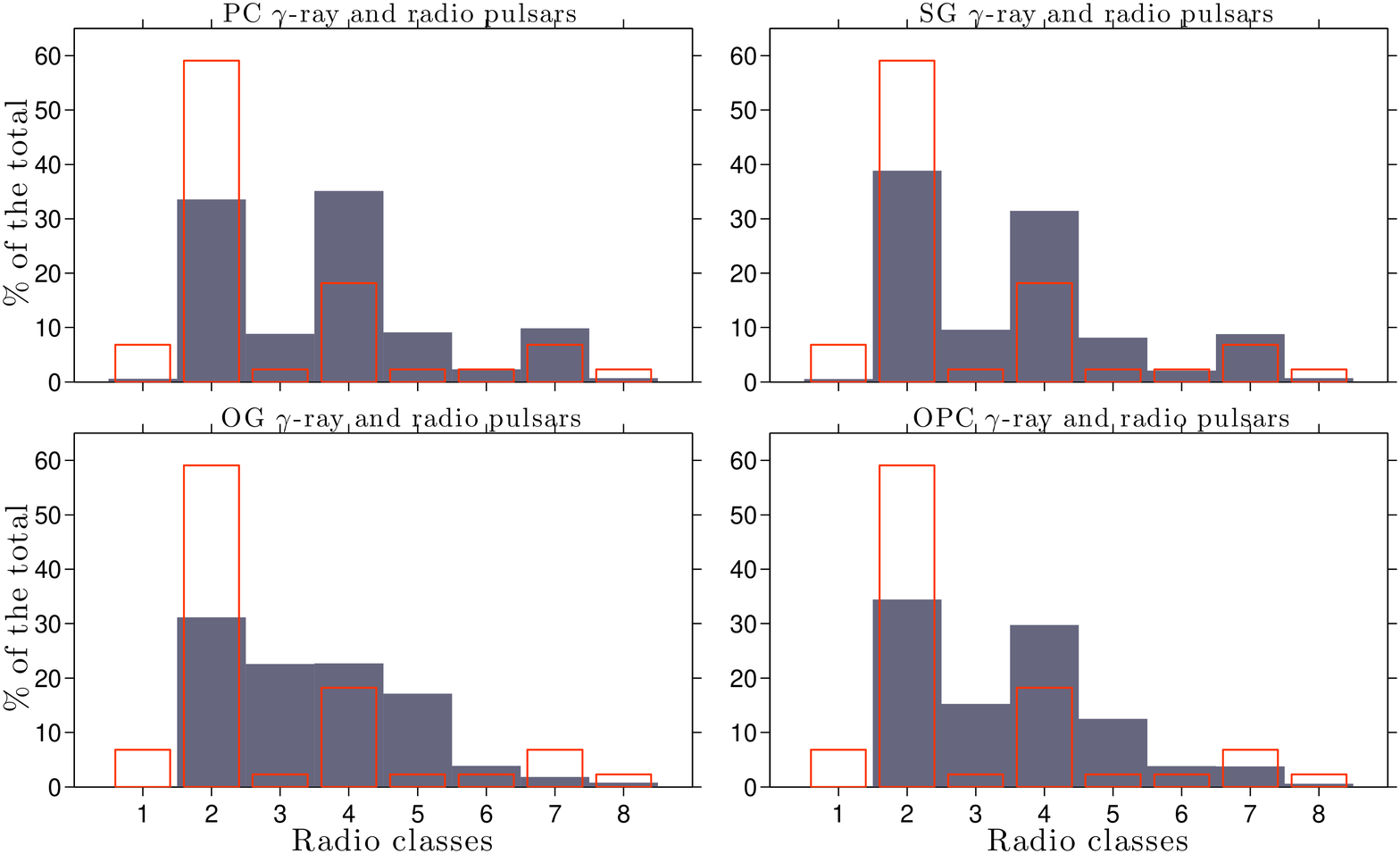} 
\caption{Recurrence shape classes for simulated $\gamma$-ray and radio pulsar populations (grey) and LAT population
(red) and each model are shown in the top and bottom panel, respectively.}
\label{G_HistoClas}
\end{center}
\end{figure*}

Figure \ref{RMulti}  compares simulated and observed multiplicities of $\gamma$-ray-selected radio profiles in the framework of each model. 
The radio model is unique, but the RL pulsars subsample changes with each model  $\gamma$-ray visibility generating different radio-peak multiplicity 
distributions.
Figure \ref{G_HistoMultRLRQ}  compares simulated and observed $\gamma$-ray peak multiplicity for RQ and RL  pulsars in the framework 
of each model. 
Figure \ref{G_HistoClas} compares the recurrence of the shape classes defined in Tables \ref{TabAssG} 
and \ref{TabAssR} and shown in Tables \ref{TabShapeG} and \ref{TabShapeR} for the simulated 
$\gamma$-ray and radio pulsars, respectively.
The statistical agreement between observed and simulated one-dimensional distributions shown in Figures \ref{RMulti} to \ref{G_HistoClas} are listed
in Table \ref{C2ksTab}.

\clearpage

\section{Multiplicities and peak separations as a function of $P$ and $w$ in the $\alpha$-$\zeta$ plane}
\label{AppD}

\begin{figure*}
\begin{center}
\includegraphics[width=0.85\textwidth]{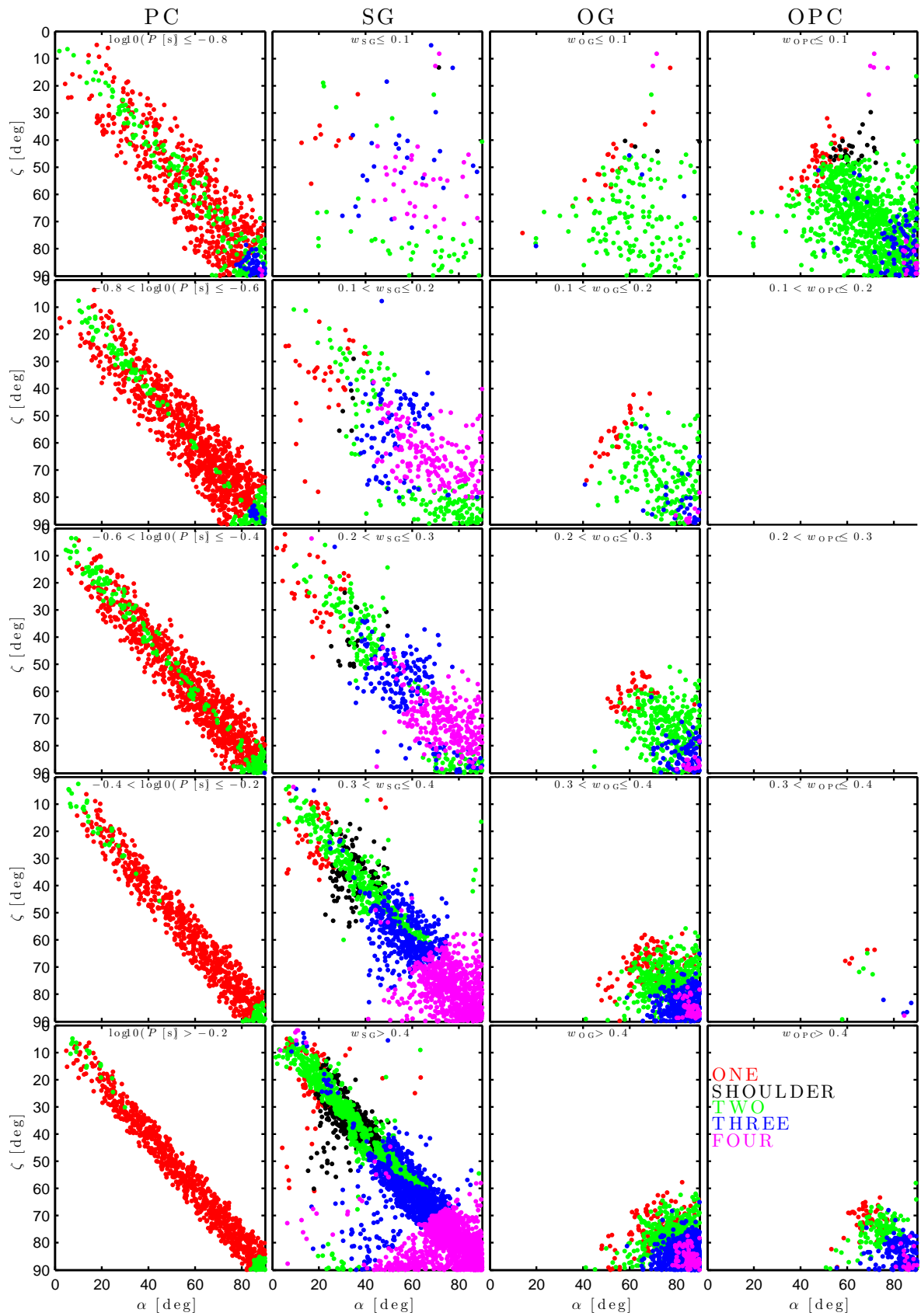}
\caption{Peak multiplicity as a function of spin period (PC) and gap width (SG, OG, OPC) in the $\alpha$-$\zeta$ plane.
Left to right columns refer to PC, SG, OG, and OPC, respectively while top to bottom rows refer to increasing period (PC) 
and gap width (SG, OG, OPC). This figure can be compared with similar figures of \cite{wrwj09}.}
\label{MultiAZ}
\end{center}
\end{figure*}

\begin{figure*}
\begin{center}
\includegraphics[width=0.85\textwidth]{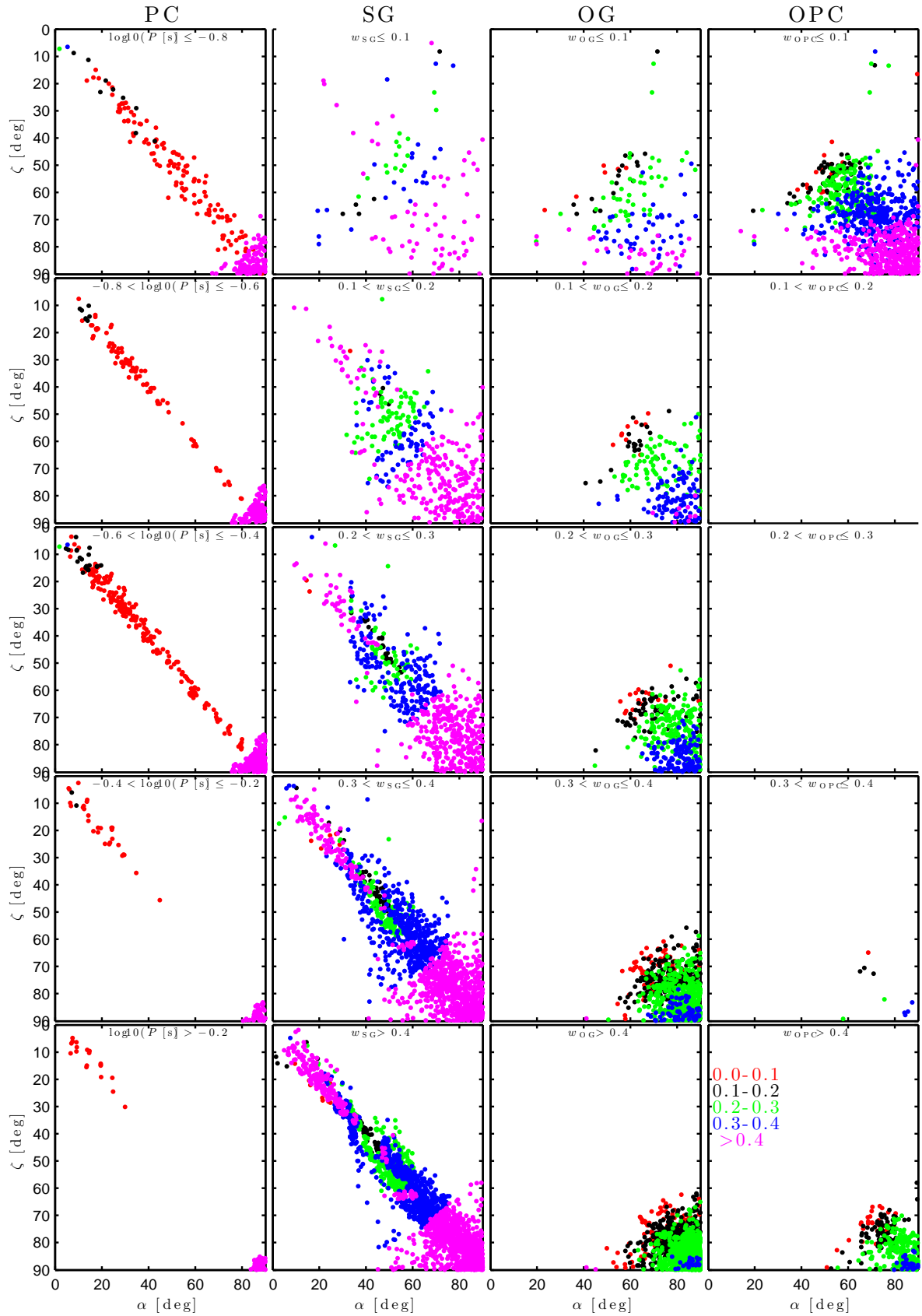}
\caption{Peak separation as a function of spin period (PC) and gap width (SG, OG, OPC) in the $\alpha$-$\zeta$ plane.
Left to right columns refer to PC, SG, OG, and OPC, respectively while top to bottom rows refer to increasing period (PC) 
and gap width (SG, OG, OPC). This figure can be compared with similar figures of \cite{wrwj09}.}
\label{SepAZ}
\end{center}
\end{figure*}

\clearpage


\section{Two-sample Kolmogorov-Smirnov test}
\label{Met2KS2}

The two-sample Kolmogorov-Smirnov test (2KS) is a non-parametric test that allows statistical quantification of the degree 
of agreement of two one-dimensional distributions. By computing the maximum distance between the cumulative density
functions (CDFs) of the tested distributions, the 2KS test allows for the rejection of the null hypothesis: that the distributions are obtained 
from the same underlying distribution at a predefined confidence level (CL). We consider two samples, 
$A$ and $B$, of sizes $a$ and $b$, respectively. If $A(x)$ and $B(x)$ are the CDF of the samples A and B, respectively, the 
2KS statistics is given by
\begin{equation}
\label{kse1}
D_{a,b} = \sup_x |A(x)-B(x)|
\end{equation}
where $\sup$ is the supremium function. The null hypothesis that the samples A and B have been obtained from the same 
underlying distribution is rejected at CL $p_\mathrm{value}$ if
\begin{equation}
\label{kse2}
D_{a,b} > c(p_\mathrm{value}) \sqrt{\frac{a+b}{ab}}
\end{equation}
with $c(p_\mathrm{value})$ a tabulated value function of the chosen CL. The D value is included between 0 and 1 indicating total 
agreement and total disagreement between the tested distributions, respectively. The $p_\mathrm{value}$ gives the probability to
observe the relative statistics $D_{a,b}$ assuming the null hypothesis to be true.
The 2KS statistic and $p_\mathrm{value}$ given in Table \ref{Tab4} and relative to one-dimension observed and simulated distributions 
shown in Figures \ref{G_VisHistoMulti} to \ref{RadLagHist} have been computed with Equations \ref{kse1} and \ref{kse1}, respectively.

\subsection{Using the 2KS to quantify the agreement between two two-dimensional distributions}
\label{ksapp}

We used the two-dimensional version of the 2KS statistics (2D-2KS) described in \cite{ptvf92} to quantify the agreement 
between observed/estimated and simulated two-dimensional distributions shown in Figures \ref{G_PeakSepBeam} to \ref{RadLagPksepG}. 
The 2D-2KS statistics described in \cite{ptvf92} gives a reasonable estimate of the distance between simulated and observed 2D 
distributions but, given the high number of possible ways to sort the 2D data plane, the 2D-2KS statistics is not distribution 
free as its 1D version, and cannot be associated with the probability that observations and simulations are obtained from the same 
parent distribution ($p_\mathrm{value}$). In order to evaluate the $p_\mathrm{value}$ associated with the 2D-2KS statistics, we 
studied the statistical distribution of the 2D-2KS statistics by paired bootstrap resampling of the observed 2D sample, 
and use it to evaluate the $p_\mathrm{value}$ corresponding to each 2D-2KS statistics.

The paired bootstrap resampling consists in resampling the observed 2D distribution by randomly rearrange the 
observed values on the 2D plane. It is possible to build a high number of bootstrapped distribution, statistically consistent with 
the observed 2D distribution, and use them to study the statistical distribution of a parameter of the observed 2D distribution (e.g.
the variance). An accurate description of the paired bootstrap resampling can be found in \cite{br93a}. 

We used the paired bootstrap method to resample each observed 2D distribution and create a large number of bootstrapped 
samples. We computed the 2D-2KS statistics between each bootstrapped sample and the simulated sample in object, and we built the 
statistical distribution of their 2D-2KS statistics for each model. Once we evaluated the statistical distribution of the 
2D-2KS statistics, we used it to read off the probability $p_\mathrm{value}$ corresponding to the 2D-2KS statistics computed 
between the simulated and observed/estimated 2D distribution in the object.

\end{document}